\documentclass[sn-mathphys]{sn-jnl}

\jyear{2022}

\newcommand{\ARIEL}{\textit{Ariel}}
\newcommand{\JWST}{\textit{JWST}}

\begin{document}

\title[Detecting molecules: \ARIEL\ Tier 1]{Detecting molecules in \ARIEL\ low resolution transmission spectra}

\author*[1]{\fnm{Andrea} \sur{Bocchieri}}\email{andrea.bocchieri@uniroma1.it}

\author[1,2,3]{\fnm{Lorenzo V.} \sur{Mugnai}}

\author[1]{\fnm{Enzo} \sur{Pascale}}

\author[4,3]{\fnm{Quentin} \sur{Changeat}}

\author[3]{\fnm{Giovanna} \sur{Tinetti}}

\affil*[1]{\orgdiv{Department of Physics}, \orgname{La Sapienza Università di Roma}, \orgaddress{\street{Piazzale Aldo Moro 2}, \city{Roma}, \postcode{00185}, \country{Italy}}}

\affil[2]{\orgdiv{INAF}, \orgname{Osservatorio Astronomico di Palermo}, \orgaddress{\street{Piazza del Parlamento 1}, \city{Palermo}, \postcode{I-90134}, \country{Italy}}}

\affil[3]{\orgdiv{Department of Physics and Astronomy}, \orgname{University College London}, \orgaddress{\street{Gower Street}, \city{London}, \postcode{WC1E 6BT}, \country{UK}}}

\affil[4]{\orgdiv{European Space Agency (ESA), ESA Office}, \orgname{Space Telescope Science Institute (STScI)}, \orgaddress{\city{Baltimore}, \postcode{MD 21218}, \country{USA}}}

\abstract
The \ARIEL\ Space Mission aims to observe a diverse sample of exoplanet atmospheres across a wide wavelength range of 0.5 to 7.8 microns. The observations are organized into four Tiers, with \textit{Tier 1} being a reconnaissance survey. This Tier is designed to achieve a sufficient signal-to-noise ratio (S/N) at low spectral resolution in order to identify featureless spectra or detect key molecular species without necessarily constraining their abundances with high confidence. We introduce a \textit{P}-statistic that uses the abundance posteriors from a spectral retrieval to infer the probability of a molecule's presence in a given planet's atmosphere in Tier 1. We find that this method predicts probabilities that correlate well with the input abundances, indicating considerable predictive power when retrieval models have comparable or higher complexity compared to the data. However, we also demonstrate that the \textit{P}-statistic loses representativity when the retrieval model has lower complexity, expressed as the inclusion of fewer than the expected molecules. The reliability and predictive power of the \textit{P}-statistic are assessed on a simulated population of exoplanets with H$_2$-He dominated atmospheres, and forecasting biases are studied and found not to adversely affect the classification of the survey.



\maketitle

\section{Introduction}
\label{sec:introduction}
During the past decade, the number of exoplanet discoveries has increased exponentially, bringing the total number of confirmed exoplanets to more than 5000 by mid-2022. Numerous space missions are contributing to the effort of detecting new exoplanets, such as Kepler~\citep{KeplerBorucki2011, KeplerBatalha2013}, TESS~\citep{Ricker2016}, CHEOPS~\citep{Cessa2017}, PLATO~\citep{Rauer2014}, GAIA~\citep{GAIA2016}, together with ground instrumentation such as HARPS~\citep{Mayor2003}, WASP~\citep{Pollacco2006}, KELT~\citep{pepper2007}, and OGLE~\citep{Udalski2015}. Over time, the field emphasis has gradually expanded from the determination of bulk planetary parameters to the search for a deeper understanding of the true nature of exoplanets and their formation-evolution histories.

Multiband photometry and spectroscopy of transiting exoplanets are currently the most promising techniques for characterizing the composition and thermodynamics of exoplanet atmospheres~\citep{Seager2000, Charbonneau2005, Tinetti2007, Madhusudhan2012, Tinetti2013, Kreidberg201, Sing2016, Line2016, Tsiaras2018, Evans_2018, pinhas2019, Welbanks2019, Mikal_Evans2020, Pluriel2020A, ares1, ares2, ares3, ares4, ares5, PC_Changeat2022}, as they allow us to effectively separate the signal of the planet from that of its host star. Observations in the near- to mid-infrared can probe the neutral atmospheres of exoplanets to study the signal from the rovibrational transitions of molecules~\citep{Tinetti2013, Encrenaz2015}.

Current instrumentation has enabled this kind of atmospheric characterization only for a few tens of planets orbiting close to their host stars over a limited wavelength range~\citep[e.g.][]{Sing2016, Tsiaras2018, Tsiaras2019, 5keyChangeat2022}. A considerable contribution to exoplanetary science will come from the James Webb Space Telescope (\JWST), launched in December 2021~\citep{Greene2016}, and \ARIEL. \JWST\ provides broadband spectroscopy in the range of $0.6$ to $28.5$ micron of the electromagnetic spectrum, sufficient to detect all molecular species~\citep{Encrenaz2015, ERS_WASP39b_JWST_NIRISS, ERS_WASP39b_JWST_NIRCam, ERS_WASP39b_JWST_NIRSpec_PRISM, ERS_WASP39b_JWST_NIRSpec_G395H, SO2_JWST_WASP39b}.

\subsection{Ariel and its Tiers}
\label{subsec:ariel-and-its-tiers}

The Atmospheric Remote-Sensing Infrared Exoplanet Large-survey, \ARIEL, will launch in $2029$ as the M$4$ ESA mission of the Cosmic Vision program~\citep[Ariel Definition Study Report\footnote{\scriptsize{https://sci.esa.int/web/ariel/-/ariel-definition-study-report-red-book}}]{Tinetti2018}. \ARIEL\ will conduct the first unbiased survey of a statistically significant sample of approximately $1000$ transiting exoplanet atmospheres in the $0.5$--$7.8$ $\mu m$ wavelength range. Three photometers (VISPhot, $0.5$--$0.6$ $\mu m$; FGS1, $0.6$--$0.80$ $\mu m$; FGS2, $0.80$--$1.1$ $\mu m$) and three spectrometers (NIRSpec, $1.1$--$1.95$ $\mu m$ and R $\geq$ 15; AIRS-CH0, $1.95$--$3.9$ $\mu m$ and R $\geq$ $100$; AIRS-CH1, $3.9$--$7.8$ $\mu m$ and R $\geq$ $30$), provide simultaneous coverage of the whole spectral band. This broad spectral range encompasses the emission peak of hot and warm exoplanets and the spectral signatures of the main expected atmospheric gases such as H$_2$O, CO$_2$, CH$_4$, NH$_3$, HCN, H$_2$S, TiO, VO~\citep[e.g.][]{Tinetti2013, Encrenaz2015}. \ARIEL\ will allow us to comprehensively understand the formation-evolution histories of exoplanets as well as to extend comparative planetology beyond the boundary of the Solar System.

After each observation, the resulting spectrum from each spectrometer is binned during data analysis to optimize the signal-to-noise ratio (S/N). Therefore, by implementing different binning options, the mission will adopt a four-Tier observation strategy, expected to produce spectra with different S/N to optimize the science return. Tier 1 is a shallow reconnaissance survey created to perform transit and eclipse spectroscopy on all targets to address questions for which a large population of objects needs to be observed. Tier 1 spectra have S/N $\geq$ $7$ when raw spectra are binned into a single spectral point in NIRSpec, two in AIRS-CH0, and one in AIRS-CH1, for a total of seven effective photometric data points. A subset of Tier 1 planets will be further observed to reach S/N $\geq$ $7$ at higher spectral resolution in Tier 2 and Tier 3 for detailed chemical and thermodynamic characterization of the atmosphere. Tier 4 is designed for bespoke or phase-curve observations~\citep{Edwards2019}.

\subsection{Detecting molecules in Tier 1 spectra}
\label{subsec:detecting-molecules-in-tier-1-spectra}

Among the main goals of Tier 1 observations is to identify planetary spectra that show no molecular absorption features (because of clouds or compact atmospheres) and to select those to be reobserved in higher Tiers for a detailed characterization of their atmospheric composition and thermodynamics. Tier 1 observations, however, have a much richer information content even though the combination of S/N and spectral resolution might not be adequate to constrain chemical abundances with high confidence using retrieval techniques. 

Adapting existing data analysis techniques or developing new methodologies can be essential to extract all relevant information from the Tier 1 data~set. In a previous study,~\cite{Mugnai2021a} were successful in demonstrating, using color-color diagrams, that Tier 1 observations can be used to infer the presence of molecules in the atmospheres of gaseous exoplanets, independently from planet parameters such as mass, size, and temperature. However, their method has an estimator bias that depends on the magnitude of the instrumental noise; a detailed characterization of instrumental uncertainties is required to remove the estimator bias before it can be used for quantitative predictions. In this follow-up paper, we develop a new method that is both reliable and unbiased to address the following question: \textit{can we use Tier 1 transmission spectra to identify the presence of a molecule, with an associated calibrated probability?}. Hence, these calibrated probabilities can also be used to inform the decision-making process to select Tier 1 targets for re-observation in \ARIEL's higher Tiers for detailed characterization.   

Section~\ref{sec:met} outlines the methodology used in this analysis. Section~\ref{met:dataanstrat} describes our data analysis strategy for detecting a molecule in these spectra. Section~\ref{met:exp} details our experimental data set, including the planetary population, forward model parameters, atmosphere randomization, and noise estimation. Section~\ref{met:retsetup} summarizes the spectral retrievals performed, discussing the optimization algorithm and the priors used. Section~\ref{met:dataantools} describes the data analysis tools used to evaluate the probability forecasts of the method. Section~\ref{sec:res} details the results obtained in terms of forecast reliability (Section~\ref{res:detrel}), predictive power (Section~\ref{res:predict}), and bias of the abundance estimator utilized (Section~\ref{res:bias}). Finally, Section~\ref{sec:disc} discusses all the results, and Section~\ref{sec:conc} summarizes the main conclusions of this analysis.

\section{Methods}
\label{sec:met}
Tier 1 transmission spectra contain sufficient information to infer the presence of several atmospheric molecules \citep{Mugnai2021a}, but Tier 1 observations are in general non-ideal for quantitative spectral retrievals in terms of molecular abundances, as they are required to achieve a S/N $\geq 7$ when binned in only seven effective photometric data points in the $0.5$–$7.8$ $\mu m$ wavelength range~\citep{Edwards2019}. Abundance posterior probabilities from retrievals can however still be informative and here we develop a new method to identify the presence of molecules in Tier 1 transmission spectra starting from these posteriors. 

\subsection{Analysis strategy}
\label{met:dataanstrat}

Given a marginalized posterior distribution of a molecular abundance, we compute an empirical probability, $P$, that the molecule is present in the atmosphere of a planet, with an abundance above some threshold, $\mathbb{T}_{Ab}$, as:

\begin{equation}
	P \simeq \int_{\mathbb{T}_{Ab}}^{\infty} \mathcal{P}(x) d x
	\label{eq:pdef}
\end{equation}

\noindent where $\mathcal{P}$ is the marginalized posterior distribution and $x$ represents the abundance values. Thus, the predicted $P$ depends on the assumed atmospheric model and the selected abundance threshold $\mathbb{T}_{Ab}$. If the assumed atmospheric model is representative of the observed atmosphere, then a clear correlation (above noise) between $P$ and the true abundance in Tier 1 data implies that $P$ can be used to identify the most likely spectra that contain a molecule, providing a preliminary classification of planets by their molecular content. Thus, this $P$-statistic can be considered robust~\citep{Wall2012}, even when $\mathcal{P}(x)$ is too broad to constrain the abundance.

To test whether this method is sensitive enough, we need to simulate transmission spectra as observed in Tier 1, using an atmospheric model that includes a certain number of molecules. Then, we need to perform a spectral retrieval with the same atmospheric model and compare each input molecular abundance with the predicted $P$ corresponding to that molecule. The test is successful if, for an agreed $\mathbb{T}_{Ab}$, we recover a high $P$ for each large input abundance and a low $P$ for each small input abundance. To understand how well the method behaves under conditions similar to the \ARIEL\ reconnaissance survey, we repeat this test on a large and diverse planetary population. 

In this study, we employ a simulated population of approximately 300 transmission spectra of H$_2$-He gaseous planets, which contain CH$_4$, H$_2$O, and CO$_2$ trace gases with randomized input abundances. Additionally, we introduce NH$_3$ with randomized abundances as a nuisance parameter since its spectral features overlap with those of water and other molecules. We utilize NH$_3$ to test the $P$-statistic's efficacy and investigate the robustness of its predictions under various assumptions, such as the exclusion of NH$_3$ from retrievals or the inclusion of additional molecules not present in the population. 

Therefore, we can study whether this method provides reliable predictions under less favorable conditions when the assumed model is not fully representative of the observed atmosphere. This might provide some insight into how robustly the method can reveal the presence of a molecule in a real observation when the atmosphere is unknown. For this, we add or remove molecules from the retrieval model (hereafter, ``fit-composition'') with respect to the simulated composition. Then, we perform different spectral retrievals, that use different fit-compositions, and compare the predictions obtained from the $P$-statistic with the input abundances.



\subsubsection{Model exploration}

We consider three cases in our analysis. In the first case (referred to as $\mathrm{R}_0$), we use an atmospheric model that includes CH$_4$, H$_2$O, CO$_2$, and NH$_3$ as trace gases, which matches the composition used in the forward model generation of the population. 

In the second case (referred to as $\mathrm{R}_1$), we consider a fit-composition that includes only CH4, CO2, and H2O, omitting NH3. In this case, there is a possibility of inadequate representation of the data because NH$_3$'s molecular features could overlap with the observed features of other molecules (hence its adoption as a nuisance), particularly H$_2$O~\citep{Encrenaz2015}. As a result, the retrieved values of $P$ may not accurately reflect the input abundances of H$_2$O, leading to decreased reliability of the predictions.

In the third case (referred to as $\mathrm{R}_2$), we expand the fit-composition beyond the input composition by including also CO, HCN, and H$_2$S. It should be noted that the spectral features of these additional molecules could also overlap with the observed features of the other molecules. For instance, CO and CO$_2$ exhibit a spectral overlap around $4.5\,\mu m$. Hence, even in this case, obtaining reliable predictions of the input composition may not be obvious.

Table~\ref{tab:rsetup} provides a summary of the molecules included in the fit-composition for each retrieval. For more detailed information on the retrievals performed, please refer to Section~\ref{met:retsetup}.

\begin{table}[htb!]
\centering
\caption{Molecules included in the fit-composition for each retrieval.\label{tab:rsetup}}%
\begin{tabular}{@{}cccccccc@{}}
    \toprule
    Retrieval & CH$_4$ & CO$_2$ & H$_2$O & NH$_3$ & CO & HCN & H$_2$S \\
    \midrule
    $\mathrm{R}_0$ & \checkmark & \checkmark & \checkmark & \checkmark & & & \\
    $\mathrm{R}_1$ & \checkmark & \checkmark & \checkmark & & & & \\
    $\mathrm{R}_2$ & \checkmark & \checkmark & \checkmark & \checkmark & \checkmark & \checkmark & \checkmark \\
    \botrule
\end{tabular}
\end{table}

\subsection{Experimental data set}
\label{met:exp}

As a simulated population, we use a planetary population generated using the Alfnoor software~\citep{Changeat2020b, Mugnai2021a}. Alfnoor is a wrapper of TauREx 3~\citep{refaie2021} and ArielRad~\citep{Mugnai2020a}. Given a list of candidate targets and a model of the \ARIEL\ payload, it automatically computes simulated exoplanet spectra as observed in each \ARIEL\ Tier.

Specifically, we use a subset of the POP-I planetary population of~\cite{Mugnai2021a}. POP-I consists of $1000$ planets from a possible realization of the \ARIEL\ Mission Reference Sample (MRS) of~\cite{Edwards2019}. That MRS (hereafter, MRS19) comprises known planets in 2019 from NASA's Exoplanet Archive and TESS forecast discoveries. Here we ignore the TESS forecasts, thus obtaining a sub-population of around $300$ planets, that we label POP-Is. Using POP-Is planets ensures that, in principle, we can compare our results with those of~\cite{Mugnai2021a}. 

\begin{figure}[htb!]
	\centering
	\includegraphics[width=\textwidth]{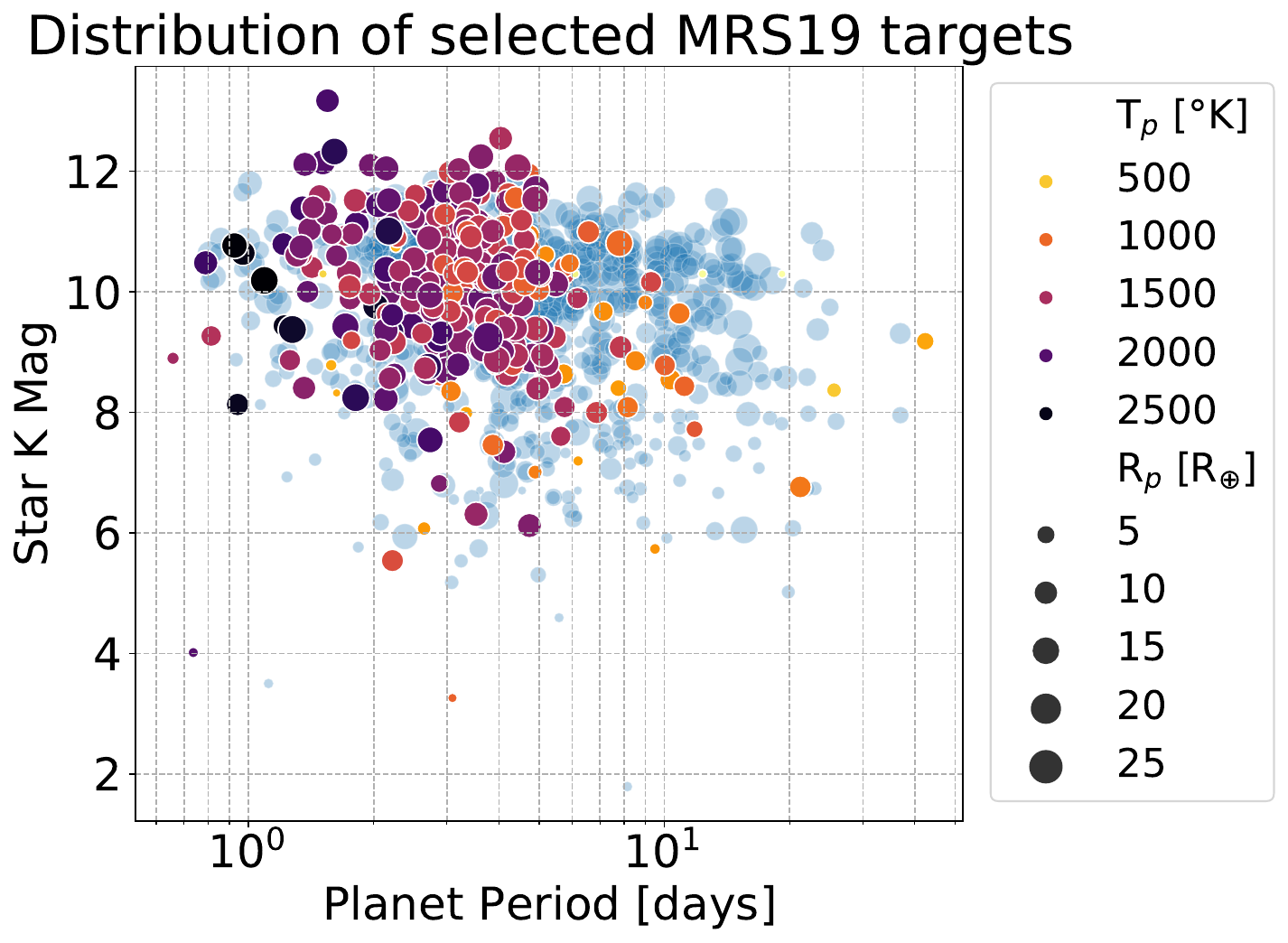}
	\caption{Parameter space distribution of the POP-Is planetary population used in this work, which comprises about $300$ selected planets from MRS19. The horizontal axis reports the planetary orbital period in days; the vertical axis reports the stellar magnitude in the K band. Each data point represents a planet; the symbol size is proportional to the planetary radius in Earth's radii; the symbol color shows the expected planetary equilibrium temperature. Light blue data points in the background show the entire MRS19/POP-I parameter space for reference.\label{fig:pop-1}}
\end{figure}

Figure~\ref{fig:pop-1} shows that POP-Is comprises a diverse sample of planets mostly with large radii ($\gtrsim$ $5$ $\mathrm{R}_{\oplus}$), short orbital periods ($\leq$ $4/5$ days), warm to hot equilibrium temperatures ($500$ -- $2500$ $^{\circ} K$) and stellar hosts with different magnitudes in the K band of the infrared spectrum ($8$ -- $12$ $m_K$). Compared to the parameter space sampled by the entire POP-I, this data set has more occasional statistics on smaller and longer-period planets around brighter stars.

The detailed properties of POP-I (and therefore POP-Is) are discussed in \cite{Mugnai2021a} and briefly summarized here. The forward model parameters are randomized to test diverse planetary atmospheres. The baseline atmosphere is a primordial atmosphere filled with H$_2$ and He with a solar mixing ratio of He/H$_2$ = 0.17. The vertical structure of the atmosphere comprises $100$ pressure layers, uniformly distributed in log space from 10$^{-4}$ to 10$^{6}$ Pa, using the plane-parallel approximation. The equilibrium temperature of each planet is randomized between $0.7 \times T_{p}$ and $1.05 \times T_{p}$, where $T_{p}$ is the equilibrium temperature of the planet listed in MRS19; the atmospheric temperature-pressure profile is isothermal. Constant vertical chemical profiles are added for H$_2$O, CO$_2$, CH$_4$, and NH$_3$, with abundances randomized according to a logarithmic uniform distribution spanning 10$^{-7}$ to 10$^{-2}$ in Vertical Mixing Ratios (VMR). Randomly generated opaque gray clouds are also added with a surface pressure varying from 5$\times$10$^{2}$ to 10$^{6}$ Pa to simulate cloudless to overcast atmospheres. Table~\ref{tab:forwmodpar} summarizes the randomized parameters of the POP-I forward models. For each planet, POP-I contains the raw spectrum binned at each \ARIEL\ Tier resolution (``noiseless spectra''), the associated noise predicted by the \ARIEL\ radiometric simulator, ArielRad, for each spectral bin, and the number of transit observations expected to reach the Tier-required S/N. To simulate an observation, we scatter the noiseless spectra according to a normal distribution with a standard deviation equal to the noise at each spectral bin. The ``observed spectra'' data set is built by repeating this process for each planet in POP-Is. As in~\cite{Mugnai2021a}, the Tier 1 data used in this work are binned on the higher resolution Tier 3 spectral grid: R = 20, 100, and 30, in NIRSpec, AIRS-CH0, and AIRS-CH1, respectively. The noise is that of Tier 1, which yields a S/N~$>$~7 if data were binned on the Tier 1 spectral grid. This is to prevent the loss of spectral information that may occur in binning.

\begin{table}[htb!]
\centering
\caption{Forward model randomized parameters in POP-I.\label{tab:forwmodpar}}%
\begin{tabular}{@{}cccc@{}}
    \toprule
    \textbf{Parameter} & \textbf{Unit} & \textbf{Range} & \textbf{Scale} \\
    \midrule
    T$_{\rm P}$ / $T_{\rm P; MRS19}$ & $^{\circ} K$ & 0.7; 1.05 & linear \\
    CH$_4$ & VMR & 10$^{-7}$; 10$^{-2}$ & log \\
    CO$_2$ & VMR & 10$^{-7}$; 10$^{-2}$ & log \\
    H$_2$O & VMR & 10$^{-7}$; 10$^{-2}$ & log \\
    NH$_3$ & VMR & 10$^{-7}$; 10$^{-2}$ & log \\
    P$_{clouds}$ & Pa & 5$\times$10$^{2}$; 10$^{6}$ & log \\
    \botrule
\end{tabular}
\end{table}

\subsection{Retrievals summary}
\label{met:retsetup}

To perform the retrievals, we use the TauREx 3 retrieval framework~\citep{refaie2021}, the same used to generate the raw POP-Is spectra. In the retrieval model, we include opaque gray clouds, pressure-dependent molecular opacities of various trace gases, Rayleigh scattering, and Collision-Induced Absorption (CIA) of H$_2$-H$_2$ and H$_2$-He. Table~\ref{tab:opacities} reports a referenced list of CIA and all molecular opacities used in this study. 

\begin{table}[htb!]
\centering
\caption{List of opacities used in this work and their references.\label{tab:opacities}}%
\begin{tabular}{@{}cc@{}}
    \toprule
    \textbf{Opacity} & \textbf{Reference(s)} \\
    \midrule
    H$_2$-H$_2$ & ~\citep{Abel2011,Fletcher2018} \\
    H$_2$-He & ~\citep{Abel2012} \\
    H$_2$O & ~\citep{Barton2017,Polyansky2018} \\
    CH$_4$ & ~\citep{Hill2013,Yurchenko2014} \\
    CO$_2$  & ~\citep{Rothman2010} \\
    NH$_3$  & ~\citep{Yurchenko2011, Tennyson2012} \\
    CO  & ~\citep{Li_2015} \\
    H$_2$S  & ~\citep{10.1093/mnras/stw1133} \\
    HCN  & ~\citep{10.1093/mnras/stt2011} \\ 
    \botrule
\end{tabular}
\end{table}

The free parameters of the retrievals are the radius and mass of the planet, as well as the molecular mixing ratios, as listed in Table~\ref{tab:priors}. We use broad logarithmic uniform priors for the molecular abundances, ranging from 10$^{-12}$ to 10$^{-1}$ in VMR. For the mass and radius of the planet, we select uniform priors of 20$\%$ and 10$\%$ around the respective values listed in MRS19. The gray cloud pressure levels are not included as free parameters in the retrieval because of their degeneracy with other parameters such as the radius \cite{changeat2020d}.

\begin{table}[htb!]
\centering
\caption{Fit parameters and their priors for the retrievals.\label{tab:priors}}%
\begin{tabular}{@{}cccc@{}}
    \toprule
    \textbf{Parameters} & \textbf{Units} & \textbf{Priors} & \textbf{Scale} \\
    \midrule
    M$_{\rm P}$ & $M_J$ & $\pm$20\% & linear \\
    $\mathrm{R}_{\rm P}$ & $R_J$ & $\pm$10\% & linear \\
    CH$_4$ & VMR & 10$^{-12}$; 10$^{-1}$ & log \\
    CO$_2$ & VMR & 10$^{-12}$; 10$^{-1}$ & log \\
    H$_2$O & VMR & 10$^{-12}$; 10$^{-1}$ & log \\
    NH$_3$ & VMR & 10$^{-12}$; 10$^{-1}$ & log \\
    CO & VMR & 10$^{-12}$; 10$^{-1}$ & log \\
    HCN & VMR & 10$^{-12}$; 10$^{-1}$ & log \\
    H$_2$S & VMR & 10$^{-12}$; 10$^{-1}$ & log \\
    \botrule
\end{tabular}
\footnotetext{We take a conservative approach by choosing larger bounds for the priors than those used for the random forward spectra generation, reported in Table~\ref{tab:forwmodpar}.}
\end{table}

We set the evidence tolerance to $0.5$ and sample the parameter space through $1500$ live points using the Multinest algorithm\footnote{v3.11, Release April 2018}~\citep{feroz2009, Buchner2021}. We disable the search for multiple modes to obtain a single marginalized posterior distribution of each molecular abundance to insert in Equation~\ref{eq:pdef}.

We then perform the three different retrievals (respectively R$_0$, R$_1$, and R$_2$) described in Section~\ref{met:dataanstrat} on each POP-Is planet. We use the Atmospheric Detectability Index (ADI)~\citep{Tsiaras2018} to assign statistical significance to the results of these retrievals. Given the Bayesian evidence of a nominal retrieval model, $E_{N}$, and of a pure-cloud/no-atmosphere model, $E_{F}$, the ADI is:

\begin{equation}
	\small
	\textrm{ADI} = \
	\begin{cases}
		log(E_{N}) - log(E_{F}), \textrm{if } log(E_{N}) > log(E_{F}) \\
		0, \textrm{otherwise}
	\end{cases}
	\label{eq:adi}
\end{equation}

ADI is a positively defined metric, equivalent to the log-Bayesian factor~\citep{kass1995, Jenkins2011} where log($E_{N}$) $>$ log($E_{F}$). To compute $E_{F}$, we perform an additional retrieval for each planet with a flat-line model with the planet radius being the only free parameter. 

\subsection{Abundance threshold}
\label{met:ab-thr-sel}

We utilized the marginalized posteriors to estimate the $P$-statistic using an abundance threshold of $\mathbb{T}_{Ab} = 10^{-5}$, which is considered ``molecular-poor'' according to the definition by~\cite{Mugnai2021a}. This threshold is higher by 1-2 orders of magnitude compared to the Tier-2 detection limits reported by~\citep{Changeat2020b}. The ``molecular-poor'' condition is met for approximately 40\% of the atmospheres due to the randomization boundaries set for each molecule (see Table~\ref{tab:forwmodpar}). The ability to detect a molecule depends on factors such as opacities, correlations among molecules, and noise in the measured spectrum. Therefore, $\mathbb{T}_{Ab}$ can be optimized for each molecule in future work, although we applied the same abundance threshold for all in this pilot study.


\subsection{Data analysis tools}
\label{met:dataantools}

The $P$-statistic can be used to reliably classify planets for the presence of a molecule with an abundance above $\mathbb{T}_{Ab}$ when $P$ correlates with the ${Ab}$ true value. The stronger the correlation above noise fluctuations, the larger the predictive power. Because this classification is binary and $P$ is defined in the range $0 \rightarrow 1$, we can use standard statistical tools such as calibration curves and ROC curves~\citep{Sanders1963, WILKS2019369} to evaluate the performance of this method in revealing the presence of molecules and in selecting Tier 1 targets for higher Tiers. These curves are routinely utilized by the Machine Learning community\footnote{In Python, the package scikit-learn~\citep{scikit-learn} (v1.0) provides the method \texttt{calibration\_curve} in \texttt{sklearn.calibration} and the method \texttt{roc\_curve} in \texttt{sklearn.metrics}.}, as they present the forecast quality of a binary classifier in a well-designed graphical format.

\subsubsection{Calibration curves}
\label{met:calcur}

\begin{figure}[htb!]
	\centering
	\includegraphics[width=\textwidth]{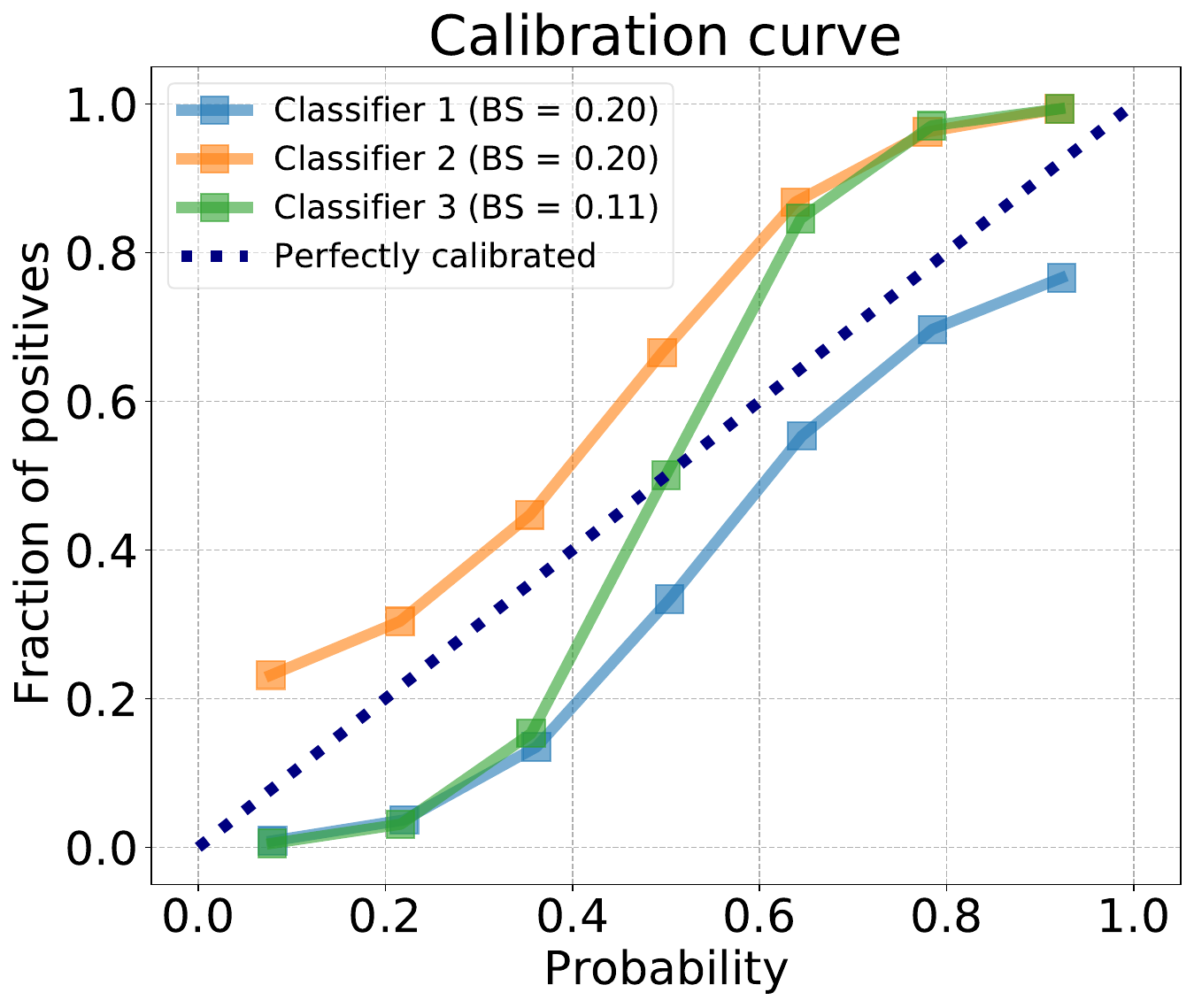}
	\caption{Calibration curves of three mock classifiers, exhibiting different forecast quality and biases. The legend reports the B-S of the forecasts of each classifier. The calibration curve for perfectly calibrated forecasts is reported for reference. \label{fig:calcurve}}
\end{figure}

A calibration curve \citep[e.g.][]{WILKS2019369} plots the forecast probability averaged in different bins on the horizontal axis and the fraction of positives, in each bin, on the vertical axis (see Figure~\ref{fig:calcurve} for a generic example). In this work, the fraction of positives is the fraction of POP-Is planets with true abundance larger than $\mathbb{T}_{Ab}$, and the forecast probability is the corresponding $P$-statistic. Calibration curves provide an immediate visual diagnosis of the quality of binary classifier forecasts and the biases that the forecasts may exhibit.

For well-calibrated predictions, the forecast probability is equal to the fraction of positives, except for deviations consistent with sampling variability. Therefore, the ideal calibration curve follows the 1:1 line. Miscalibrated forecasts can be biased differently depending on whether the calibration curve lies on the left or on the right of the 1:1 line. A curve entirely to the right of the 1:1 line indicates an over-forecasting bias, as the forecasts are consistently too large relative to the fraction of positives, as seen in the calibration curve of Classifier 1 in Figure~\ref{fig:calcurve}. On the contrary, the calibration curve of Classifier 2 shows the characteristic signature of under-forecasting, being entirely on the left of the 1:1 line, indicating that the forecasts are consistently too small relative to the fraction of positives. There may also be more subtle deficiencies in forecast performance, such as an under-confident forecast, with over-forecasting biases associated with lower probabilities and under-forecasting biases associated with higher probabilities, as seen in the calibration curve of Classifier~3.

Calibration curves paint a detailed picture of forecast performance, often summarized in a scalar metric known as the Brier Score \citep[B-S,][]{BrierScore1950}, which is defined as the mean square difference between probability forecasts and true class labels (positive or negative); the lower the B-S, the better the predictions are calibrated. From Figure~\ref{fig:calcurve}, we see that Classifier 3 achieves the best B-S, although the forecasts are not well calibrated. In general, uncalibrated forecasts can be calibrated using calibration methods such as Platt scaling and Isotonic regression~\citep{Platt1999, isotonic2002, Niculescu-Mizil2005}. 

\subsubsection{ROC curves}
\label{met:roccur}

Given the predicted probabilities of a classifier, and a selected probability threshold $\mathbb{P}$, the number of True Positives (TP), True Negatives (TN), False Positives (FP), and False Negatives (FN), are defined in Table~\ref{tab:class}.

\begin{table}[htb!]
\centering
\caption{Contingency table formulating all four possible outcomes of a binary classification problem. \label{tab:class}}%
\begin{tabular}{@{}cccc@{}}
    \toprule
     & & \multicolumn{2}{c}{True label} \\ \cmidrule{3-4} Forecast & Forecast label & Yes & No \\
    \midrule
    $\textrm{P} \geq \mathbb{P}$  & Yes & TP & FP \\
    $\textrm{P} < \mathbb{P}$  & No  & FN & TN \\
    \botrule
\end{tabular}
\end{table}

A binary classifier with high predictive power assigns larger $P$ to positive observations (true label ``Yes") and smaller $P$ to negative (true label ``No"). This maximizes TP and TN, and minimizes FP and FN\@.

\begin{figure}[htb!]
	\centering
	\includegraphics[width=\textwidth]{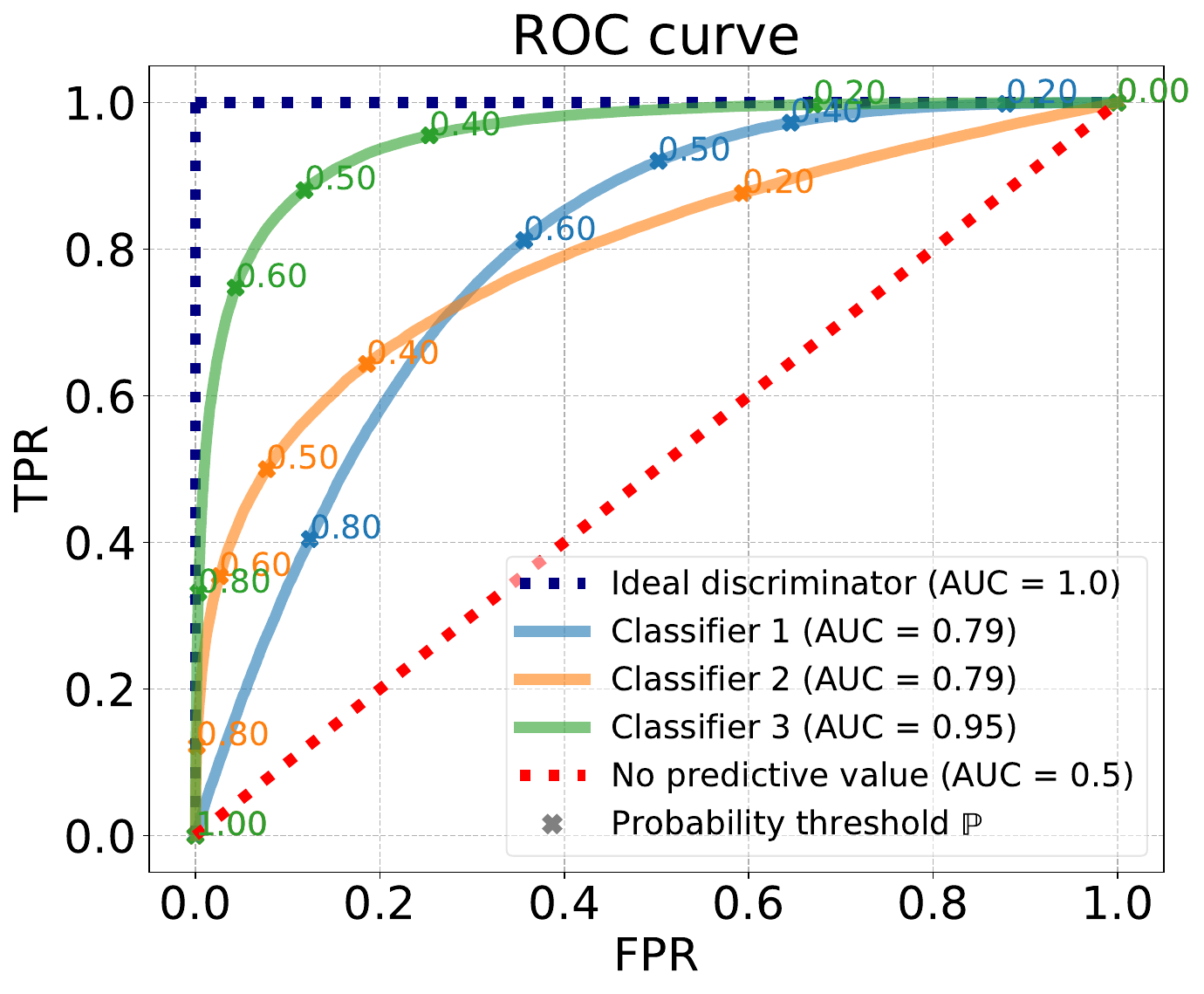}
	\caption{ROC curves of the same mock classifiers shown in Figure~\ref{fig:calcurve}, exhibiting different predictive powers. The legend reports the AUC associated with each ROC curve. The ideal and worst possible classifier ROC curves are reported for reference. Several probability thresholds $\mathbb{P}$ at regularly spaced intervals are also displayed on each curve. \label{fig:roccurve}}
\end{figure}

A ROC curve~\citep[e.g.][]{WILKS2019369} is a square diagram that illustrates the predictive power at different values of the probability threshold $\mathbb{P}$. It plots the False Positive Rate (FPR) on the horizontal axis and the True Positive Rate (TPR) on the vertical axis (see Figure~\ref{fig:roccurve} for a generic example), defined as:

\begin{subequations}
	\begin{align}
		\textrm{FPR} = \frac{\textrm{FP}}{\textrm{Negatives}} = \frac{\textrm{FP}}{\textrm{FP} + \textrm{TN}} \\
		\textrm{TPR} = \frac{\textrm{TP}}{\textrm{Positives}} = \frac{\textrm{TP}}{\textrm{TP} + \textrm{FN}}
	\end{align}
	\label{eq:fprandtpr}
\end{subequations}

FPR and TPR are commonly known as ``false alarm'' and ``hit'' rates. ROC curves are constructed by calculating the TPR and FPR from the number of TP, TN, FP, and FN as $\mathbb{P}$ decreases from 1 to 0. The ideal classifier minimizes the FPR while maximizing the TPR; thus, its ROC curve is the unit step function. On the other hand, the worst possible classifier is a random classifier with a ROC curve along the 1:1 line. Real-world classifiers have intermediate ROC curves ranked by how close they are to the unit step function. As seen in Figure~\ref{fig:roccurve}, Classifier 3 exhibits the highest predictive power, as the corresponding ROC curve arcs everywhere above the ROC curves for Classifiers 1 and~2.

ROC curves portray a detailed picture of predictive power, often summarized in a scalar metric known as the Area Under the Curve (AUC), the fraction of the unit square area subtended by a ROC curve. The higher the AUC, the higher the predictive power. The ideal classifier has $\textrm{AUC} = 1.0$; the random one has $\textrm{AUC} = 0.5$. From Figure~\ref{fig:roccurve}, we see that, as expected, Classifier 3 also achieves the largest AUC\@.

ROC curves can also be used to select the optimal classification threshold $\mathbb{P}$, which roughly corresponds to the position on the curve where the TPR cannot be raised without significantly increasing the FPR. For example, as seen in Figure~\ref{fig:roccurve}, the optimal $\mathbb{P}$ for Classifier 3 is around $0.5$, where it achieves a TPR of nearly $0.9$ at a low FPR of approximately $0.1$. Reducing $\mathbb{P}$ to $0.4$ is not advantageous, as it only increases the TPR to approximately $0.95$, at the expense of increasing the FPR to almost $0.3$.

\subsection{Using calibration and ROC curves}
\label{met:using-calibration-and-roc-curves}

Using calibration curves and the B-S metric, we can immediately diagnose the forecast quality of the $P$-statistic and its potential biases. Suppose that the forecast probability $P$ matches the fraction of planets with input abundances greater than $\mathbb{T}_{Ab}$ (fraction of positives) in each probability bin. In that case, the prediction of the method is well-calibrated. Moreover, we can compare the forecast quality achieved for different molecules using the B-S metric. If the forecasts are not well calibrated, we can infer which kind of bias affects the predictions of the method by inspecting the shape of the calibration curve. If the forecasts show an over-forecasting bias (as in the example of Classifier~1, Fig.~\ref{fig:calcurve}) and therefore incorrectly classify a fraction of planets as bearing a molecule, too many Tier 1 planets may be selected for re-observation in higher Tiers, resulting in less optimal scheduling of observations. On the contrary, an under-forecasting bias (as in the example of Classifier 2, Fig.~\ref{fig:calcurve}) may imply that fewer Tier 1 planets than possible would be scheduled for re-observing in higher Tiers. 

Using ROC curves and the AUC metric, the power of the $P$-statistic to predict the presence of molecules can be assessed. The closer the ROC curve approaches the unit step function (AUC $\simeq 1$, Fig.~\ref{fig:roccurve}), the higher the predictive power. Moreover, we can directly compare the predictive power achieved for different molecules by analyzing the shape of the corresponding ROC curves and the AUC values. 

The shape of the ROC curve provides a way to select the optimal classification threshold, $\mathbb{P_*}$, for the problem under study. For instance, $\mathbb{P_*}$ can be chosen in a trade-off process that maximizes the TPR while keeping the FPR at an acceptable low value.

This choice can aid the selection of Tier 1 targets for re-observation in a higher Tier: a large FPR would result in a poor allocation of observing time while a low TPR would result in a reduction of observational opportunities. It can also benefit population studies where one might need to track the presence of certain molecules across families of planets and extrasolar systems. These types of studies are outside the scope of this work, but can profit from the methodology developed here. 

\section{Results}
\label{sec:res}
As detailed in Section~\ref{met:dataanstrat}, we designed a method based on the $P$-statistic to reveal the presence of a molecule in Tier 1 spectra. In the following sections, we use the statistical tools described in Section~\ref{met:dataantools} to show the performance of the $P$-statistic in predicting the presence of several molecules in our simulated planetary population. In particular, in Section~\ref{res:detrel}, we use calibration curves to assess the reliability of the predictions of the method and related biases, while in Section~\ref{res:predict}, we use ROC curves to assess the predictive power of the method and discuss the optimal classification threshold, $\mathbb{P_*}$. In Section~\ref{res:bias}, we use the median abundance as an estimator of the true abundance and investigate its biases in the low S/N regime to explain the biases observed in the calibration curves.

\subsection{Detection reliability}
\label{res:detrel}

\subsubsection{Retrieval \texorpdfstring{$\mathrm{R}_0$}{R0}}
\label{res:detrel:retrieval-r0}

\begin{figure}[htb!]
\centering
\includegraphics[width=\textwidth]{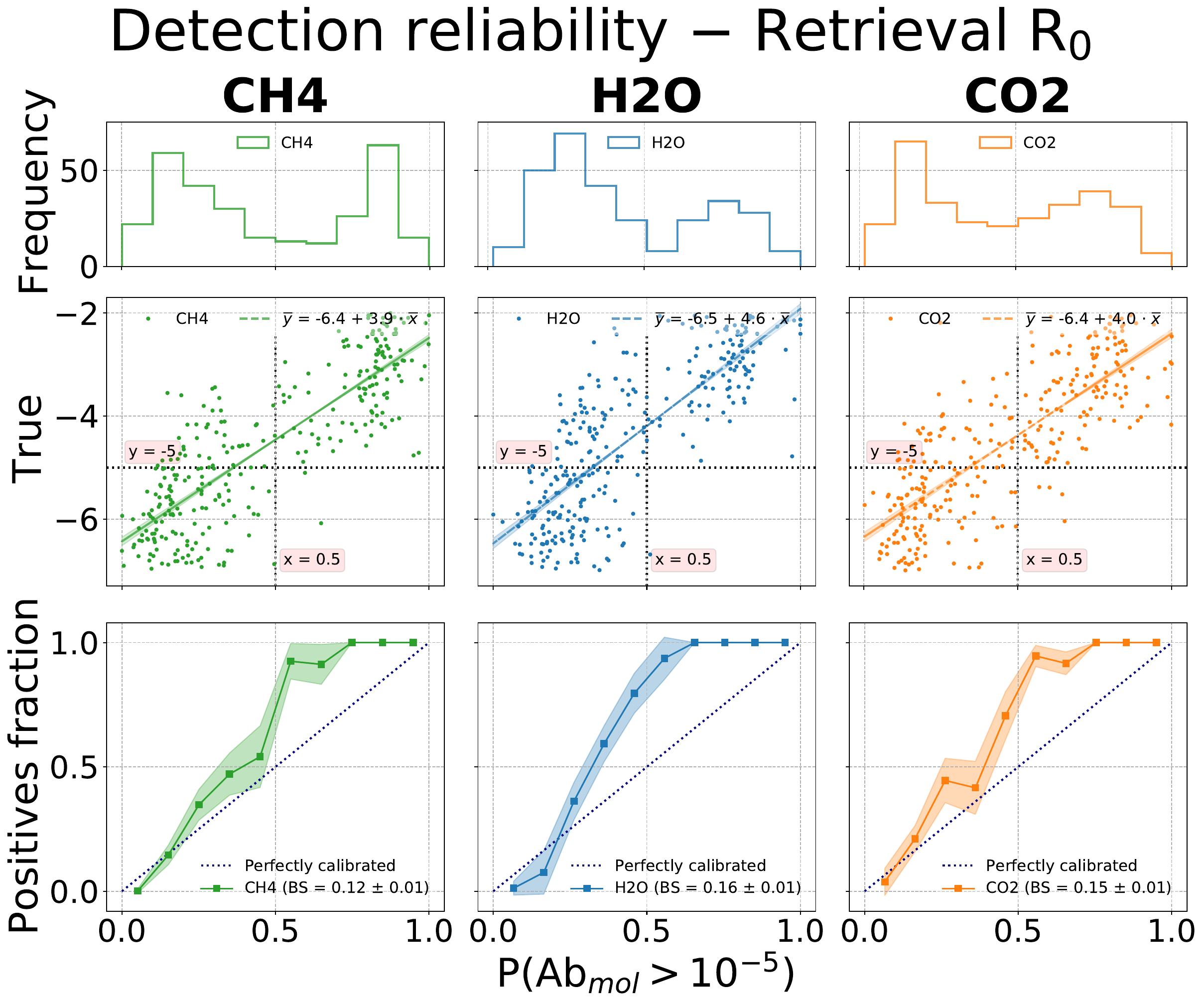}
\caption{Detection reliability analysis for CH$_4$, H$_2$O, and CO$_2$ from the $\mathrm{R}_0$ retrievals, that implement a model that is fully representative of the simulated atmospheres. All plots in the same column share the same horizontal axis with the predicted probabilities, $P(Ab_{mol}$~$>$~$10^{-5})$, that a molecule is present in the atmosphere of a planet, with an abundance above the selected abundance threshold, $\mathbb{T}_{Ab} = 10^{-5}$. Top row: histogram with the frequency of the $P$ forecasts. Middle row: diagrams showing the correlation between $P$ values on the horizontal axis and input abundances on the vertical axis. The linear fit parameters of the data points are reported on each legend. For visual reference, the dotted horizontal lines show the position of $\mathbb{T}_{Ab}$ and the dotted vertical lines the value 0.5 on the x-axis. Bottom row: calibration curves with associated bootstrap confidence intervals; each legend shows the B-S of the forecasts. \label{fig:detrel0}}
\end{figure}

Figure~\ref{fig:detrel0} shows the analysis performed to evaluate the reliability of the method when using the abundance posteriors of the retrieval $\mathrm{R}_0$, which uses the same atmospheric composition as the one used in the generation of the simulated atmospheres (see Table~\ref{tab:rsetup}). The subplots in each column share the same horizontal axis with the predicted probability $P$ that a molecule is present with an input abundance, $Ab_{mol}$, above the selected abundance threshold $\mathbb{T}_{Ab} = 10^{-5}$ (see Section~\ref{met:ab-thr-sel}). The figure reports the results for CH$_4$, H$_2$O, and CO$_2$, shown from left to right, respectively.

The top row displays histograms of the $P$-statistic realizations, which exhibit a bimodal distribution. Two peaks are observed in the distribution, with one located at $P \approx 0.2$ and the other at $P \approx 0.8$, with the former being more prominent. Additionally, a valley is observed at intermediate values, with $P \approx 0.5$.

The middle row shows the correlation between the predicted probabilities on the horizontal axis and the input abundances of each molecule on the vertical axis. We take a rough measure of the correlation by calculating the angular coefficient of the data points from a linear fit. These coefficients are listed in Table~\ref{tab:mandBS}. The lower right quadrant of these diagrams ($P \gtrsim 0.5$ and $Ab_{mol} < 10^{-5}$) is almost empty of data points, indicating that whenever the method predicts a high $P$, the corresponding input abundance is likely higher than $\mathbb{T}_{Ab}$. However, not all planets with an input abundance greater than $\mathbb{T}_{Ab}$ are associated with a high $P$, as the upper left quadrants of these diagrams ($P\lesssim 0.5$ and $Ab_{mol} > 10^{-5}$) are not empty of data points.

The bottom row shows the calibration curves computed for each molecule; each curve is shown with a bootstrap confidence interval calculated using $1000$ bootstrap samples. That is, following~\cite{Press2007}, we randomly remove $\sim 1/e \approx 36\%$ of the data from each of these samples and replace them by repeating some randomly chosen instances of the ones kept. For each molecule, we calculate the B-S using the $\texttt{brier\_score\_loss}$ method of $\texttt{sklearn.metrics}$~\citep{scikit-learn}, with the associated uncertainty estimated from the same bootstrap samples. Table~\ref{tab:mandBS} lists the B-S values obtained.

\begin{table}[htb!]
\centering
\caption{Best-fit value for the angular coefficient m from the linear fit $\log(Ab_{mol}) \propto m \ P(Ab_{mol} > \mathbb{T}_{Ab})$, with $\mathbb{T}_{Ab} = 10^{-5}$, and Brier Score for the calibration curves for all possible combinations of retrievals and molecules.\label{tab:mandBS}}%
\begin{tabular}{@{}cccc@{}}
    \toprule
    \textbf{Retrieval} & \textbf{molecule} & \textbf{m} & \textbf{B-S [\%]} \\
    \midrule
    $\mathrm{R}_0$ & CH$_4$ & 3.9 & 12 $\pm$ 1  \\
    & H$_2$O & 4.6 & 16 $\pm$ 1 \\
    & CO$_2$ & 4.0 & 15 $\pm$ 1  \\ 
    \midrule
    $\mathrm{R}_1$ & CH$_4$ & 3.2 & 15 $\pm$ 1  \\
    & H$_2$O & 3.8 & 17 $\pm$ 1  \\
    & CO$_2$ & 3.7 & 14 $\pm$ 1  \\ 
    \midrule
    $\mathrm{R}_2$ & CH$_4$ & 3.9 & 13 $\pm$ 1  \\
    & H$_2$O & 4.4 & 16 $\pm$ 1  \\
    & CO$_2$ & 3.9 & 16 $\pm$ 1  \\
    \botrule
\end{tabular}
\end{table}

\noindent The calibration curves show an under-forecasting bias (curve to the left of the 1:1 line; see Section~\ref{met:calcur}) especially associated with larger forecast probabilities, giving a fraction of positives $\approx 1.0$ for $P \gtrsim 0.6$. On the contrary, the probabilities are better calibrated for $P \lesssim 0.4$. From the B-S values (less accurate forecasts receive higher B-S), we see that CH$_4$ is the best-scoring molecule, probably due to its strong absorption spectral features.

It is possible that the observed under-forecasting of the calibration curves and the bimodality of the $P$-statistic distribution are both related to the sampling of the parameter space. This is briefly discussed further in Section~\ref{disc:priors}. 



\subsubsection{Retrieval \texorpdfstring{$\mathrm{R}_1$}{R1}}
\label{res:detrel:retrieval-r1}

\begin{figure}[htb!]
	\centering
	\includegraphics[width=\textwidth]{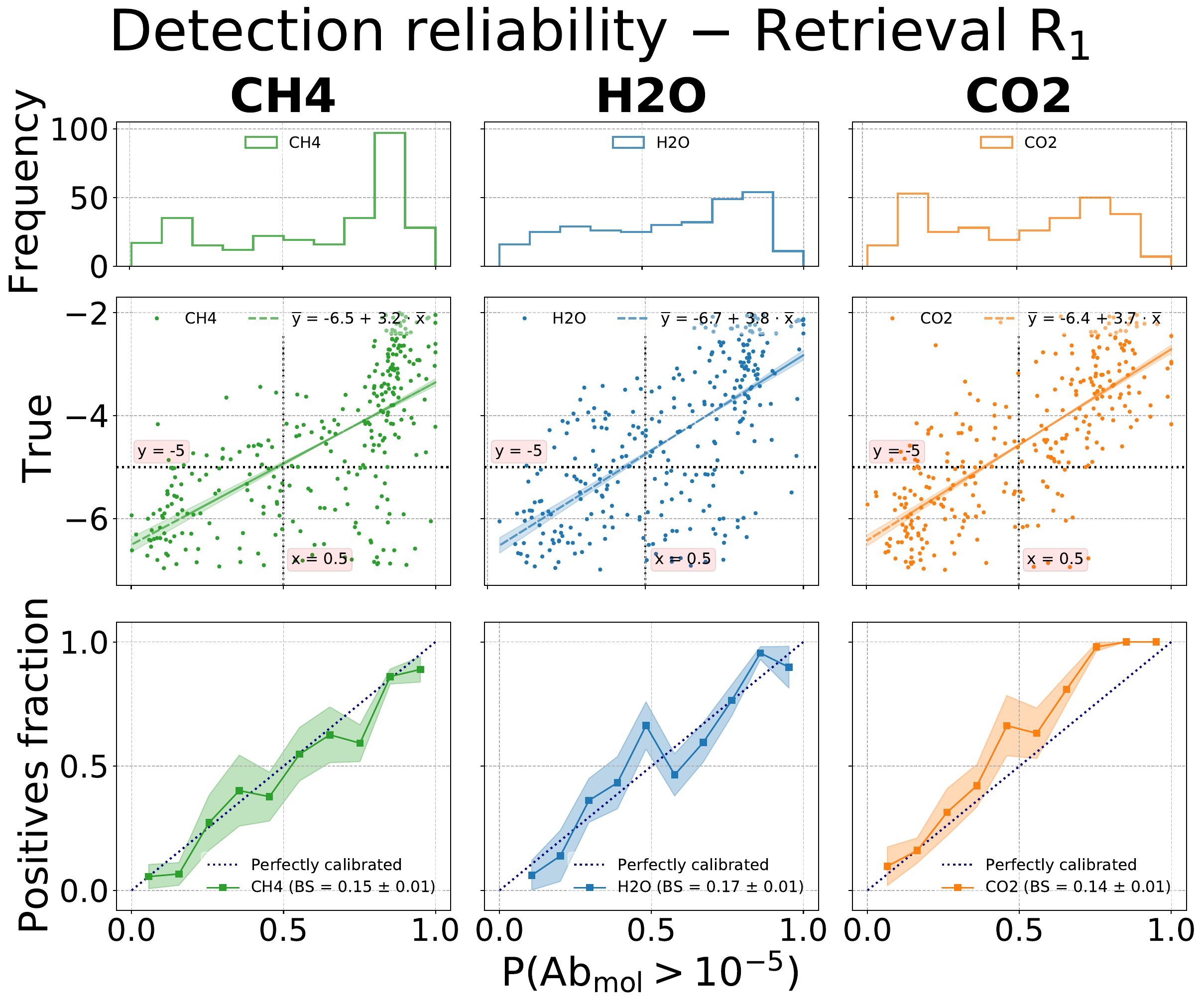}
	\caption{Same as Figure~\ref{fig:detrel0}. Detection reliability for the $\mathrm{R}_1$ retrievals, implementing a model that excludes  NH$_3$ from the fit-composition.}\label{fig:detrel1}
\end{figure}

Figure~\ref{fig:detrel1} shows the same analysis for the retrieval $\mathrm{R}_1$, which includes only CH$_4$, CO$_2$, and H$_2$O in the fit-composition and excludes NH$_3$, although this molecule is present in the data set (see Table~\ref{tab:rsetup}).
Comparing the histograms from the top row of this figure with those obtained for the retrieval $\mathrm{R}_0$ (Figure~\ref{fig:detrel0}), we notice a decrease in the forecast frequency at low $P$, especially for CH$_4$ and H$_2$O, with a reduced peak at $P$ around $0.2$. On the contrary, high values of $P$ are more frequent, enhancing the peak at $P$ around $0.8$: for CH$_4$, more than $30 \%$ of the data set receives $P$ between $0.8$ and $0.9$. These are samples with high input abundance. 

The plots in the middle row show an increase in the scatter in the data points compared to $\mathrm{R}_0$. In this case, we find a decrease in the correlation between $P$ and the input abundances, and the angular coefficients of the linear fit are reported in Table~\ref{tab:mandBS}. Planets that receive $P \gtrsim 0.8$ have high input abundance, $Ab_{mol} > 10^{-5}$. 

The calibration curves for H$_2$O and CH$_4$ in the bottom row are, within the uncertainties, closer to the 1:1 line than for $\mathrm{R}_0$, both for high and low forecast probabilities. Although this might appear closer to the ideal behavior, it could be misleading. The B-S is higher than for $\mathrm{R}_0$, because the mean squared difference between the forecasts and true class labels is larger. This is visualized in the middle plots: for $Ab_{mol} < 10^{-5}$ (negative true class label), there are many forecast values with $P > 0.5$. In other words, the correlation between the $P$-statistic and the true input abundances is weaker. In contrast, the entire CO$_2$ calibration curve shows the signature of under-forecasting. The curve for CO$_2$ is almost the same as for $\mathrm{R}_0$, likely because the missing NH$_3$ affects less the CO$_2$ abundance posteriors. On the other hand, the overlap of NH$_3$ with H$_2$O but also CH$_4$ makes the model used in the retrieval less suitable to describe the data. 


The reduced correlation between probability forecasts and input abundances, as well as the higher B-S values, suggest that excluding NH$_3$, despite its presence in the data set, leads to less representative abundance posteriors. However, predictions for CO$_2$ are less affected, possibly because this trace gas has less spectral overlap with NH$_3$ compared to H$_2$O or CH$_4$.

\subsubsection{Retrieval \texorpdfstring{$\mathrm{R}_2$}{R2}}
\label{res:detrel:retrieval-r2}

The results of the same analysis for the retrieval $\mathrm{R}_2$, which includes CO, HCN, and H$_2$S as additional molecules to the fit-composition (see Table~\ref{tab:rsetup}) are very similar to those of $\mathrm{R}_0$ (see Section~\ref{res:detrel:retrieval-r0}). Therefore, we refer the reader to Table~\ref{tab:mandBS} that summarizes the results for the correlation between predicted probabilities and input abundances, along with the B-S values, and to Figure~\ref{fig:detrel2} in Section~\ref{sec:complfig} of the Appendix.

\subsection{Predictor assessment}
\label{res:predict}

\subsubsection{Retrieval \texorpdfstring{$\mathrm{R}_0$}{R0}}
\label{res:predict:retrieval-r0}

\begin{figure}[htb!]
	\centering
	\includegraphics[width=\textwidth]{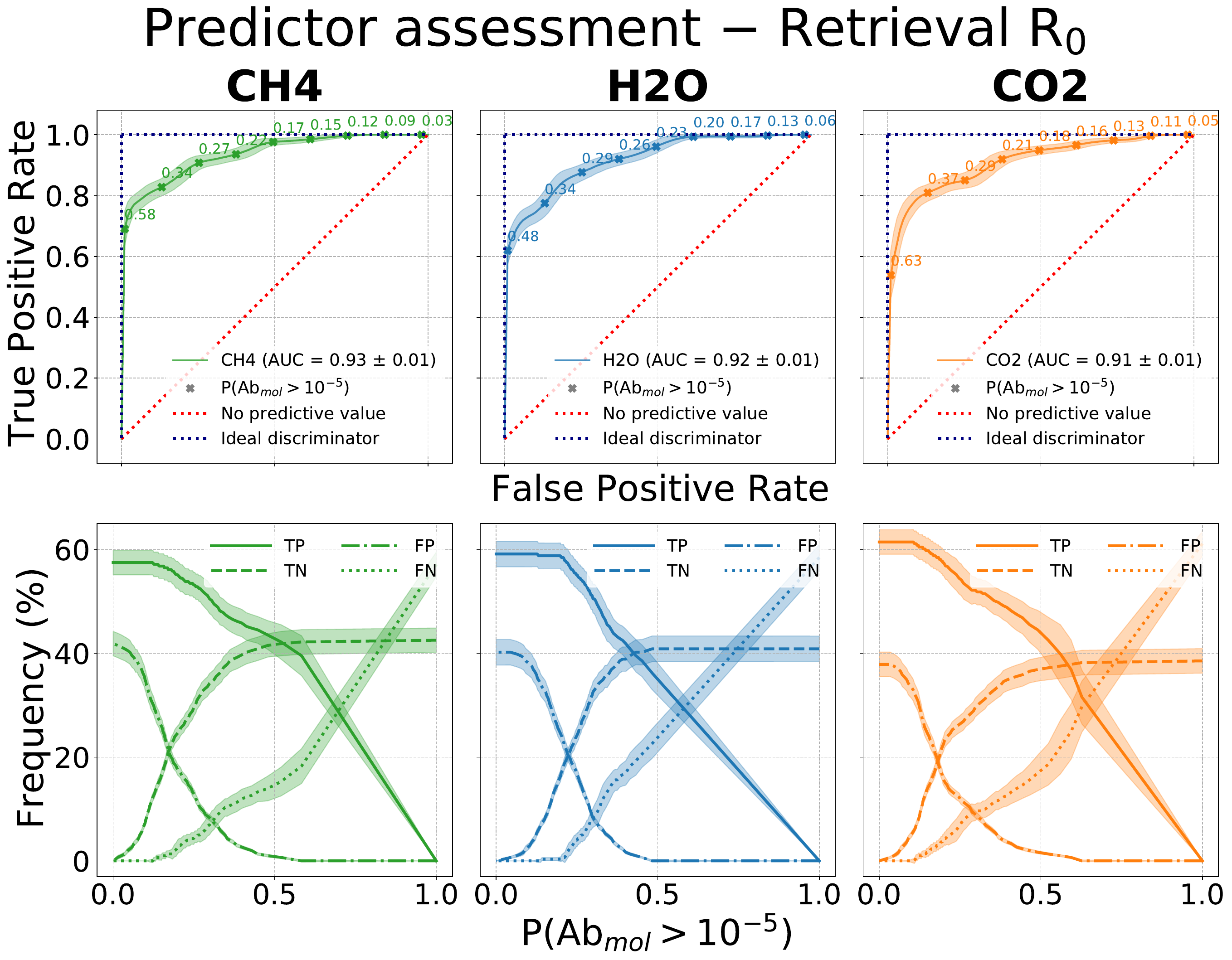}
	\caption{Predictor assessment analysis for CH$_4$, H$_2$O, and CO$_2$ from the $\mathrm{R}_0$ retrievals, that implement a model that is fully representative of the simulated atmospheres. Top row: ROC curves with associated bootstrap confidence intervals. The ideal and worst possible classifier ROC curves are reported for reference. The legends report the AUC associated with each ROC curve. Several probability thresholds $\mathbb{P}$ at regularly spaced intervals are also displayed on each curve. Bottom row: TP, TN, FP, and FN curves plotted as a function of the probability threshold $\mathbb{P}$, with confidence intervals from the same bootstrap estimation. \label{fig:predass0}}
\end{figure}

Figure~\ref{fig:predass0} shows the analysis performed to assess the predictive power of the $P$-statistic (ability to maximize TP and TN while minimizing FP and FN) when using the abundance posteriors from the retrieval $\mathrm{R}_0$. The figure reports the results for CH$_4$, H$_2$O, and CO$_2$, shown in different columns from left to right, respectively.

The upper row shows the calculated ROC curves for each molecule. Each curve is reported with a bootstrap confidence interval calculated using $1000$ bootstrap samples, with the same random removal and replacement of the data as discussed in Section~\ref{res:detrel}, involving $1/e \approx 36\%$ of the data. For each molecule, we calculate the AUC using the $\texttt{roc\_auc\_score}$ method of $\texttt{sklearn.metrics}$~\citep{scikit-learn}, with the associated uncertainty estimated from the same bootstrap samples. The AUC values thus obtained are collected in Table~\ref{tab:AUCandOdds}. For all molecules, the ROC curves are close to ideal behavior (curve near the unit step function, see Section~\ref{met:roccur}), showcasing that the $P$-statistic has significant predictive power. Consequently, the corresponding AUC values are $> 0.9$, with no considerable variation between molecules, implying similar predictive power.

\begin{table}[htb!]
\centering
\caption{AUC of the ROC curves and probability odds at the probability threshold $P = 0.5$ for all possible combinations of retrievals and molecules.\label{tab:AUCandOdds}}%
\begin{tabular}{@{}ccccc@{}}
    \toprule
    \textbf{Retrieval} & \textbf{molecule} & \textbf{AUC [\%]} & \textbf{TP [\%]: FP [\%]} & \textbf{TN [\%]: FN [\%]} \\
    \midrule
    $\mathrm{R}_0$ & CH$_4$ & 93 $\pm$ 1 & 43 $\pm$ 3 : $<$ 1 & 42 $\pm$ 2 : 15 $\pm$ 3 \\
    & H$_2$O & 92 $\pm$ 1 & 37 $\pm$ 3 : $<$ 1 & 41 $\pm$ 3 : 23 $\pm$ 3 \\
    & CO$_2$ & 91 $\pm$ 1 & 45 $\pm$ 4 : 1.7 $\pm$ 0.3 & 37 $\pm$ 2 : 17 $\pm$ 3 \\ 
    \midrule
    $\mathrm{R}_1$ & CH$_4$ & 86 $\pm$ 2 & 51 $\pm$ 3 : 16 $\pm$ 1 & 27 $\pm$ 2 : 7 $\pm$ 2 \\
    & H$_2$O & 82 $\pm$ 2 & 47 $\pm$ 3 : 15 $\pm$ 1 & 26 $\pm$ 2 : 13 $\pm$ 3 \\
    & CO$_2$ & 90 $\pm$ 1 & 48 $\pm$ 3 : 5.6 $\pm$ 0.5 & 33 $\pm$ 2 : 14 $\pm$ 2 \\ 
    \midrule
    $\mathrm{R}_2$ & CH$_4$ & 93 $\pm$ 1 & 41 $\pm$ 3 : $<$ 1 & 42 $\pm$ 2 : 17 $\pm$ 3 \\
    & H$_2$O & 92 $\pm$ 1 & 37 $\pm$ 4 : $<$ 1 & 41 $\pm$ 2 : 23 $\pm$ 3 \\
    & CO$_2$ & 91 $\pm$ 1 & 45 $\pm$ 3 : 1.7 $\pm$ 0.3 & 37 $\pm$ 2 : 17 $\pm$ 3 \\
    \botrule
\end{tabular}
\end{table}

For each molecule, the bottom row shows the number of TP, TN, FP, and FN (see Table~\ref{tab:class}), used to construct the ROC, versus the probability threshold $\mathbb{P}$. Also shown are the associated confidence intervals estimated from the same bootstrap samples. These diagrams provide information on how the predictive power of the method changes as $\mathbb{P}$ varies from $1$ to $0$ and aid in the selection of the optimal classification threshold $\mathbb{P_*}$ (see Section~\ref{met:using-calibration-and-roc-curves}).

Given the randomization of trace gas abundances in the forward model ($10^{-7}$ to $10^{-2}$ on a uniform logarithmic scale, see Table~\ref{tab:forwmodpar}), and the selected abundance threshold ($\mathbb{T}_{Ab} = 10^{-5}$), the data set contains $\sim 60 \%$ positive observations and $\sim 40 \%$ negative observations. By definition, for $\mathbb{P} = 1$, the number of positive forecasts, $\mathrm{N_{P} = TP + FP}$, is zero, and the number of negative forecasts, $\mathrm{N_{N} = TN + FN}$, is equal to the size of the data set. Therefore, at this probability threshold, $\mathrm{TN} \simeq 40 \%$ and $\mathrm{FN} \simeq 60 \%$. As $\mathbb{P}$ decreases, $\mathrm{N_{P}}$ increases (TP and FP increase), while $\mathrm{N_{N}}$ decreases (TN and FN decrease). For $\mathbb{P} = 0$, $\mathrm{N_{N}}$ is zero and $\mathrm{N_{P}}$ is equal to the data set size; at this classification threshold, $\mathrm{TP} \simeq 60 \%$ and $\mathrm{FP} \simeq 40 \%$.

\bigskip In those cases where there are no external constraints on which misclassification is more bearable (FP or FN), the intersection of their curves gives an optimized classification threshold $\mathbb{P_*}$. \bigskip

From this intersection, we obtain $\mathbb{P_*} \approx 0.3$ for all molecules. For confirmation, we can trace this $\mathbb{P_*}$ on the ROC curves. As expected, it roughly corresponds to the point where we cannot significantly increase TPR without increasing FPR, which is at TPR $ \approx 0.8$. If, instead, we need a more conservative number of FP, we can choose a higher $\mathbb{P_*}$, for example $\mathbb{P_*} = 0.5$, the default classification threshold for a binary classifier.

A concise way to demonstrate the effectiveness of the $P$-statistic in rejecting misclassifications is by computing the odds TP:FP and TN:FN, estimated from the curves in the bottom row of Figure~\ref{fig:predass0}. Odds relate to the probability that a molecule is correctly identified at the selected $\mathbb{P}$, with an example shown in Table~\ref{tab:AUCandOdds}, estimated at $\mathbb{P_*} = 0.5$. The table shows that the $P$-statistic is quite effective in rejecting FP, as they are negligible for all molecules at this threshold. Moreover, TPR at $\mathbb{P_*} = 0.5$ indicates that more than 60\% of the positives in the dataset is correctly identified, with TP values of approximately 45\%, 35\%, and 45\% for CH$_4$, H$_2$O, and CO$_2$, respectively (rounded to the nearest 5\% from the odds values listed in the table). However, at this $\mathbb{P}$, FN increases to approximately 15-25\% of the dataset (as seen in the bottom row of Figure~\ref{fig:predass0} at $\mathbb{P_*} = 0.5$), resulting in TN:FN odds of less than 3:1.

\subsubsection{Retrieval \texorpdfstring{$\mathrm{R}_1$}{R1}}
\label{res:predict:retrieval-r1}

\begin{figure}[htb!]
	\centering
	\includegraphics[width=\textwidth]{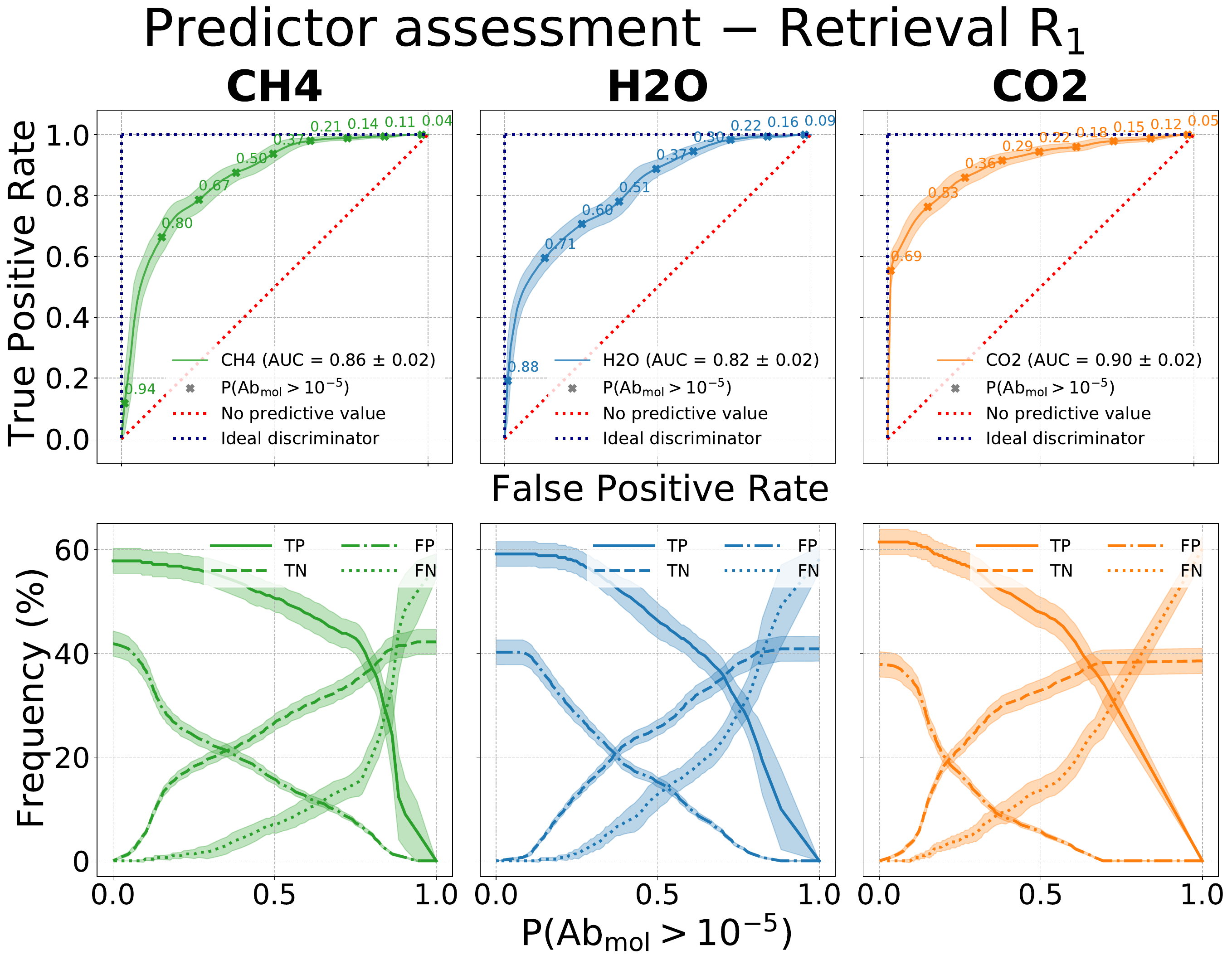}
	\caption{Same as Figure~\ref{fig:predass0}. Predictor assessment for the $\mathrm{R}_1$ retrievals, implementing a model that excludes NH$_3$ from the fit-composition. \label{fig:predass1} }
\end{figure}

Figure~\ref{fig:predass1} shows the same analysis for the retrieval $\mathrm{R}_1$.

Comparing the ROC curves in the top row with those obtained for the retrieval $\mathrm{R}_0$ (see Section~\ref{res:predict:retrieval-r0}), we notice a decrease in the predictive power of the method, measured by a reduction in AUC for CH$_4$ and H$_2$O, as reported in Table~\ref{tab:AUCandOdds}. On the contrary, the CO$_2$ ROC achieves the highest AUC, similar to that of $\mathrm{R}_0$, possibly caused by the limited overlap between NH$_3$ and CO$_2$, when compared to the case of CH$_4$ and H$_2$O.

The plots in the bottom row show a significant reduction in the performance of the FP curve compared to that achieved for $\mathrm{R}_0$: for CH$_4$ and H$_2$O, it is above $10 \%$ up to $\mathbb{P} \simeq 0.6$, instead of $<1 \%$ at $\mathbb{P} \simeq 0.5$. The TN curve also shows a decrease in performance: it remains below $30 \%$ to $\mathbb{P} \simeq 0.6$, instead of reaching $40 \%$ at $\mathbb{P} \simeq 0.4$ in $\mathrm{R}_0$. Although the TP and FN curves demonstrate relatively better performance, the optimal classification threshold denoted as $\mathbb{P}_{*}$, determined at the intersection of the FP and FN curves, increases to approximately $\mathbb{P}_* \sim 0.65, 0.5, 0.4$ for CH$_4$, H$_2$O, and CO$_2$, respectively. Tracing these $\mathbb{P}{*}$ values on the ROC curves reveals that they correspond to a TPR of approximately 0.8 for all molecules, similar to $\mathrm{R}_0$, but with a significantly worse FPR, as a consequence of the reduced predictive power.


Table~\ref{tab:AUCandOdds} reflects this, showing the odds of TP:FP and TN:FN at the same probability threshold $\mathbb{P}_{*} = 0.5$, which was used for $\mathrm{R}_0$. In this case, the method is less efficient in rejecting FP, despite having TP of approximately 50\% and 45\% for CH$_4$ and H$_2$O, respectively, resulting in only about 3:1 odds for TP:FP. However, the method is still effective in correctly identifying planets with CO$_2$, with TP:FP odds of about 9:1. As for TN:FN, the results are similar to $\mathrm{R}_0$, with a slightly better rejection of FN in the case of CH$_4$ (4:1 instead of 3:1).

\subsubsection{Retrieval \texorpdfstring{$\mathrm{R}_2$}{R2}}
\label{res:predict:retrieval-r2}
The results from the same analysis for the retrieval $\mathrm{R}_2$ are very similar to $\mathrm{R}_0$'s (see Section~\ref{res:predict:retrieval-r0}). Therefore, we refer the reader to Table~\ref{tab:AUCandOdds} that summarizes the AUC values obtained and the odds TP:FP and TN:FN at the probability threshold $\mathbb{P_*} = 0.5$, and to Figure~\ref{fig:predass2} in Section~\ref{sec:complfig} of the Appendix.

\subsection{Abundance estimates}
\label{res:bias}
Tier 1 might not be adequate for reliable abundance retrieval, for which higher \ARIEL\ Tiers are better suited. Therefore, we study the retrieved Tier 1 abundances to investigate trends in their distribution that may clarify some of the behavior observed in the calibration and ROC curves seen in the previous sections. The abundance estimator used is obtained from the median of the marginalized posterior distribution of the $\log Ab_{mol}$ with asymmetric error bars estimated from the $68.3 \%$ confidence level around the median. In particular, we are interested in investigating the regime of input abundances under which this median-based estimator is unbiased.

\subsubsection{Retrieval \texorpdfstring{$\mathrm{R}_0$}{R0}}
\label{res:bias:retrieval-r0}

\begin{figure}[htb!]
\centering
\includegraphics[width=\textwidth]{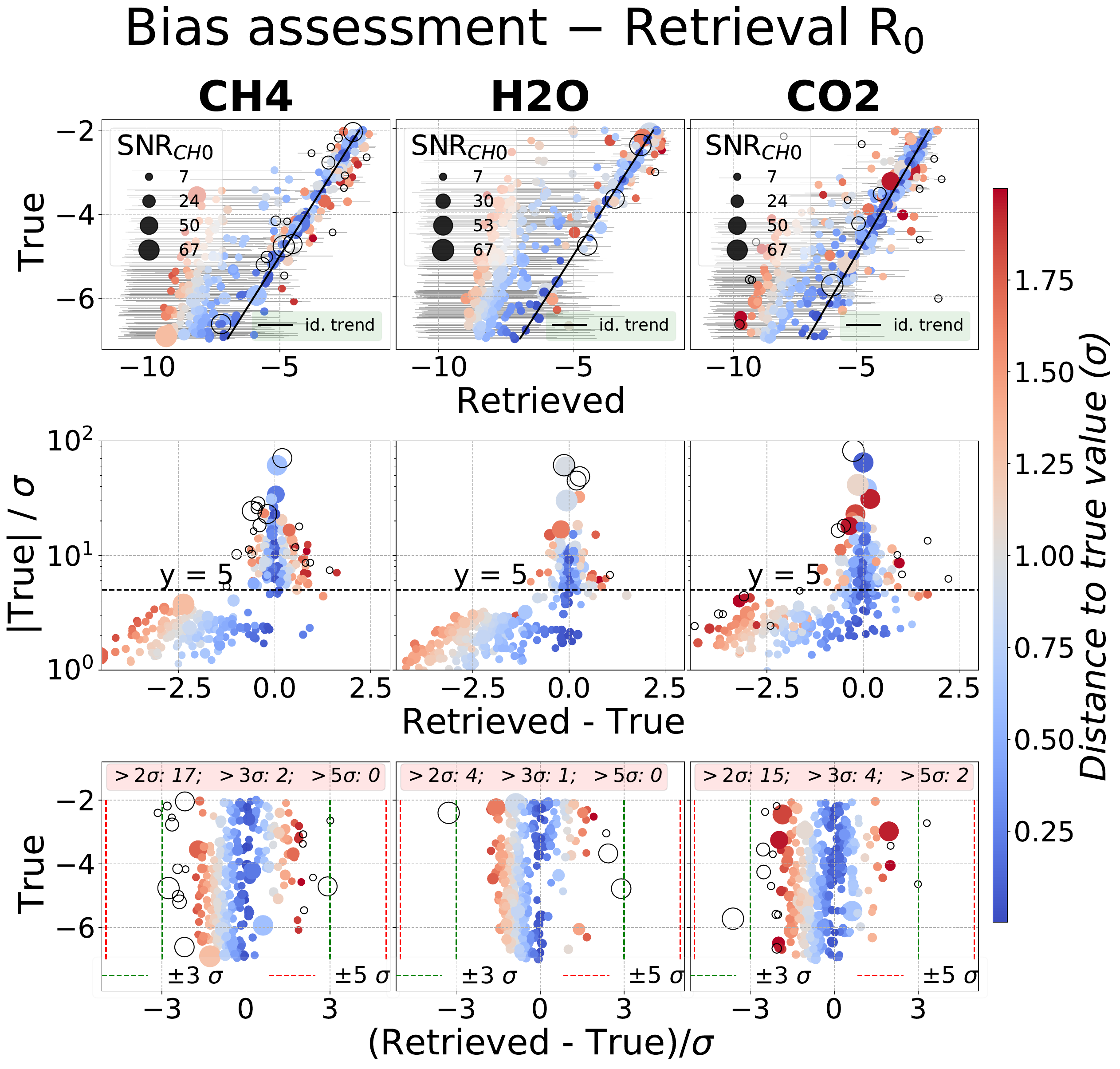}
\caption{Comparison between the retrieved molecular abundances and their true values is shown from the $\mathrm{R}_0$ retrievals. The estimator for the retrieved log-abundances is the median of the posterior distributions from the retrievals. Top row: retrieved vs. input molecular abundances. The solid black line represents the ideal trend, and the color bar visualizes the distance between input and retrieved abundances in units of uncertainty $\sigma$. The symbol size is proportional to the S/N in the AIRS-CH0 spectroscopic channel. Middle row: log-abundance S/N vs. the difference between the retrieved and input log-abundances. A black dashed line is drawn at a value of 5 on the vertical axis for visual reference. Bottom row: true abundances vs. the difference between the retrieved and true log-abundances, in units of $\sigma$. Dashed vertical lines are drawn at $3$ and $5$-$\sigma$. Text boxes show the number of 2-, 3-, and 5-$\sigma$ outliers. \label{fig:bias0}}
\end{figure}

Figure~\ref{fig:bias0} reports the analysis performed to investigate potential biases affecting the median of the marginalized posteriors when used as an estimator of the log-abundances. The figure reports the results for CH$_4$, H$_2$O, and CO$_2$, shown in different columns from left to right, respectively. NH$_3$ exhibits similar behavior to the other three molecules, but it is not included in the figure in line with the decision to treat it as a nuisance in this study. 

Panels in the top row show the molecular log-abundance input vs. the retrieved with the error bar. A solid black line serves as the ideal trend (1:1 line) for visual reference. The color bar indicates the distances between the input and retrieved log-abundance, expressed in units of the uncertainty $\sigma$ on $\log Ab_{mol}$, estimated by averaging the asymmetric error bars. Blue colors denote distances up to $1 \sigma$; red colors represent distances in the range of $1 \rightarrow 2 \sigma$. Larger distances are marked with black circles, which serve to diagnose potential trends and biases that may affect the retrieval results. In addition, the symbol size reflects the signal-to-noise ratio (S/N) of each observation as estimated in the AIRS-CH0 spectroscopic channel, providing insight into possible trends between the distance to the input abundance and the S/N condition.

The retrieved abundances exhibit good agreement with the input abundances in the large abundance regime, characterized by limited scatter around the ideal trend and by low retrieved uncertainties. This regime is generally observed for $Ab_{mol} \gtrsim 10^{-4}$, but starts to break down at $10^{-5} \lesssim Ab_{mol} \lesssim 10^{-4}$. For $Ab_{mol} \lesssim 10^{-5}$, the input abundances are rarely retrieved accurately. This analysis can provide insights into the detection limits of CH$_4$, H$_2$O, and CO$_2$ in \ARIEL\ Tier 1, which are estimated to be around $10^{-4}$. These values can be compared with the expected detection limits of the same molecules in \ARIEL\ Tier~2, which are anticipated to be significantly lower, with previous studies~\citep{Changeat2020b} reporting limits between $10^{-7}$ and $10^{-6.5}$.

Let the log-abundance S/N be defined as $\frac{1}{\sigma}\mid\log{Ab_{mol}}\mid$, where $Ab_{mol}$ is the true value of the molecular abundance. The middle row panels in Figure~\ref{fig:bias0} show the plot of log-abundance S/N vs. the difference between the retrieved and input log abundances. It can be observed that the distribution of data points is broadly separated into two sub-populations at a S/N of about 5. Data points with high S/N correspond to cases where the input is confidently retrieved and aligned along the 1:1 line in the upper row diagrams, indicating unbiased estimation. On the other hand, data points with low S/N cluster in the bottom left portion of the diagram. In these cases, the median is no longer an unbiased estimator of the true value, as the corresponding data points lie to the left of the 1:1 line in the upper row diagrams. As discussed further in Section~\ref{disc:priors}, these cases have posteriors dominated by the prior imposed in the retrieval and are best treated as upper limits.


In the bottom row of Figure~\ref{fig:bias0}, the true abundances are shown vs. the difference between the retrieved and true abundances, in units of $\sigma$. The diagrams provide a visualization of how many samples are 2-, 3-, and 5-$\sigma$ outliers, allowing verification that the distribution is compatible with the tail of the abundance posteriors. The number of outliers is shown in the text box inserted in the diagrams and (converted into percentages) in Table~\ref{tab:Outliers}. Assuming that the abundance posteriors are representative of the data, the fraction of expected outliers outside is $5\%$, $0.3\%$, and $\ll 1 \%$, respectively at 2-, 3-, and 5-$\sigma$. We find good agreement between the percentages reported in Table~\ref{tab:Outliers} and these values, with minor deviations compatible with the statistical fluctuations of a random variable.

\begin{table}[htb!]
\centering
\caption{Percentage of data points counted outside three confidence intervals for all possible combinations of retrievals and molecules.\label{tab:Outliers}}%
\begin{tabular}{@{}ccccc@{}}
    \toprule
    \textbf{Retrieval} & \textbf{molecule} & \textbf{$> 2\sigma$ [$\%$]} & \textbf{$> 3\sigma$ [$\%$]} & \textbf{$> 5\sigma$ [$\%$]} \\
    \midrule
    $\mathrm{R}_0$ & CH$_4$ & 5.6 & 0.7 & $\ll 1$ \\
    & H$_2$O & 1.3 & 0.3 & $\ll 1$ \\
    & CO$_2$ & 5.0 & 1.3 & 0.7 \\ 
    \midrule
    $\mathrm{R}_1$ & CH$_4$ & 32.9 & 19.6 & 11.6 \\
    & H$_2$O & 17.9 & 13.6 & 9.6 \\
    & CO$_2$ & 16.6 & 10.3 & 6.6 \\ 
    \midrule
    $\mathrm{R}_2$ & CH$_4$ & 6.0 & 0.7 & $\ll 1$ \\
    & H$_2$O & 1.3 & 0.3 & $\ll 1$ \\
    & CO$_2$ & 5.3 & 1.7 & 1.3 \\
    \botrule
\end{tabular}
\end{table}

\subsubsection{Retrieval \texorpdfstring{$\mathrm{R}_1$}{R1}}
\label{res:bias:retrieval-r1}

\begin{figure}[htb!]
	\centering
	\includegraphics[width=\textwidth]{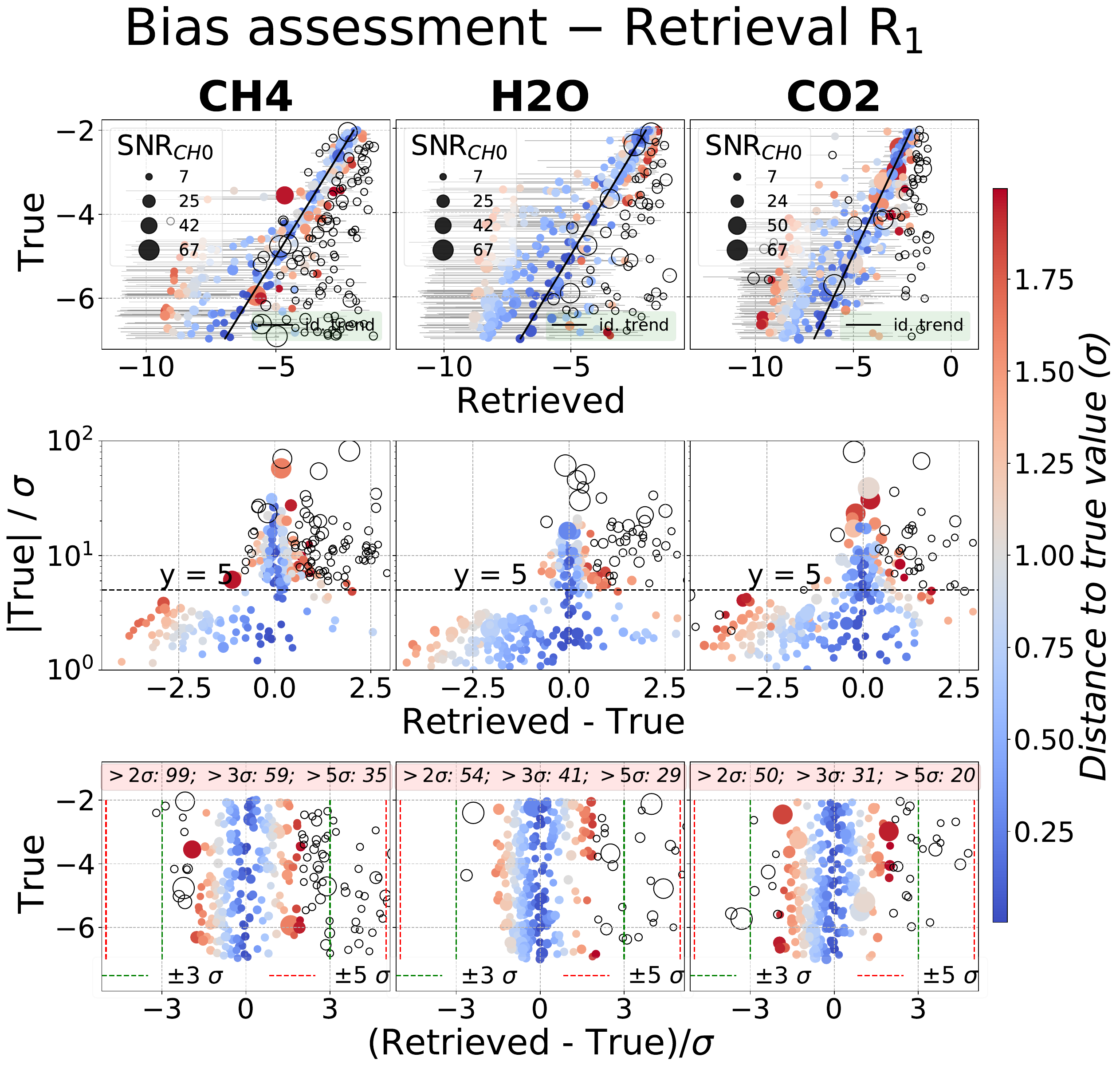}
	\caption{Same as Figure~\ref{fig:bias0} for the $\mathrm{R}_1$ retrievals, implementing a model that excludes NH$_3$ from the fit-composition. \label{fig:bias1}}
\end{figure}

Figure~\ref{fig:bias1} shows the same analysis for the retrieval $\mathrm{R}_1$.
The top row shows that, although there is still a correlation between the retrieved and input abundances, it is less significant than for $\mathrm{R}_0$. Furthermore, comparing the retrieved and input abundances yields different regimes for each molecule. However, the main difference from $\mathrm{R}_0$ is the significant number of data points at distances greater than $2 \sigma$ (marked by black circles), corresponding to 2-$\sigma$ outliers. In particular, for all molecules, most of these points are located to the right of the ideal trend, indicating the presence of an overestimation bias for the retrieved abundances. 
These data points are located in the region y $\gtrsim 5$ and x $>$ 0 in the plots in the middle row. Therefore, in addition to the overestimation bias for the abundances, their retrieved uncertainties are underestimated. Furthermore, the bottom-row diagrams show a larger number of outliers compared to the $\mathrm{R}_0$ case: too many for the posterior to be considered representative. This is a consequence of an atmospheric model which is not representative of the data, biasing the likelihood, the abundance posteriors, and the median estimator of the abundances.

\subsubsection{Retrieval \texorpdfstring{$\mathrm{R}_2$}{R2}}
\label{res:bias:retrieval-r2}

The results of the same analysis for the retrieval $\mathrm{R}_2$ are very similar to those of $\mathrm{R}_0$, including the number of outliers that are compatible with the expectations for a model that is representative of the data. Therefore, we refer the reader to Table~\ref{tab:Outliers}, and to Figure~\ref{fig:bias2} in Section~\ref{sec:complfig} of the Appendix. Here, we only stress that adding molecules to the fit-composition that are not present in the data set does not appear to significantly bias the abundance posteriors, compared to $\mathrm{R}_0$. This is further discussed in Section~\ref{disc:priors}.

\section{Discussion}
\label{sec:disc}
In this section, we first discuss the similarities between the results from the retrievals $\mathrm{R}_0$ and $\mathrm{R}_2$, shown in Sections~\ref{res:detrel} and~\ref{res:predict}. Then we apply the ADI metric to compare all retrievals from the point of view of the Bayesian evidence (Section~\ref{disc:ADI-comp}). Finally, we expand the discussion to the role of the priors in the retrieved abundance posteriors (Section~\ref{disc:priors}).

\begin{figure}[htb!]
	\centering
	\includegraphics[width=\textwidth]{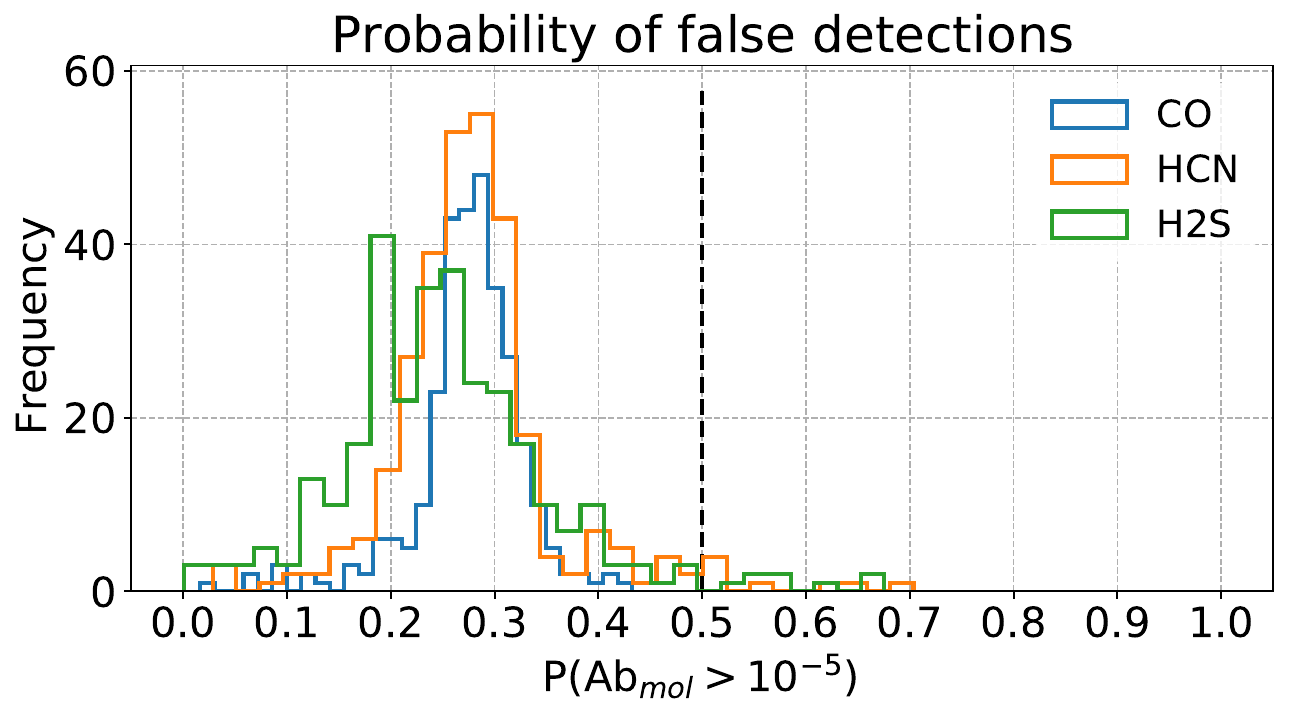}
	\caption{Histogram of the frequency of use of each possible $P$ forecast for CO, HCN, and H$_2$S, using the abundance posteriors from the retrieval $\mathrm{R}_2$. 
 The dotted vertical line marks the default binary classification threshold $P = 0.5$ for reference. \label{fig:absentmol}}
\end{figure}

The results of Sections~\ref{res:detrel} and~\ref{res:predict} show that the predictions of the $P$-statistic for the retrievals $\mathrm{R}_0$ and $\mathrm{R}_2$ are comparable, despite the quite different fit-compositions, while the reliability of the $P$-statistic is lower in the $\mathrm{R}_1$ case. The $\mathrm{R}_0$ model and its parameters are identical to those used to generate the POP-Is population, and the $\mathrm{R}_2$ extends the parameter space with new molecules. In $\mathrm{R}_2$, the abundance posteriors for CH$_4$, H$_2$O, and CO$_2$ do not appear to be significantly affected by the addition of CO, HCN, and H$_2$S in $\mathrm{R}_2$, despite that the latter three spectral signatures partially overlap with those of CH$_4$, H$_2$O, and CO$_2$~\citep{Encrenaz2015}. It should be noted that the absence of the three molecules from the simulated atmospheres is correctly revealed in $\mathrm{R}_2$ by their low $P$-statistic, shown in Figure~\ref{fig:absentmol}, that take values smaller than $40 \%$ for CO, HCN, and H$_2$S, respectively. The extension of the analysis to include the calibration and ROC curves to these molecules is left to future work.

The analysis, therefore, suggests that the $P$-statistic is robust (that means, provides reliable results) against retrieval models that are over-representative of the observed atmosphere. However, the $P$-statistic can no longer be considered robust when the retrieval models are under-representative of the observed atmosphere. 

In the current study, the threshold abundance used to estimate the $P$-statistic remains constant for all molecules. While it is possible to optimize this threshold for individual molecules, we leave this aspect for future research as discussed in Section \ref{met:ab-thr-sel}. Lowering the threshold reduces the information provided by the ROC curves. To achieve the optimal point of operation, one must balance the True and False Positive Rates, which is necessary to promote a Tier-1 target to higher Tiers. It is important to note that ROC curves calculated at different threshold levels provide a statistical estimation of the sample's completeness, enabling the inference of population-wide properties such as the fraction of planets containing certain molecules. While this aspect requires further investigation in future research, it should be noted that the fraction of positive, $\Sigma$ (planets with true abundance in excess of $\mathbb{T}_{Ab}$) is related to the fraction of Tier-1 targets, $\Tilde{\Sigma}$, selected with $P(>\mathbb{T}_{Ab}) > \mathbb{P}$ by
\[
\Sigma = \frac{\Tilde{\Sigma} - FPR}{TPR - FPR} .
\]

The similarities between the $\mathrm{R}_0$ and $\mathrm{R}_2$ models are further discussed in the next section.

\subsection{ADI comparison}
\label{disc:ADI-comp}

The ADI metric, described in Section~\ref{met:retsetup}, is used to assess the statistical significance of a model atmosphere with respect to a featureless spectrum using the log-Bayesian factor. A large ADI suggests that a featureless spectrum is less favored by the data. 
From the ADI definition, the log-Bayesian factor of two competing models is the difference between their respective ADI. 

\begin{figure}[htb!]
	\centering
	\includegraphics[width=\textwidth]{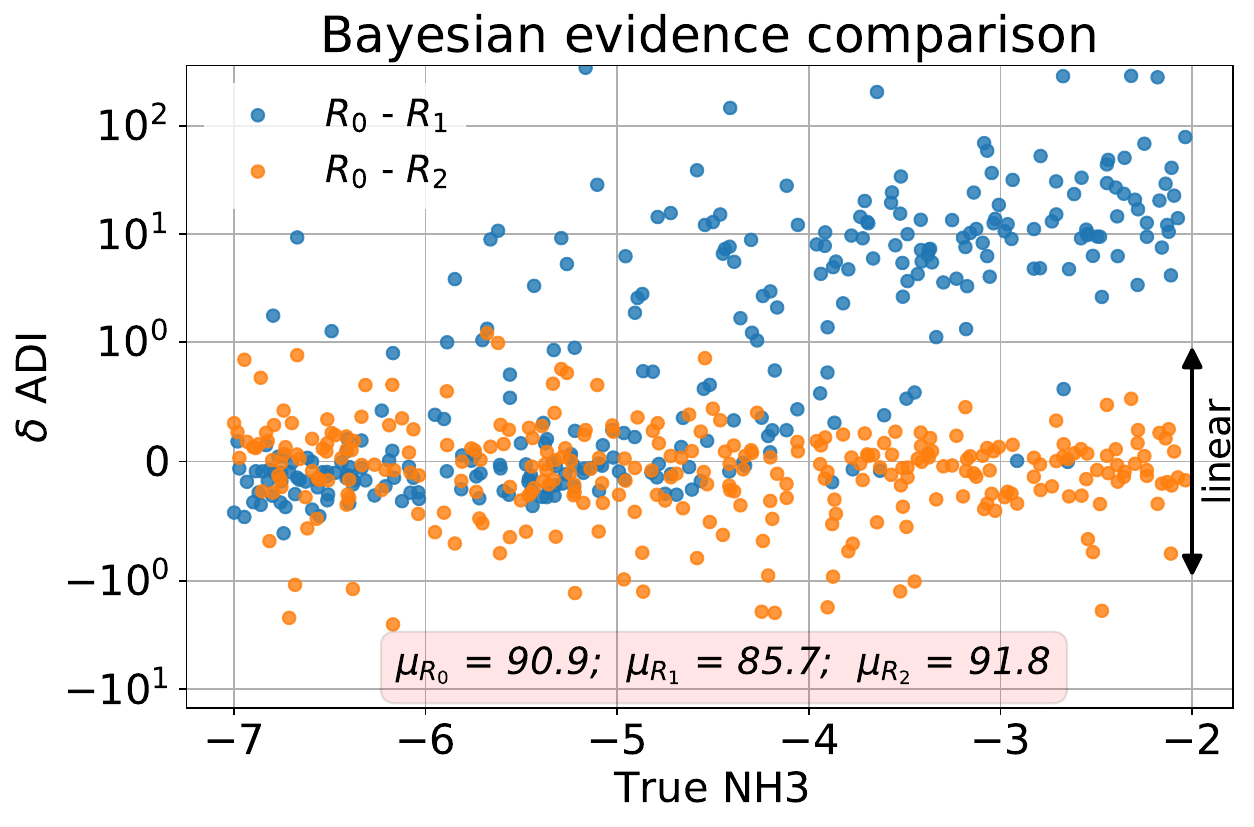}
	\caption{Bayesian evidence comparison of the retrievals $\mathrm{R}_0$, $\mathrm{R}_1$, and $\mathrm{R}_2$, measured in	ADI. The horizontal axis plots the input abundances of NH$_3$; the vertical axis reports the ADI difference between $\mathrm{R}_0$ and the other two retrievals, $\mathrm{R}_1$ and $\mathrm{R}_2$. The y-axis uses a \texttt{matplotlib} ``symlog'' scale with the linear threshold set at 1 for better visualization. The text box on the bottom shows the median ADI reported by each retrieval. \label{fig:adidiff}}
\end{figure}

Figure~\ref{fig:adidiff} shows the ADI differences between the $\mathrm{R}_0$ model and the two competing models, $\mathrm{R}_1$ and $\mathrm{R}_2$, plotted against NH$_3$ abundances. A large, positive difference indicates that the competing models are less representative of the data compared to $\mathrm{R}_0$. The median ADI values for all retrievals are approximately 91, 86, and 92 for $\mathrm{R}_0$, $\mathrm{R}_1$, and $\mathrm{R}_2$, respectively, as shown in the text box within Figure~\ref{fig:adidiff}. This suggests that a featureless atmospheric model is not favored by the data, and $\mathrm{R}_1$ is the least representative, as expected. This is further supported by the fact that the ADI difference between $\mathrm{R}_0$ and $\mathrm{R}_1$ increases with increasing NH$_3$ abundance, indicating that higher NH$_3$ abundances make $\mathrm{R}_1$ less representative compared to $\mathrm{R}_0$, in agreement with the analysis of Section \ref{sec:res}. In contrast, the ADI difference between $\mathrm{R}_0$ and $\mathrm{R}_2$ is close to zero, with a scatter described by a standard deviation of approximately 0.5, which is independent of NH$_3$ abundance. This confirms that $\mathrm{R}_2$ is similarly representative of the data compared to $\mathrm{R}_0$, despite describing a wider parameter space.

\subsection{Priors}
\label{disc:priors}

\begin{figure}[htb!]
\centering
\includegraphics[width=\textwidth]{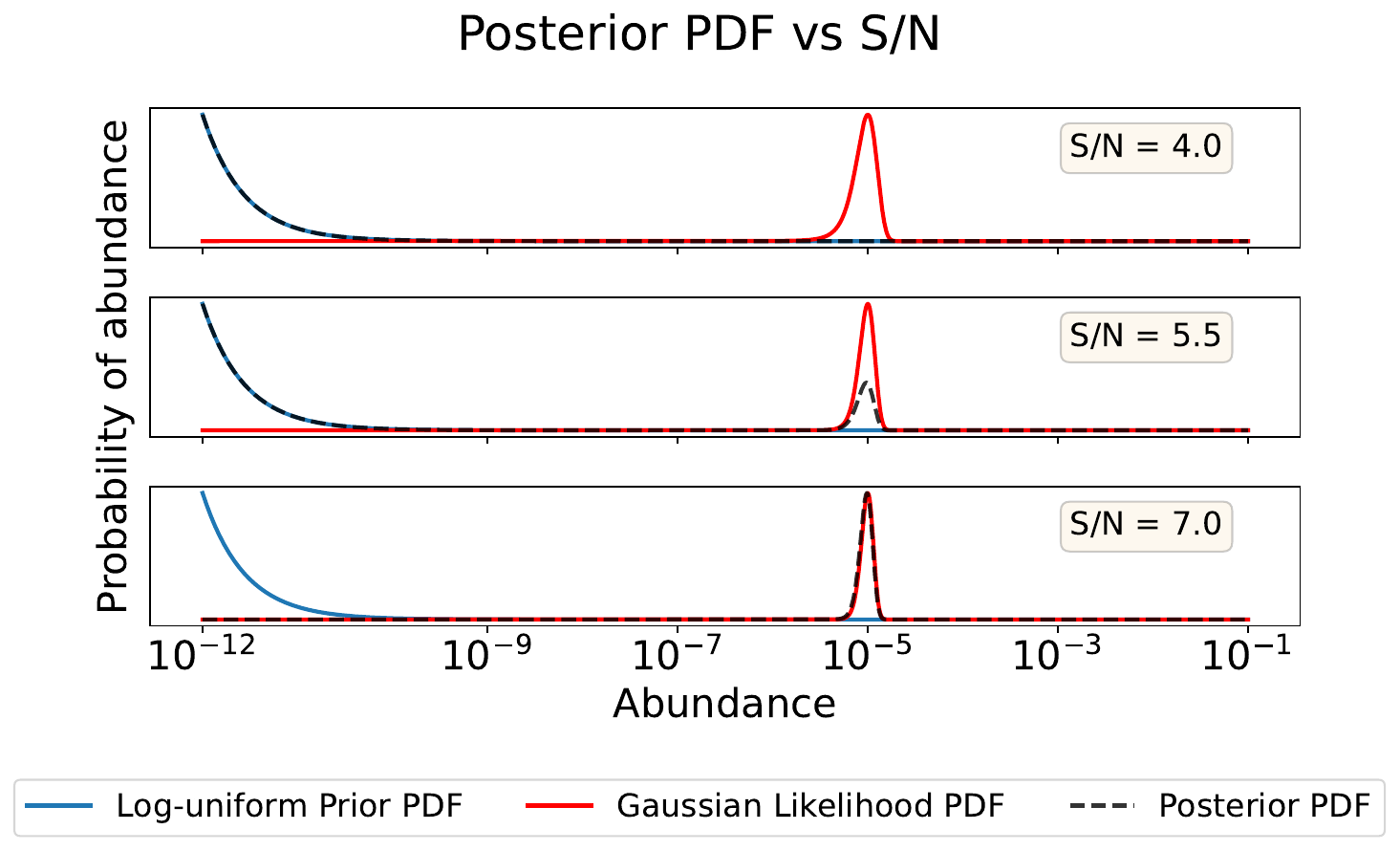}
\caption{
The probability density functions (PDF) of the likelihood, prior and posterior are shown by the red, blue, and black lines, respectively. 
The PDFs are normalized to 1 at their peak. 
The assumed abundance S/N is 4.0, 5.5, and 7.0, respectively, from the top to the bottom panel.
An input abundance of $10^{-5}$ is assumed.
    \label{fig:priors}}
\end{figure}

In this section, we discuss the impact of the log-uniform priors adopted in the analysis on the results presented. The consequence is a non-Gaussian posterior distribution, and the mean, mode, and median are not equivalent moments of the distribution. In particular, the median is not an unbiased estimator of the true abundance as shown in Figure~\ref{fig:bias0} for low log-abundance S/N (hereafter, ``abundance S/N''). This can be explained in terms of the Bayesian formulation of the posterior, $\mathcal{P}$, which is proportional to the product of the likelihood, $\mathcal{L}$, and the prior, $\Pi$. 
\begin{equation}
	\mathcal{P} \propto \mathcal{L} \times \Pi
	\label{eq:bayes}
\end{equation}
Because $\Pi(\log x)$ is uniform, $\Pi(x) \sim 1/x$, for large abundance S/N, the likelihood dominates, the posterior is Gaussian (because of the central limit theorem), and the median estimator is unbiased. For low abundances, the prior dominates, $\mathcal{P}(x) \propto 1/x$, and the median is an estimator of the molecular abundance that is biased towards low abundances. This is shown in Figure~\ref{fig:priors}. 
Each panel shows the probability density function (PDF) of the likelihood, prior and posterior normalized to 1 at the peak, for three cases where the abundance S/N is 4.0, 5.5, and 7.0, respectively, from the top to the bottom panel, assuming an input abundance of $10^{-5}$. 
The posterior is likelihood-dominated when the abundance S/N is 7 and is prior-dominated when the abundance S/N is 4. 

Although logarithmic uniform priors are often assumed in spectral retrieval studies, they are certainly not ``uninformative priors''~\citep{Trotta2008, Oreshenko_2017}. Clearly, using these priors biases the median estimator of the molecular abundance in the low S/N regime, explaining the trends seen in Figure~\ref{fig:bias0}. As a side note, log-priors on molecular abundances could as well introduce biases on the derived elemental abundances, therefore the issue has to be investigated carefully in future studies.  

The low abundance S/N targets are those that contribute to the leftmost peak in the bimodal distribution of the $P$-statistic (Figure \ref{fig:detrel0}). Further investigation is however needed to fully understand the origin of the $P$-statistic bimodality and its under-forecasting properties. 


%

\section{Conclusion}
\label{sec:conc}

The \ARIEL\ Tier 1 is a shallow reconnaissance survey of a large and diverse sample of approximately 1000 exoplanet atmospheres. It is designed to achieve a signal-to-noise ratio (S/N) greater than 7 when the target exoplanet atmospheric spectra are binned into 7 photometric bands. Tier 1 enables rapid and broad characterization of planets to prioritize re-observations in higher Tiers for detailed chemical and physical characterization. However, Tier 1 may not have sufficient S/N at the spectral resolution required for high-confidence abundance retrieval of chemical species. Nonetheless, it contains a wealth of spectral information that can be extracted to address questions requiring population studies. 

In this study, we have introduced a $P$-statistic, which is a function of the data that is sensitive enough to reveal the presence of molecules from transit spectroscopy observations of exoplanet atmospheres and can be used as a binary classifier. The $P$-statistic is estimated from the marginalized retrieval posterior distribution and provides an estimate of the probability that a molecule is present with an abundance exceeding a threshold, fixed at $\mathbb{T}_{Ab} \sim 10^{-5}$ in this study, but can be optimized in future analyses.

We have tested the performance of the $P$-statistic on a simulated population of gaseous exoplanets, POP-Is, with traces of H$_2$O, CH$_4$, and CO$_2$ of randomized abundances, in a H$_2$-He dominated atmosphere. NH$_3$ is also included as a disturbance parameter to test the robustness of the $P$-statistic. For this, three models are used in the retrievals: R$_0$, which is representative of the data; R$_1$, which is under-representative as it excludes NH$_3$; and R$_2$, which is over-representative as it includes additional molecules not considered in the simulated POP-Is.

We find that the $P$-statistic estimated from R$_0$ posteriors shows a clear, above-noise correlation with the input abundances, allowing us to infer the presence of molecules. The $P$-statistic appears to follow a bimodal distribution, where targets with low abundance S/N are likely contributors to the peak at low $P$ values. This is supported by the distribution of the median of the abundance posterior, which is an unbiased estimator of the true value only when the abundance S/N is sufficiently large (typically above 5). The $P$-statistic is affected by an under-forecasting bias, but this is not expected to adversely affect the classification of the planets in the survey as it can be calibrated in principle. This is further evidenced by ROC curves with large AUC, indicating that the $P$-statistic can be used to implement a reliable classifier for the presence of molecules. However, further investigation is needed to fully understand the origin of the $P$-statistic bimodality and its under-forecasting properties.

The results obtained appear not to be affected by the increase in complexity of the assumed atmospheric model, implemented in this study with the R$_2$ retrieval model, as indicated by similar calibration and ROC curves. We find that the predictive power of the $P$-statistic is adversely affected by an under-representative model, as implemented in the R$_1$ retrieval model, which is evident from a weaker correlation between the $P$-statistic and the input abundances, and the median of the posterior abundance no longer being a reliable unbiased estimator of the true value, even in the high abundance S/N regime.

Based on our findings, we conclude that the $P$-statistic is a reliable predictor of the presence of molecules within the parameter space explored, as long as the retrieval model matches the complexity of the data. Models that are under-representative can result in poor predictive power, while the investigated over-representative model does not seem to adversely affect classification. Further investigations are needed to test the robustness of the $P$-statistic over a wider parameter space, particularly including a wider set of molecules in both the simulated population and retrievals.


\backmatter

\bmhead{Acknowledgments}
This version of the article has been accepted for publication, after peer review, but is not the Version of Record and does not reflect post-acceptance improvements, or any corrections. The Version of Record is available online at: \url{https://doi.org/10.1007/s10686-023-09911-x}

\bmhead{Software}
ArielRad~\citep{Mugnai2020a}, TauREx 3~\citep{refaie2021}, Alfnoor~\citep{Changeat2020b, Mugnai2021a}, Astropy~\citep{astropy}, h5py~\citep{hdf5_collette}, Matplotlib~\citep{Hunter_matplotlib}, Numpy~\citep{oliphant_numpy}.

\section*{Declarations}

\bmhead{Funding}
The authors acknowledge that this work has been supported by the ASI grant n. 2021.5.HH.0.

\bmhead{Conflict of interest}
The authors declare they have no conflict of interest.

\bmhead{Authors' contributions}
Andrea Bocchieri wrote the main manuscript text and prepared all the figures. Lorenzo V. Mugnai provided the forward models for the analysis. All authors provided comments on the analysis. Andrea Bocchieri and Enzo Pascale edited the final manuscript. All authors read and approved the final manuscript.







\begin{appendices}
	
\section{Complementary figures}
\label{sec:complfig}


\begin{figure}[htb!]
	\centering
	\includegraphics[width=\textwidth]{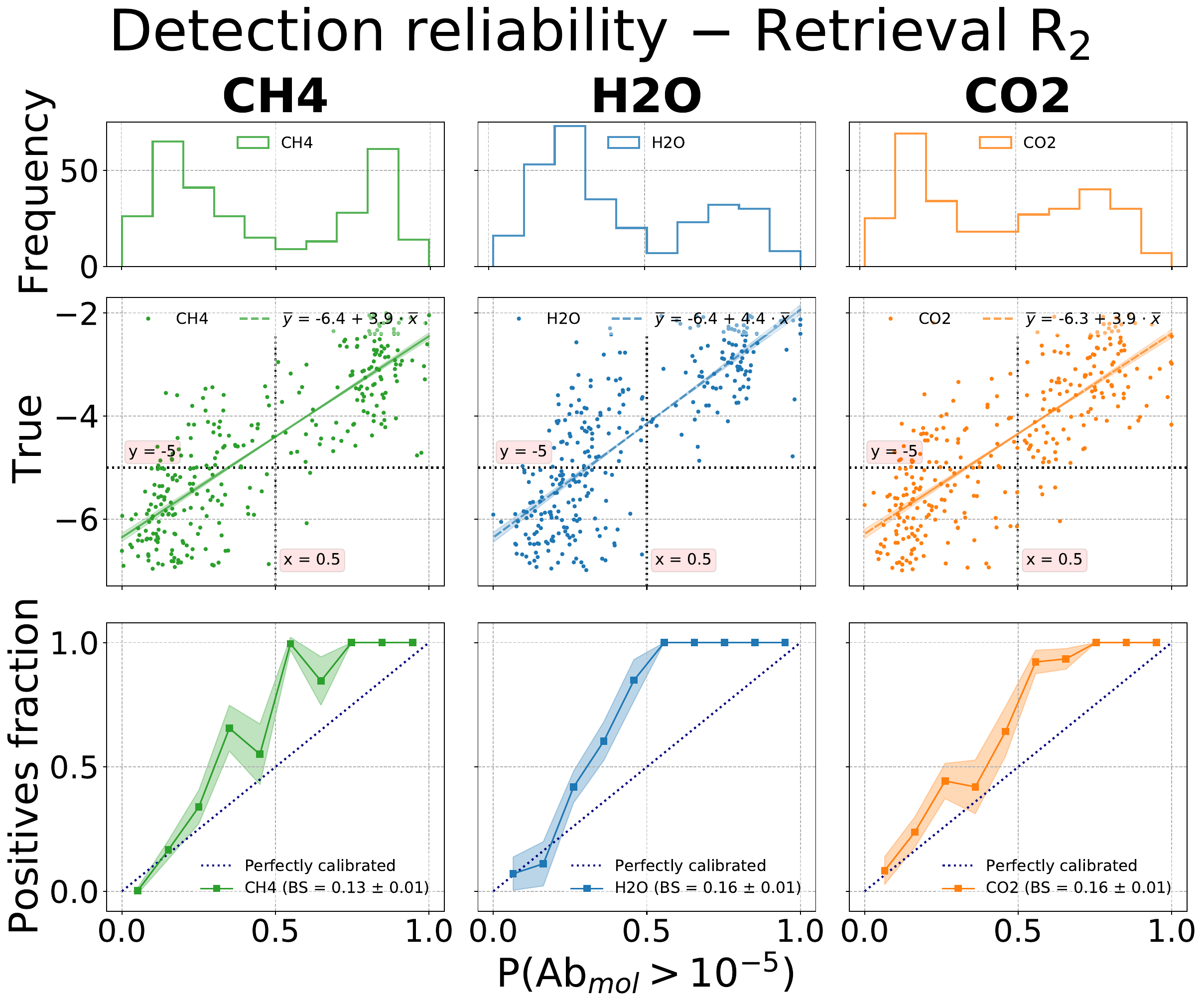}
	\caption{Same as Figure~\ref{fig:detrel0}. Detection reliability for the $\mathrm{R}_2$ retrievals, that implement a model that is over-representative of the simulated atmospheres, by including CO, HCN, and H$_2$S as additional trace gases.
		\label{fig:detrel2}}
\end{figure}

\begin{figure}[htb!]
	\centering
	\includegraphics[width=\textwidth]{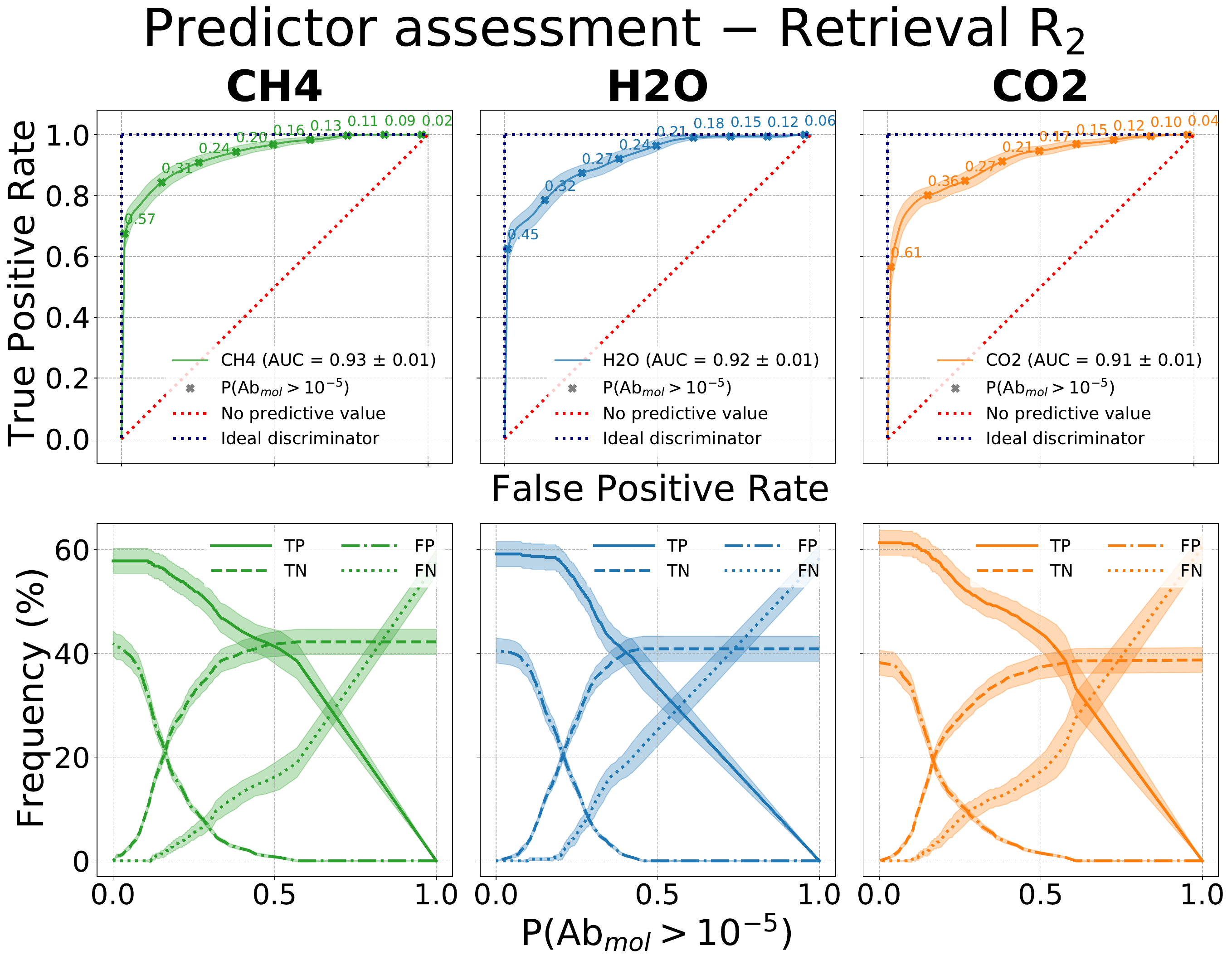}
	\caption{Same as Figure~\ref{fig:predass0}. Predictor assessment for the $\mathrm{R}_2$ retrievals, that implement a model that is over-representative of the simulated atmospheres, by including CO, HCN, and H$_2$S as additional trace gases. 
            \label{fig:predass2}}
\end{figure}

\begin{figure}[htb!]
	\centering
	\includegraphics[width=\textwidth]{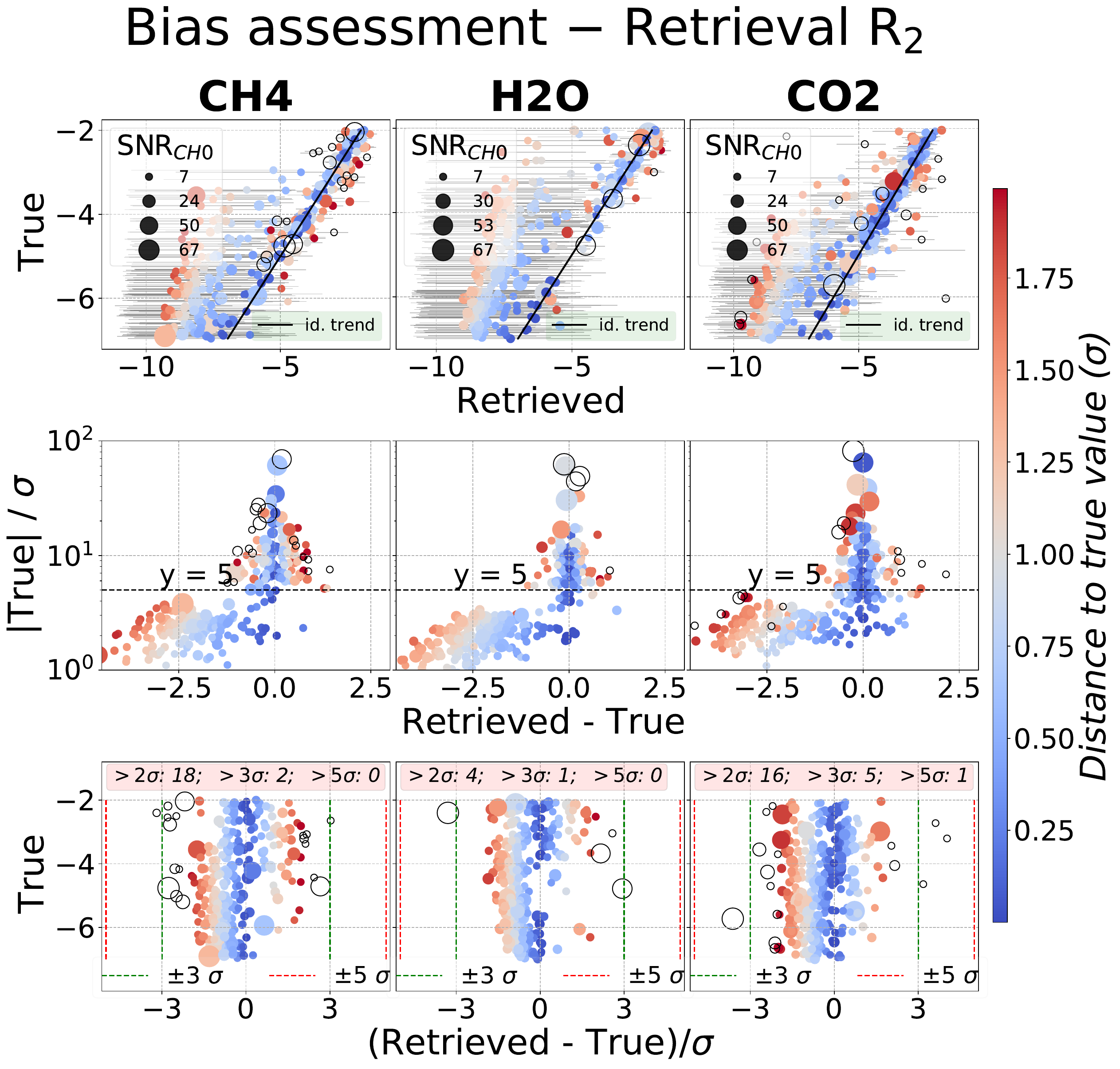}
	\caption{Same as Figure~\ref{fig:bias0} for the $\mathrm{R}_2$ retrievals, that implement a model that is over-representative of the simulated atmospheres, by including CO, HCN, and H$_2$S as additional trace gases.
		\label{fig:bias2}}
\end{figure}

\end{appendices}
\clearpage

\bibliography{biblio}


\begin{thebibliography}{78}
\ifx \bisbn   \undefined \def \bisbn  #1{ISBN #1}\fi
\ifx \binits  \undefined \def \binits#1{#1}\fi
\ifx \bauthor  \undefined \def \bauthor#1{#1}\fi
\ifx \batitle  \undefined \def \batitle#1{#1}\fi
\ifx \bjtitle  \undefined \def \bjtitle#1{#1}\fi
\ifx \bvolume  \undefined \def \bvolume#1{\textbf{#1}}\fi
\ifx \byear  \undefined \def \byear#1{#1}\fi
\ifx \bissue  \undefined \def \bissue#1{#1}\fi
\ifx \bfpage  \undefined \def \bfpage#1{#1}\fi
\ifx \blpage  \undefined \def \blpage #1{#1}\fi
\ifx \burl  \undefined \def \burl#1{\textsf{#1}}\fi
\ifx \doiurl  \undefined \def \doiurl#1{\url{https://doi.org/#1}}\fi
\ifx \betal  \undefined \def \betal{\textit{et al.}}\fi
\ifx \binstitute  \undefined \def \binstitute#1{#1}\fi
\ifx \binstitutionaled  \undefined \def \binstitutionaled#1{#1}\fi
\ifx \bctitle  \undefined \def \bctitle#1{#1}\fi
\ifx \beditor  \undefined \def \beditor#1{#1}\fi
\ifx \bpublisher  \undefined \def \bpublisher#1{#1}\fi
\ifx \bbtitle  \undefined \def \bbtitle#1{#1}\fi
\ifx \bedition  \undefined \def \bedition#1{#1}\fi
\ifx \bseriesno  \undefined \def \bseriesno#1{#1}\fi
\ifx \blocation  \undefined \def \blocation#1{#1}\fi
\ifx \bsertitle  \undefined \def \bsertitle#1{#1}\fi
\ifx \bsnm \undefined \def \bsnm#1{#1}\fi
\ifx \bsuffix \undefined \def \bsuffix#1{#1}\fi
\ifx \bparticle \undefined \def \bparticle#1{#1}\fi
\ifx \barticle \undefined \def \barticle#1{#1}\fi
\bibcommenthead
\ifx \bconfdate \undefined \def \bconfdate #1{#1}\fi
\ifx \botherref \undefined \def \botherref #1{#1}\fi
\ifx \url \undefined \def \url#1{\textsf{#1}}\fi
\ifx \bchapter \undefined \def \bchapter#1{#1}\fi
\ifx \bbook \undefined \def \bbook#1{#1}\fi
\ifx \bcomment \undefined \def \bcomment#1{#1}\fi
\ifx \oauthor \undefined \def \oauthor#1{#1}\fi
\ifx \citeauthoryear \undefined \def \citeauthoryear#1{#1}\fi
\ifx \endbibitem  \undefined \def \endbibitem {}\fi
\ifx \bconflocation  \undefined \def \bconflocation#1{#1}\fi
\ifx \arxivurl  \undefined \def \arxivurl#1{\textsf{#1}}\fi
\csname PreBibitemsHook\endcsname

\bibitem{KeplerBorucki2011}
\begin{barticle}
\bauthor{\bsnm{{Borucki}}, \binits{W.J.}},
\bauthor{\bsnm{{Koch}}, \binits{D.G.}},
\bauthor{\bsnm{{Basri}}, \binits{G.}},
\bauthor{\bsnm{{Batalha}}, \binits{N.}},
\bauthor{\bsnm{{Brown}}, \binits{T.M.}},
\bauthor{\bsnm{{Bryson}}, \binits{S.T.}},
\bauthor{\bsnm{{Caldwell}}, \binits{D.}},
\bauthor{\bsnm{{Christensen-Dalsgaard}}, \binits{J.}},
\bauthor{\bsnm{{Cochran}}, \binits{W.D.}},
\bauthor{\bsnm{{DeVore}}, \binits{E.}},
\bauthor{\bsnm{{Dunham}}, \binits{E.W.}},
\bauthor{\bsnm{{Gautier}}, \binits{I.} \bsuffix{Thomas~N.}},
\bauthor{\bsnm{{Geary}}, \binits{J.C.}},
\bauthor{\bsnm{{Gilliland}}, \binits{R.}},
\bauthor{\bsnm{{Gould}}, \binits{A.}},
\bauthor{\bsnm{{Howell}}, \binits{S.B.}},
\bauthor{\bsnm{{Jenkins}}, \binits{J.M.}},
\bauthor{\bsnm{{Latham}}, \binits{D.W.}},
\bauthor{\bsnm{{Lissauer}}, \binits{J.J.}},
\bauthor{\bsnm{{Marcy}}, \binits{G.W.}},
\bauthor{\bsnm{{Rowe}}, \binits{J.}},
\bauthor{\bsnm{{Sasselov}}, \binits{D.}},
\bauthor{\bsnm{{Boss}}, \binits{A.}},
\bauthor{\bsnm{{Charbonneau}}, \binits{D.}},
\bauthor{\bsnm{{Ciardi}}, \binits{D.}},
\bauthor{\bsnm{{Doyle}}, \binits{L.}},
\bauthor{\bsnm{{Dupree}}, \binits{A.K.}},
\bauthor{\bsnm{{Ford}}, \binits{E.B.}},
\bauthor{\bsnm{{Fortney}}, \binits{J.}},
\bauthor{\bsnm{{Holman}}, \binits{M.J.}},
\bauthor{\bsnm{{Seager}}, \binits{S.}},
\bauthor{\bsnm{{Steffen}}, \binits{J.H.}},
\bauthor{\bsnm{{Tarter}}, \binits{J.}},
\bauthor{\bsnm{{Welsh}}, \binits{W.F.}},
\bauthor{\bsnm{{Allen}}, \binits{C.}},
\bauthor{\bsnm{{Buchhave}}, \binits{L.A.}},
\bauthor{\bsnm{{Christiansen}}, \binits{J.L.}},
\bauthor{\bsnm{{Clarke}}, \binits{B.D.}},
\bauthor{\bsnm{{Das}}, \binits{S.}},
\bauthor{\bsnm{{D{\'e}sert}}, \binits{J.-M.}},
\bauthor{\bsnm{{Endl}}, \binits{M.}},
\bauthor{\bsnm{{Fabrycky}}, \binits{D.}},
\bauthor{\bsnm{{Fressin}}, \binits{F.}},
\bauthor{\bsnm{{Haas}}, \binits{M.}},
\bauthor{\bsnm{{Horch}}, \binits{E.}},
\bauthor{\bsnm{{Howard}}, \binits{A.}},
\bauthor{\bsnm{{Isaacson}}, \binits{H.}},
\bauthor{\bsnm{{Kjeldsen}}, \binits{H.}},
\bauthor{\bsnm{{Kolodziejczak}}, \binits{J.}},
\bauthor{\bsnm{{Kulesa}}, \binits{C.}},
\bauthor{\bsnm{{Li}}, \binits{J.}},
\bauthor{\bsnm{{Lucas}}, \binits{P.W.}},
\bauthor{\bsnm{{Machalek}}, \binits{P.}},
\bauthor{\bsnm{{McCarthy}}, \binits{D.}},
\bauthor{\bsnm{{MacQueen}}, \binits{P.}},
\bauthor{\bsnm{{Meibom}}, \binits{S.}},
\bauthor{\bsnm{{Miquel}}, \binits{T.}},
\bauthor{\bsnm{{Prsa}}, \binits{A.}},
\bauthor{\bsnm{{Quinn}}, \binits{S.N.}},
\bauthor{\bsnm{{Quintana}}, \binits{E.V.}},
\bauthor{\bsnm{{Ragozzine}}, \binits{D.}},
\bauthor{\bsnm{{Sherry}}, \binits{W.}},
\bauthor{\bsnm{{Shporer}}, \binits{A.}},
\bauthor{\bsnm{{Tenenbaum}}, \binits{P.}},
\bauthor{\bsnm{{Torres}}, \binits{G.}},
\bauthor{\bsnm{{Twicken}}, \binits{J.D.}},
\bauthor{\bsnm{{Van Cleve}}, \binits{J.}},
\bauthor{\bsnm{{Walkowicz}}, \binits{L.}},
\bauthor{\bsnm{{Witteborn}}, \binits{F.C.}},
\bauthor{\bsnm{{Still}}, \binits{M.}}:
\batitle{{Characteristics of Planetary Candidates Observed by Kepler. II.
  Analysis of the First Four Months of Data}}.
\bjtitle{\apj}
\bvolume{736}(\bissue{1}),
\bfpage{19}
(\byear{2011})
{\href{https://arxiv.org/abs/1102.0541}{{arXiv:1102.0541}}}
{[astro-ph.EP]}.
\doiurl{10.1088/0004-637X/736/1/19}
\end{barticle}
\endbibitem

\bibitem{KeplerBatalha2013}
\begin{barticle}
\bauthor{\bsnm{{Batalha}}, \binits{N.M.}},
\bauthor{\bsnm{{Rowe}}, \binits{J.F.}},
\bauthor{\bsnm{{Bryson}}, \binits{S.T.}},
\bauthor{\bsnm{{Barclay}}, \binits{T.}},
\bauthor{\bsnm{{Burke}}, \binits{C.J.}},
\bauthor{\bsnm{{Caldwell}}, \binits{D.A.}},
\bauthor{\bsnm{{Christiansen}}, \binits{J.L.}},
\bauthor{\bsnm{{Mullally}}, \binits{F.}},
\bauthor{\bsnm{{Thompson}}, \binits{S.E.}},
\bauthor{\bsnm{{Brown}}, \binits{T.M.}},
\bauthor{\bsnm{{Dupree}}, \binits{A.K.}},
\bauthor{\bsnm{{Fabrycky}}, \binits{D.C.}},
\bauthor{\bsnm{{Ford}}, \binits{E.B.}},
\bauthor{\bsnm{{Fortney}}, \binits{J.J.}},
\bauthor{\bsnm{{Gilliland}}, \binits{R.L.}},
\bauthor{\bsnm{{Isaacson}}, \binits{H.}},
\bauthor{\bsnm{{Latham}}, \binits{D.W.}},
\bauthor{\bsnm{{Marcy}}, \binits{G.W.}},
\bauthor{\bsnm{{Quinn}}, \binits{S.N.}},
\bauthor{\bsnm{{Ragozzine}}, \binits{D.}},
\bauthor{\bsnm{{Shporer}}, \binits{A.}},
\bauthor{\bsnm{{Borucki}}, \binits{W.J.}},
\bauthor{\bsnm{{Ciardi}}, \binits{D.R.}},
\bauthor{\bsnm{{Gautier}}, \binits{I.} \bsuffix{Thomas~N.}},
\bauthor{\bsnm{{Haas}}, \binits{M.R.}},
\bauthor{\bsnm{{Jenkins}}, \binits{J.M.}},
\bauthor{\bsnm{{Koch}}, \binits{D.G.}},
\bauthor{\bsnm{{Lissauer}}, \binits{J.J.}},
\bauthor{\bsnm{{Rapin}}, \binits{W.}},
\bauthor{\bsnm{{Basri}}, \binits{G.S.}},
\bauthor{\bsnm{{Boss}}, \binits{A.P.}},
\bauthor{\bsnm{{Buchhave}}, \binits{L.A.}},
\bauthor{\bsnm{{Carter}}, \binits{J.A.}},
\bauthor{\bsnm{{Charbonneau}}, \binits{D.}},
\bauthor{\bsnm{{Christensen-Dalsgaard}}, \binits{J.}},
\bauthor{\bsnm{{Clarke}}, \binits{B.D.}},
\bauthor{\bsnm{{Cochran}}, \binits{W.D.}},
\bauthor{\bsnm{{Demory}}, \binits{B.-O.}},
\bauthor{\bsnm{{Desert}}, \binits{J.-M.}},
\bauthor{\bsnm{{Devore}}, \binits{E.}},
\bauthor{\bsnm{{Doyle}}, \binits{L.R.}},
\bauthor{\bsnm{{Esquerdo}}, \binits{G.A.}},
\bauthor{\bsnm{{Everett}}, \binits{M.}},
\bauthor{\bsnm{{Fressin}}, \binits{F.}},
\bauthor{\bsnm{{Geary}}, \binits{J.C.}},
\bauthor{\bsnm{{Girouard}}, \binits{F.R.}},
\bauthor{\bsnm{{Gould}}, \binits{A.}},
\bauthor{\bsnm{{Hall}}, \binits{J.R.}},
\bauthor{\bsnm{{Holman}}, \binits{M.J.}},
\bauthor{\bsnm{{Howard}}, \binits{A.W.}},
\bauthor{\bsnm{{Howell}}, \binits{S.B.}},
\bauthor{\bsnm{{Ibrahim}}, \binits{K.A.}},
\bauthor{\bsnm{{Kinemuchi}}, \binits{K.}},
\bauthor{\bsnm{{Kjeldsen}}, \binits{H.}},
\bauthor{\bsnm{{Klaus}}, \binits{T.C.}},
\bauthor{\bsnm{{Li}}, \binits{J.}},
\bauthor{\bsnm{{Lucas}}, \binits{P.W.}},
\bauthor{\bsnm{{Meibom}}, \binits{S.}},
\bauthor{\bsnm{{Morris}}, \binits{R.L.}},
\bauthor{\bsnm{{Pr{\v{s}}a}}, \binits{A.}},
\bauthor{\bsnm{{Quintana}}, \binits{E.}},
\bauthor{\bsnm{{Sanderfer}}, \binits{D.T.}},
\bauthor{\bsnm{{Sasselov}}, \binits{D.}},
\bauthor{\bsnm{{Seader}}, \binits{S.E.}},
\bauthor{\bsnm{{Smith}}, \binits{J.C.}},
\bauthor{\bsnm{{Steffen}}, \binits{J.H.}},
\bauthor{\bsnm{{Still}}, \binits{M.}},
\bauthor{\bsnm{{Stumpe}}, \binits{M.C.}},
\bauthor{\bsnm{{Tarter}}, \binits{J.C.}},
\bauthor{\bsnm{{Tenenbaum}}, \binits{P.}},
\bauthor{\bsnm{{Torres}}, \binits{G.}},
\bauthor{\bsnm{{Twicken}}, \binits{J.D.}},
\bauthor{\bsnm{{Uddin}}, \binits{K.}},
\bauthor{\bsnm{{Van Cleve}}, \binits{J.}},
\bauthor{\bsnm{{Walkowicz}}, \binits{L.}},
\bauthor{\bsnm{{Welsh}}, \binits{W.F.}}:
\batitle{{Planetary Candidates Observed by Kepler. III. Analysis of the First
  16 Months of Data}}.
\bjtitle{\apjs}
\bvolume{204}(\bissue{2}),
\bfpage{24}
(\byear{2013})
{\href{https://arxiv.org/abs/1202.5852}{{arXiv:1202.5852}}}
{[astro-ph.EP]}.
\doiurl{10.1088/0067-0049/204/2/24}
\end{barticle}
\endbibitem

\bibitem{Ricker2016}
\begin{bchapter}
\bauthor{\bsnm{{Ricker}}, \binits{G.R.}},
\bauthor{\bsnm{{Winn}}, \binits{J.N.}},
\bauthor{\bsnm{{Vanderspek}}, \binits{R.}},
\bauthor{\bsnm{{Latham}}, \binits{D.W.}},
\bauthor{\bsnm{{Bakos}}, \binits{G.{\'A}.}},
\bauthor{\bsnm{{Bean}}, \binits{J.L.}},
\bauthor{\bsnm{{Berta-Thompson}}, \binits{Z.K.}},
\bauthor{\bsnm{{Brown}}, \binits{T.M.}},
\bauthor{\bsnm{{Buchhave}}, \binits{L.}},
\bauthor{\bsnm{{Butler}}, \binits{N.R.}},
\bauthor{\bsnm{{Butler}}, \binits{R.P.}},
\bauthor{\bsnm{{Chaplin}}, \binits{W.J.}},
\bauthor{\bsnm{{Charbonneau}}, \binits{D.}},
\bauthor{\bsnm{{Christensen-Dalsgaard}}, \binits{J.}},
\bauthor{\bsnm{{Clampin}}, \binits{M.}},
\bauthor{\bsnm{{Deming}}, \binits{D.}},
\bauthor{\bsnm{{Doty}}, \binits{J.}},
\bauthor{\bsnm{{De Lee}}, \binits{N.}},
\bauthor{\bsnm{{Dressing}}, \binits{C.}},
\bauthor{\bsnm{{Dunham}}, \binits{E.W.}},
\bauthor{\bsnm{{Endl}}, \binits{M.}},
\bauthor{\bsnm{{Fressin}}, \binits{F.}},
\bauthor{\bsnm{{Ge}}, \binits{J.}},
\bauthor{\bsnm{{Henning}}, \binits{T.}},
\bauthor{\bsnm{{Holman}}, \binits{M.J.}},
\bauthor{\bsnm{{Howard}}, \binits{A.W.}},
\bauthor{\bsnm{{Ida}}, \binits{S.}},
\bauthor{\bsnm{{Jenkins}}, \binits{J.}},
\bauthor{\bsnm{{Jernigan}}, \binits{G.}},
\bauthor{\bsnm{{Johnson}}, \binits{J.A.}},
\bauthor{\bsnm{{Kaltenegger}}, \binits{L.}},
\bauthor{\bsnm{{Kawai}}, \binits{N.}},
\bauthor{\bsnm{{Kjeldsen}}, \binits{H.}},
\bauthor{\bsnm{{Laughlin}}, \binits{G.}},
\bauthor{\bsnm{{Levine}}, \binits{A.M.}},
\bauthor{\bsnm{{Lin}}, \binits{D.}},
\bauthor{\bsnm{{Lissauer}}, \binits{J.J.}},
\bauthor{\bsnm{{MacQueen}}, \binits{P.}},
\bauthor{\bsnm{{Marcy}}, \binits{G.}},
\bauthor{\bsnm{{McCullough}}, \binits{P.R.}},
\bauthor{\bsnm{{Morton}}, \binits{T.D.}},
\bauthor{\bsnm{{Narita}}, \binits{N.}},
\bauthor{\bsnm{{Paegert}}, \binits{M.}},
\bauthor{\bsnm{{Palle}}, \binits{E.}},
\bauthor{\bsnm{{Pepe}}, \binits{F.}},
\bauthor{\bsnm{{Pepper}}, \binits{J.}},
\bauthor{\bsnm{{Quirrenbach}}, \binits{A.}},
\bauthor{\bsnm{{Rinehart}}, \binits{S.A.}},
\bauthor{\bsnm{{Sasselov}}, \binits{D.}},
\bauthor{\bsnm{{Sato}}, \binits{B.}},
\bauthor{\bsnm{{Seager}}, \binits{S.}},
\bauthor{\bsnm{{Sozzetti}}, \binits{A.}},
\bauthor{\bsnm{{Stassun}}, \binits{K.G.}},
\bauthor{\bsnm{{Sullivan}}, \binits{P.}},
\bauthor{\bsnm{{Szentgyorgyi}}, \binits{A.}},
\bauthor{\bsnm{{Torres}}, \binits{G.}},
\bauthor{\bsnm{{Udry}}, \binits{S.}},
\bauthor{\bsnm{{Villasenor}}, \binits{J.}}:
\bctitle{{Transiting Exoplanet Survey Satellite (TESS)}}.
In: \beditor{\bsnm{{Oschmann}}, \binits{J.} \bsuffix{Jacobus~M.}},
\beditor{\bsnm{{Clampin}}, \binits{M.}},
\beditor{\bsnm{{Fazio}}, \binits{G.G.}},
\beditor{\bsnm{{MacEwen}}, \binits{H.A.}} (eds.)
\bbtitle{Space Telescopes and Instrumentation 2014: Optical, Infrared, and
  Millimeter Wave}.
\bsertitle{Society of Photo-Optical Instrumentation Engineers (SPIE) Conference
  Series},
vol. \bseriesno{9143},
p. \bfpage{914320}
(\byear{2014}).
\doiurl{10.1117/12.2063489}
\end{bchapter}
\endbibitem

\bibitem{Cessa2017}
\begin{bchapter}
\bauthor{\bsnm{{Fortier}}, \binits{A.}},
\bauthor{\bsnm{{Beck}}, \binits{T.}},
\bauthor{\bsnm{{Benz}}, \binits{W.}},
\bauthor{\bsnm{{Broeg}}, \binits{C.}},
\bauthor{\bsnm{{Cessa}}, \binits{V.}},
\bauthor{\bsnm{{Ehrenreich}}, \binits{D.}},
\bauthor{\bsnm{{Thomas}}, \binits{N.}}:
\bctitle{{CHEOPS: a space telescope for ultra-high precision photometry of
  exoplanet transits}}.
In: \beditor{\bsnm{{Oschmann}}, \binits{J.} \bsuffix{Jacobus~M.}},
\beditor{\bsnm{{Clampin}}, \binits{M.}},
\beditor{\bsnm{{Fazio}}, \binits{G.G.}},
\beditor{\bsnm{{MacEwen}}, \binits{H.A.}} (eds.)
\bbtitle{Space Telescopes and Instrumentation 2014: Optical, Infrared, and
  Millimeter Wave}.
\bsertitle{Society of Photo-Optical Instrumentation Engineers (SPIE) Conference
  Series},
vol. \bseriesno{9143},
p. \bfpage{91432}
(\byear{2014}).
\doiurl{10.1117/12.2056687}
\end{bchapter}
\endbibitem

\bibitem{Rauer2014}
\begin{barticle}
\bauthor{\bsnm{{Rauer}}, \binits{H.}},
\bauthor{\bsnm{{Catala}}, \binits{C.}},
\bauthor{\bsnm{{Aerts}}, \binits{C.}},
\bauthor{\bsnm{{Appourchaux}}, \binits{T.}},
\bauthor{\bsnm{{Benz}}, \binits{W.}},
\bauthor{\bsnm{{Brandeker}}, \binits{A.}},
\bauthor{\bsnm{{Christensen-Dalsgaard}}, \binits{J.}},
\bauthor{\bsnm{{Deleuil}}, \binits{M.}},
\bauthor{\bsnm{{Gizon}}, \binits{L.}},
\bauthor{\bsnm{{Goupil}}, \binits{M.-J.}},
\bauthor{\bsnm{{G{\"u}del}}, \binits{M.}},
\bauthor{\bsnm{{Janot-Pacheco}}, \binits{E.}},
\bauthor{\bsnm{{Mas-Hesse}}, \binits{M.}},
\bauthor{\bsnm{{Pagano}}, \binits{I.}},
\bauthor{\bsnm{{Piotto}}, \binits{G.}},
\bauthor{\bsnm{{Pollacco}}, \binits{D.}},
\bauthor{\bsnm{{Santos}}, \binits{{\. {C}}.}},
\bauthor{\bsnm{{Smith}}, \binits{A.}},
\bauthor{\bsnm{{Su{\'a}rez}}, \binits{J.-C.}},
\bauthor{\bsnm{{Szab{\'o}}}, \binits{R.}},
\bauthor{\bsnm{{Udry}}, \binits{S.}},
\bauthor{\bsnm{{Adibekyan}}, \binits{V.}},
\bauthor{\bsnm{{Alibert}}, \binits{Y.}},
\bauthor{\bsnm{{Almenara}}, \binits{J.-M.}},
\bauthor{\bsnm{{Amaro-Seoane}}, \binits{P.}},
\bauthor{\bsnm{{Eiff}}, \binits{M.A.-v.}},
\bauthor{\bsnm{{Asplund}}, \binits{M.}},
\bauthor{\bsnm{{Antonello}}, \binits{E.}},
\bauthor{\bsnm{{Barnes}}, \binits{S.}},
\bauthor{\bsnm{{Baudin}}, \binits{F.}},
\bauthor{\bsnm{{Belkacem}}, \binits{K.}},
\bauthor{\bsnm{{Bergemann}}, \binits{M.}},
\bauthor{\bsnm{{Bihain}}, \binits{G.}},
\bauthor{\bsnm{{Birch}}, \binits{A.C.}},
\bauthor{\bsnm{{Bonfils}}, \binits{X.}},
\bauthor{\bsnm{{Boisse}}, \binits{I.}},
\bauthor{\bsnm{{Bonomo}}, \binits{A.S.}},
\bauthor{\bsnm{{Borsa}}, \binits{F.}},
\bauthor{\bsnm{{Brand{\~a}o}}, \binits{I.M.}},
\bauthor{\bsnm{{Brocato}}, \binits{E.}},
\bauthor{\bsnm{{Brun}}, \binits{S.}},
\bauthor{\bsnm{{Burleigh}}, \binits{M.}},
\bauthor{\bsnm{{Burston}}, \binits{R.}},
\bauthor{\bsnm{{Cabrera}}, \binits{J.}},
\bauthor{\bsnm{{Cassisi}}, \binits{S.}},
\bauthor{\bsnm{{Chaplin}}, \binits{W.}},
\bauthor{\bsnm{{Charpinet}}, \binits{S.}},
\bauthor{\bsnm{{Chiappini}}, \binits{C.}},
\bauthor{\bsnm{{Church}}, \binits{R.P.}},
\bauthor{\bsnm{{Csizmadia}}, \binits{S.}},
\bauthor{\bsnm{{Cunha}}, \binits{M.}},
\bauthor{\bsnm{{Damasso}}, \binits{M.}},
\bauthor{\bsnm{{Davies}}, \binits{M.B.}},
\bauthor{\bsnm{{Deeg}}, \binits{H.J.}},
\bauthor{\bsnm{{D{\'\i}az}}, \binits{R.F.}},
\bauthor{\bsnm{{Dreizler}}, \binits{S.}},
\bauthor{\bsnm{{Dreyer}}, \binits{C.}},
\bauthor{\bsnm{{Eggenberger}}, \binits{P.}},
\bauthor{\bsnm{{Ehrenreich}}, \binits{D.}},
\bauthor{\bsnm{{Eigm{\"u}ller}}, \binits{P.}},
\bauthor{\bsnm{{Erikson}}, \binits{A.}},
\bauthor{\bsnm{{Farmer}}, \binits{R.}},
\bauthor{\bsnm{{Feltzing}}, \binits{S.}},
\bauthor{\bsnm{{de Oliveira Fialho}}, \binits{F.}},
\bauthor{\bsnm{{Figueira}}, \binits{P.}},
\bauthor{\bsnm{{Forveille}}, \binits{T.}},
\bauthor{\bsnm{{Fridlund}}, \binits{M.}},
\bauthor{\bsnm{{Garc{\'\i}a}}, \binits{R.A.}},
\bauthor{\bsnm{{Giommi}}, \binits{P.}},
\bauthor{\bsnm{{Giuffrida}}, \binits{G.}},
\bauthor{\bsnm{{Godolt}}, \binits{M.}},
\bauthor{\bsnm{{Gomes da Silva}}, \binits{J.}},
\bauthor{\bsnm{{Granzer}}, \binits{T.}},
\bauthor{\bsnm{{Grenfell}}, \binits{J.L.}},
\bauthor{\bsnm{{Grotsch-Noels}}, \binits{A.}},
\bauthor{\bsnm{{G{\"u}nther}}, \binits{E.}},
\bauthor{\bsnm{{Haswell}}, \binits{C.A.}},
\bauthor{\bsnm{{Hatzes}}, \binits{A.P.}},
\bauthor{\bsnm{{H{\'e}brard}}, \binits{G.}},
\bauthor{\bsnm{{Hekker}}, \binits{S.}},
\bauthor{\bsnm{{Helled}}, \binits{R.}},
\bauthor{\bsnm{{Heng}}, \binits{K.}},
\bauthor{\bsnm{{Jenkins}}, \binits{J.M.}},
\bauthor{\bsnm{{Johansen}}, \binits{A.}},
\bauthor{\bsnm{{Khodachenko}}, \binits{M.L.}},
\bauthor{\bsnm{{Kislyakova}}, \binits{K.G.}},
\bauthor{\bsnm{{Kley}}, \binits{W.}},
\bauthor{\bsnm{{Kolb}}, \binits{U.}},
\bauthor{\bsnm{{Krivova}}, \binits{N.}},
\bauthor{\bsnm{{Kupka}}, \binits{F.}},
\bauthor{\bsnm{{Lammer}}, \binits{H.}},
\bauthor{\bsnm{{Lanza}}, \binits{A.F.}},
\bauthor{\bsnm{{Lebreton}}, \binits{Y.}},
\bauthor{\bsnm{{Magrin}}, \binits{D.}},
\bauthor{\bsnm{{Marcos-Arenal}}, \binits{P.}},
\bauthor{\bsnm{{Marrese}}, \binits{P.M.}},
\bauthor{\bsnm{{Marques}}, \binits{J.P.}},
\bauthor{\bsnm{{Martins}}, \binits{J.}},
\bauthor{\bsnm{{Mathis}}, \binits{S.}},
\bauthor{\bsnm{{Mathur}}, \binits{S.}},
\bauthor{\bsnm{{Messina}}, \binits{S.}},
\bauthor{\bsnm{{Miglio}}, \binits{A.}},
\bauthor{\bsnm{{Montalban}}, \binits{J.}},
\bauthor{\bsnm{{Montalto}}, \binits{M.}},
\bauthor{\bsnm{{Monteiro}}, \binits{M.J.P.F.G.}},
\bauthor{\bsnm{{Moradi}}, \binits{H.}},
\bauthor{\bsnm{{Moravveji}}, \binits{E.}},
\bauthor{\bsnm{{Mordasini}}, \binits{C.}},
\bauthor{\bsnm{{Morel}}, \binits{T.}},
\bauthor{\bsnm{{Mortier}}, \binits{A.}},
\bauthor{\bsnm{{Nascimbeni}}, \binits{V.}},
\bauthor{\bsnm{{Nelson}}, \binits{R.P.}},
\bauthor{\bsnm{{Nielsen}}, \binits{M.B.}},
\bauthor{\bsnm{{Noack}}, \binits{L.}},
\bauthor{\bsnm{{Norton}}, \binits{A.J.}},
\bauthor{\bsnm{{Ofir}}, \binits{A.}},
\bauthor{\bsnm{{Oshagh}}, \binits{M.}},
\bauthor{\bsnm{{Ouazzani}}, \binits{R.-M.}},
\bauthor{\bsnm{{P{\'a}pics}}, \binits{P.}},
\bauthor{\bsnm{{Parro}}, \binits{V.C.}},
\bauthor{\bsnm{{Petit}}, \binits{P.}},
\bauthor{\bsnm{{Plez}}, \binits{B.}},
\bauthor{\bsnm{{Poretti}}, \binits{E.}},
\bauthor{\bsnm{{Quirrenbach}}, \binits{A.}},
\bauthor{\bsnm{{Ragazzoni}}, \binits{R.}},
\bauthor{\bsnm{{Raimondo}}, \binits{G.}},
\bauthor{\bsnm{{Rainer}}, \binits{M.}},
\bauthor{\bsnm{{Reese}}, \binits{D.R.}},
\bauthor{\bsnm{{Redmer}}, \binits{R.}},
\bauthor{\bsnm{{Reffert}}, \binits{S.}},
\bauthor{\bsnm{{Rojas-Ayala}}, \binits{B.}},
\bauthor{\bsnm{{Roxburgh}}, \binits{I.W.}},
\bauthor{\bsnm{{Salmon}}, \binits{S.}},
\bauthor{\bsnm{{Santerne}}, \binits{A.}},
\bauthor{\bsnm{{Schneider}}, \binits{J.}},
\bauthor{\bsnm{{Schou}}, \binits{J.}},
\bauthor{\bsnm{{Schuh}}, \binits{S.}},
\bauthor{\bsnm{{Schunker}}, \binits{H.}},
\bauthor{\bsnm{{Silva-Valio}}, \binits{A.}},
\bauthor{\bsnm{{Silvotti}}, \binits{R.}},
\bauthor{\bsnm{{Skillen}}, \binits{I.}},
\bauthor{\bsnm{{Snellen}}, \binits{I.}},
\bauthor{\bsnm{{Sohl}}, \binits{F.}},
\bauthor{\bsnm{{Sousa}}, \binits{S.G.}},
\bauthor{\bsnm{{Sozzetti}}, \binits{A.}},
\bauthor{\bsnm{{Stello}}, \binits{D.}},
\bauthor{\bsnm{{Strassmeier}}, \binits{K.G.}},
\bauthor{\bsnm{{{\v{S}}vanda}}, \binits{M.}},
\bauthor{\bsnm{{Szab{\'o}}}, \binits{G.M.}},
\bauthor{\bsnm{{Tkachenko}}, \binits{A.}},
\bauthor{\bsnm{{Valencia}}, \binits{D.}},
\bauthor{\bsnm{{Van Grootel}}, \binits{V.}},
\bauthor{\bsnm{{Vauclair}}, \binits{S.D.}},
\bauthor{\bsnm{{Ventura}}, \binits{P.}},
\bauthor{\bsnm{{Wagner}}, \binits{F.W.}},
\bauthor{\bsnm{{Walton}}, \binits{N.A.}},
\bauthor{\bsnm{{Weingrill}}, \binits{J.}},
\bauthor{\bsnm{{Werner}}, \binits{S.C.}},
\bauthor{\bsnm{{Wheatley}}, \binits{P.J.}},
\bauthor{\bsnm{{Zwintz}}, \binits{K.}}:
\batitle{{The PLATO 2.0 mission}}.
\bjtitle{Experimental Astronomy}
\bvolume{38}(\bissue{1-2}),
\bfpage{249}--\blpage{330}
(\byear{2014})
{\href{https://arxiv.org/abs/1310.0696}{{arXiv:1310.0696}}}
{[astro-ph.EP]}.
\doiurl{10.1007/s10686-014-9383-4}
\end{barticle}
\endbibitem

\bibitem{GAIA2016}
\begin{barticle}
\bauthor{\bsnm{{Gaia Collaboration}}},
\bauthor{\bsnm{{Prusti, T.}}},
\bauthor{\bsnm{{de Bruijne, J. H. J.}}},
\bauthor{\bsnm{{Brown, A. G. A.}}},
\bauthor{\bsnm{{Vallenari, A.}}},
\bauthor{\bsnm{{Babusiaux, C.}}},
\bauthor{\bsnm{{Bailer-Jones, C. A. L.}}},
\bauthor{\bsnm{{Bastian, U.}}},
\bauthor{\bsnm{{Biermann, M.}}},
\bauthor{\bsnm{{Evans, D. W.}}},
\bauthor{\bsnm{{Eyer, L.}}},
\bauthor{\bsnm{{Jansen, F.}}},
\bauthor{\bsnm{{Jordi, C.}}},
\bauthor{\bsnm{{Klioner, S. A.}}},
\bauthor{\bsnm{{Lammers, U.}}},
\bauthor{\bsnm{{Lindegren, L.}}},
\bauthor{\bsnm{{Luri, X.}}},
\bauthor{\bsnm{{Mignard, F.}}},
\bauthor{\bsnm{{Milligan, D. J.}}},
\bauthor{\bsnm{{Panem, C.}}},
\bauthor{\bsnm{{Poinsignon, V.}}},
\bauthor{\bsnm{{Pourbaix, D.}}},
\bauthor{\bsnm{{Randich, S.}}},
\bauthor{\bsnm{{Sarri, G.}}},
\bauthor{\bsnm{{Sartoretti, P.}}},
\bauthor{\bsnm{{Siddiqui, H. I.}}},
\bauthor{\bsnm{{Soubiran, C.}}},
\bauthor{\bsnm{{Valette, V.}}},
\bauthor{\bsnm{{van Leeuwen, F.}}},
\bauthor{\bsnm{{Walton, N. A.}}},
\bauthor{\bsnm{{Aerts, C.}}},
\bauthor{\bsnm{{Arenou, F.}}},
\bauthor{\bsnm{{Cropper, M.}}},
\bauthor{\bsnm{{Drimmel, R.}}},
\bauthor{\bsnm{{H\o{}g, E.}}},
\bauthor{\bsnm{{Katz, D.}}},
\bauthor{\bsnm{{Lattanzi, M. G.}}},
\bauthor{\bsnm{{O\'{}Mullane, W.}}},
\bauthor{\bsnm{{Grebel, E. K.}}},
\bauthor{\bsnm{{Holland, A. D.}}},
\bauthor{\bsnm{{Huc, C.}}},
\bauthor{\bsnm{{Passot, X.}}},
\bauthor{\bsnm{{Bramante, L.}}},
\bauthor{\bsnm{{Cacciari, C.}}},
\bauthor{\bsnm{{Casta\~neda, J.}}},
\bauthor{\bsnm{{Chaoul, L.}}},
\bauthor{\bsnm{{Cheek, N.}}},
\bauthor{\bsnm{{De Angeli, F.}}},
\bauthor{\bsnm{{Fabricius, C.}}},
\bauthor{\bsnm{{Guerra, R.}}},
\bauthor{\bsnm{{Hern\'andez, J.}}},
\bauthor{\bsnm{{Jean-Antoine-Piccolo, A.}}},
\bauthor{\bsnm{{Masana, E.}}},
\bauthor{\bsnm{{Messineo, R.}}},
\bauthor{\bsnm{{Mowlavi, N.}}},
\bauthor{\bsnm{{Nienartowicz, K.}}},
\bauthor{\bsnm{{Ord\'o\~nez-Blanco, D.}}},
\bauthor{\bsnm{{Panuzzo, P.}}},
\bauthor{\bsnm{{Portell, J.}}},
\bauthor{\bsnm{{Richards, P. J.}}},
\bauthor{\bsnm{{Riello, M.}}},
\bauthor{\bsnm{{Seabroke, G. M.}}},
\bauthor{\bsnm{{Tanga, P.}}},
\bauthor{\bsnm{{Th\'evenin, F.}}},
\bauthor{\bsnm{{Torra, J.}}},
\bauthor{\bsnm{{Els, S. G.}}},
\bauthor{\bsnm{{Gracia-Abril, G.}}},
\bauthor{\bsnm{{Comoretto, G.}}},
\bauthor{\bsnm{{Garcia-Reinaldos, M.}}},
\bauthor{\bsnm{{Lock, T.}}},
\bauthor{\bsnm{{Mercier, E.}}},
\bauthor{\bsnm{{Altmann, M.}}},
\bauthor{\bsnm{{Andrae, R.}}},
\bauthor{\bsnm{{Astraatmadja, T. L.}}},
\bauthor{\bsnm{{Bellas-Velidis, I.}}},
\bauthor{\bsnm{{Benson, K.}}},
\bauthor{\bsnm{{Berthier, J.}}},
\bauthor{\bsnm{{Blomme, R.}}},
\bauthor{\bsnm{{Busso, G.}}},
\bauthor{\bsnm{{Carry, B.}}},
\bauthor{\bsnm{{Cellino, A.}}},
\bauthor{\bsnm{{Clementini, G.}}},
\bauthor{\bsnm{{Cowell, S.}}},
\bauthor{\bsnm{{Creevey, O.}}},
\bauthor{\bsnm{{Cuypers, J.}}},
\bauthor{\bsnm{{Davidson, M.}}},
\bauthor{\bsnm{{De Ridder, J.}}},
\bauthor{\bsnm{{de Torres, A.}}},
\bauthor{\bsnm{{Delchambre, L.}}},
\bauthor{\bsnm{{Dell\'{}Oro, A.}}},
\bauthor{\bsnm{{Ducourant, C.}}},
\bauthor{\bsnm{{Fr\'emat, Y.}}},
\bauthor{\bsnm{{Garc\'{\i}a-Torres, M.}}},
\bauthor{\bsnm{{Gosset, E.}}},
\bauthor{\bsnm{{Halbwachs, J.-L.}}},
\bauthor{\bsnm{{Hambly, N. C.}}},
\bauthor{\bsnm{{Harrison, D. L.}}},
\bauthor{\bsnm{{Hauser, M.}}},
\bauthor{\bsnm{{Hestroffer, D.}}},
\bauthor{\bsnm{{Hodgkin, S. T.}}},
\bauthor{\bsnm{{Huckle, H. E.}}},
\bauthor{\bsnm{{Hutton, A.}}},
\bauthor{\bsnm{{Jasniewicz, G.}}},
\bauthor{\bsnm{{Jordan, S.}}},
\bauthor{\bsnm{{Kontizas, M.}}},
\bauthor{\bsnm{{Korn, A. J.}}},
\bauthor{\bsnm{{Lanzafame, A. C.}}},
\bauthor{\bsnm{{Manteiga, M.}}},
\bauthor{\bsnm{{Moitinho, A.}}},
\bauthor{\bsnm{{Muinonen, K.}}},
\bauthor{\bsnm{{Osinde, J.}}},
\bauthor{\bsnm{{Pancino, E.}}},
\bauthor{\bsnm{{Pauwels, T.}}},
\bauthor{\bsnm{{Petit, J.-M.}}},
\bauthor{\bsnm{{Recio-Blanco, A.}}},
\bauthor{\bsnm{{Robin, A. C.}}},
\bauthor{\bsnm{{Sarro, L. M.}}},
\bauthor{\bsnm{{Siopis, C.}}},
\bauthor{\bsnm{{Smith, M.}}},
\bauthor{\bsnm{{Smith, K. W.}}},
\bauthor{\bsnm{{Sozzetti, A.}}},
\bauthor{\bsnm{{Thuillot, W.}}},
\bauthor{\bsnm{{van Reeven, W.}}},
\bauthor{\bsnm{{Viala, Y.}}},
\bauthor{\bsnm{{Abbas, U.}}},
\bauthor{\bsnm{{Abreu Aramburu, A.}}},
\bauthor{\bsnm{{Accart, S.}}},
\bauthor{\bsnm{{Aguado, J. J.}}},
\bauthor{\bsnm{{Allan, P. M.}}},
\bauthor{\bsnm{{Allasia, W.}}},
\bauthor{\bsnm{{Altavilla, G.}}},
\bauthor{\bsnm{{\'Alvarez, M. A.}}},
\bauthor{\bsnm{{Alves, J.}}},
\bauthor{\bsnm{{Anderson, R. I.}}},
\bauthor{\bsnm{{Andrei, A. H.}}},
\bauthor{\bsnm{{Anglada Varela, E.}}},
\bauthor{\bsnm{{Antiche, E.}}},
\bauthor{\bsnm{{Antoja, T.}}},
\bauthor{\bsnm{{Ant\'on, S.}}},
\bauthor{\bsnm{{Arcay, B.}}},
\bauthor{\bsnm{{Atzei, A.}}},
\bauthor{\bsnm{{Ayache, L.}}},
\bauthor{\bsnm{{Bach, N.}}},
\bauthor{\bsnm{{Baker, S. G.}}},
\bauthor{\bsnm{{Balaguer-N\'u\~nez, L.}}},
\bauthor{\bsnm{{Barache, C.}}},
\bauthor{\bsnm{{Barata, C.}}},
\bauthor{\bsnm{{Barbier, A.}}},
\bauthor{\bsnm{{Barblan, F.}}},
\bauthor{\bsnm{{Baroni, M.}}},
\bauthor{\bsnm{{Barrado y Navascu\'es, D.}}},
\bauthor{\bsnm{{Barros, M.}}},
\bauthor{\bsnm{{Barstow, M. A.}}},
\bauthor{\bsnm{{Becciani, U.}}},
\bauthor{\bsnm{{Bellazzini, M.}}},
\bauthor{\bsnm{{Bellei, G.}}},
\bauthor{\bsnm{{Bello Garc\'{\i}a, A.}}},
\bauthor{\bsnm{{Belokurov, V.}}},
\bauthor{\bsnm{{Bendjoya, P.}}},
\bauthor{\bsnm{{Berihuete, A.}}},
\bauthor{\bsnm{{Bianchi, L.}}},
\bauthor{\bsnm{{Bienaym\'e, O.}}},
\bauthor{\bsnm{{Billebaud, F.}}},
\bauthor{\bsnm{{Blagorodnova, N.}}},
\bauthor{\bsnm{{Blanco-Cuaresma, S.}}},
\bauthor{\bsnm{{Boch, T.}}},
\bauthor{\bsnm{{Bombrun, A.}}},
\bauthor{\bsnm{{Borrachero, R.}}},
\bauthor{\bsnm{{Bouquillon, S.}}},
\bauthor{\bsnm{{Bourda, G.}}},
\bauthor{\bsnm{{Bouy, H.}}},
\bauthor{\bsnm{{Bragaglia, A.}}},
\bauthor{\bsnm{{Breddels, M. A.}}},
\bauthor{\bsnm{{Brouillet, N.}}},
\bauthor{\bsnm{{Br\"usemeister, T.}}},
\bauthor{\bsnm{{Bucciarelli, B.}}},
\bauthor{\bsnm{{Budnik, F.}}},
\bauthor{\bsnm{{Burgess, P.}}},
\bauthor{\bsnm{{Burgon, R.}}},
\bauthor{\bsnm{{Burlacu, A.}}},
\bauthor{\bsnm{{Busonero, D.}}},
\bauthor{\bsnm{{Buzzi, R.}}},
\bauthor{\bsnm{{Caffau, E.}}},
\bauthor{\bsnm{{Cambras, J.}}},
\bauthor{\bsnm{{Campbell, H.}}},
\bauthor{\bsnm{{Cancelliere, R.}}},
\bauthor{\bsnm{{Cantat-Gaudin, T.}}},
\bauthor{\bsnm{{Carlucci, T.}}},
\bauthor{\bsnm{{Carrasco, J. M.}}},
\bauthor{\bsnm{{Castellani, M.}}},
\bauthor{\bsnm{{Charlot, P.}}},
\bauthor{\bsnm{{Charnas, J.}}},
\bauthor{\bsnm{{Charvet, P.}}},
\bauthor{\bsnm{{Chassat, F.}}},
\bauthor{\bsnm{{Chiavassa, A.}}},
\bauthor{\bsnm{{Clotet, M.}}},
\bauthor{\bsnm{{Cocozza, G.}}},
\bauthor{\bsnm{{Collins, R. S.}}},
\bauthor{\bsnm{{Collins, P.}}},
\bauthor{\bsnm{{Costigan, G.}}},
\bauthor{\bsnm{{Crifo, F.}}},
\bauthor{\bsnm{{Cross, N. J. G.}}},
\bauthor{\bsnm{{Crosta, M.}}},
\bauthor{\bsnm{{Crowley, C.}}},
\bauthor{\bsnm{{Dafonte, C.}}},
\bauthor{\bsnm{{Damerdji, Y.}}},
\bauthor{\bsnm{{Dapergolas, A.}}},
\bauthor{\bsnm{{David, P.}}},
\bauthor{\bsnm{{David, M.}}},
\bauthor{\bsnm{{De Cat, P.}}},
\bauthor{\bsnm{{de Felice, F.}}},
\bauthor{\bsnm{{de Laverny, P.}}},
\bauthor{\bsnm{{De Luise, F.}}},
\bauthor{\bsnm{{De March, R.}}},
\bauthor{\bsnm{{de Martino, D.}}},
\bauthor{\bsnm{{de Souza, R.}}},
\bauthor{\bsnm{{Debosscher, J.}}},
\bauthor{\bsnm{{del Pozo, E.}}},
\bauthor{\bsnm{{Delbo, M.}}},
\bauthor{\bsnm{{Delgado, A.}}},
\bauthor{\bsnm{{Delgado, H. E.}}},
\bauthor{\bsnm{{di Marco, F.}}},
\bauthor{\bsnm{{Di Matteo, P.}}},
\bauthor{\bsnm{{Diakite, S.}}},
\bauthor{\bsnm{{Distefano, E.}}},
\bauthor{\bsnm{{Dolding, C.}}},
\bauthor{\bsnm{{Dos Anjos, S.}}},
\bauthor{\bsnm{{Drazinos, P.}}},
\bauthor{\bsnm{{Dur\'an, J.}}},
\bauthor{\bsnm{{Dzigan, Y.}}},
\bauthor{\bsnm{{Ecale, E.}}},
\bauthor{\bsnm{{Edvardsson, B.}}},
\bauthor{\bsnm{{Enke, H.}}},
\bauthor{\bsnm{{Erdmann, M.}}},
\bauthor{\bsnm{{Escolar, D.}}},
\bauthor{\bsnm{{Espina, M.}}},
\bauthor{\bsnm{{Evans, N. W.}}},
\bauthor{\bsnm{{Eynard Bontemps, G.}}},
\bauthor{\bsnm{{Fabre, C.}}},
\bauthor{\bsnm{{Fabrizio, M.}}},
\bauthor{\bsnm{{Faigler, S.}}},
\bauthor{\bsnm{{Falc\~ao, A. J.}}},
\bauthor{\bsnm{{Farr\`as Casas, M.}}},
\bauthor{\bsnm{{Faye, F.}}},
\bauthor{\bsnm{{Federici, L.}}},
\bauthor{\bsnm{{Fedorets, G.}}},
\bauthor{\bsnm{{Fern\'andez-Hern\'andez, J.}}},
\bauthor{\bsnm{{Fernique, P.}}},
\bauthor{\bsnm{{Fienga, A.}}},
\bauthor{\bsnm{{Figueras, F.}}},
\bauthor{\bsnm{{Filippi, F.}}},
\bauthor{\bsnm{{Findeisen, K.}}},
\bauthor{\bsnm{{Fonti, A.}}},
\bauthor{\bsnm{{Fouesneau, M.}}},
\bauthor{\bsnm{{Fraile, E.}}},
\bauthor{\bsnm{{Fraser, M.}}},
\bauthor{\bsnm{{Fuchs, J.}}},
\bauthor{\bsnm{{Furnell, R.}}},
\bauthor{\bsnm{{Gai, M.}}},
\bauthor{\bsnm{{Galleti, S.}}},
\bauthor{\bsnm{{Galluccio, L.}}},
\bauthor{\bsnm{{Garabato, D.}}},
\bauthor{\bsnm{{Garc\'{\i}a-Sedano, F.}}},
\bauthor{\bsnm{{Gar\'e, P.}}},
\bauthor{\bsnm{{Garofalo, A.}}},
\bauthor{\bsnm{{Garralda, N.}}},
\bauthor{\bsnm{{Gavras, P.}}},
\bauthor{\bsnm{{Gerssen, J.}}},
\bauthor{\bsnm{{Geyer, R.}}},
\bauthor{\bsnm{{Gilmore, G.}}},
\bauthor{\bsnm{{Girona, S.}}},
\bauthor{\bsnm{{Giuffrida, G.}}},
\bauthor{\bsnm{{Gomes, M.}}},
\bauthor{\bsnm{{Gonz\'alez-Marcos, A.}}},
\bauthor{\bsnm{{Gonz\'alez-N\'u\~nez, J.}}},
\bauthor{\bsnm{{Gonz\'alez-Vidal, J. J.}}},
\bauthor{\bsnm{{Granvik, M.}}},
\bauthor{\bsnm{{Guerrier, A.}}},
\bauthor{\bsnm{{Guillout, P.}}},
\bauthor{\bsnm{{Guiraud, J.}}},
\bauthor{\bsnm{{G\'urpide, A.}}},
\bauthor{\bsnm{{Guti\'errez-S\'anchez, R.}}},
\bauthor{\bsnm{{Guy, L. P.}}},
\bauthor{\bsnm{{Haigron, R.}}},
\bauthor{\bsnm{{Hatzidimitriou, D.}}},
\bauthor{\bsnm{{Haywood, M.}}},
\bauthor{\bsnm{{Heiter, U.}}},
\bauthor{\bsnm{{Helmi, A.}}},
\bauthor{\bsnm{{Hobbs, D.}}},
\bauthor{\bsnm{{Hofmann, W.}}},
\bauthor{\bsnm{{Holl, B.}}},
\bauthor{\bsnm{{Holland, G.}}},
\bauthor{\bsnm{{Hunt, J. A. S.}}},
\bauthor{\bsnm{{Hypki, A.}}},
\bauthor{\bsnm{{Icardi, V.}}},
\bauthor{\bsnm{{Irwin, M.}}},
\bauthor{\bsnm{{Jevardat de Fombelle, G.}}},
\bauthor{\bsnm{{Jofr\'e, P.}}},
\bauthor{\bsnm{{Jonker, P. G.}}},
\bauthor{\bsnm{{Jorissen, A.}}},
\bauthor{\bsnm{{Julbe, F.}}},
\bauthor{\bsnm{{Karampelas, A.}}},
\bauthor{\bsnm{{Kochoska, A.}}},
\bauthor{\bsnm{{Kohley, R.}}},
\bauthor{\bsnm{{Kolenberg, K.}}},
\bauthor{\bsnm{{Kontizas, E.}}},
\bauthor{\bsnm{{Koposov, S. E.}}},
\bauthor{\bsnm{{Kordopatis, G.}}},
\bauthor{\bsnm{{Koubsky, P.}}},
\bauthor{\bsnm{{Kowalczyk, A.}}},
\bauthor{\bsnm{{Krone-Martins, A.}}},
\bauthor{\bsnm{{Kudryashova, M.}}},
\bauthor{\bsnm{{Kull, I.}}},
\bauthor{\bsnm{{Bachchan, R. K.}}},
\bauthor{\bsnm{{Lacoste-Seris, F.}}},
\bauthor{\bsnm{{Lanza, A. F.}}},
\bauthor{\bsnm{{Lavigne, J.-B.}}},
\bauthor{\bsnm{{Le Poncin-Lafitte, C.}}},
\bauthor{\bsnm{{Lebreton, Y.}}},
\bauthor{\bsnm{{Lebzelter, T.}}},
\bauthor{\bsnm{{Leccia, S.}}},
\bauthor{\bsnm{{Leclerc, N.}}},
\bauthor{\bsnm{{Lecoeur-Taibi, I.}}},
\bauthor{\bsnm{{Lemaitre, V.}}},
\bauthor{\bsnm{{Lenhardt, H.}}},
\bauthor{\bsnm{{Leroux, F.}}},
\bauthor{\bsnm{{Liao, S.}}},
\bauthor{\bsnm{{Licata, E.}}},
\bauthor{\bsnm{{Lindstr\o{}m, H. E. P.}}},
\bauthor{\bsnm{{Lister, T. A.}}},
\bauthor{\bsnm{{Livanou, E.}}},
\bauthor{\bsnm{{Lobel, A.}}},
\bauthor{\bsnm{{L\"offler, W.}}},
\bauthor{\bsnm{{L\'opez, M.}}},
\bauthor{\bsnm{{Lopez-Lozano, A.}}},
\bauthor{\bsnm{{Lorenz, D.}}},
\bauthor{\bsnm{{Loureiro, T.}}},
\bauthor{\bsnm{{MacDonald, I.}}},
\bauthor{\bsnm{{Magalh\~aes Fernandes, T.}}},
\bauthor{\bsnm{{Managau, S.}}},
\bauthor{\bsnm{{Mann, R. G.}}},
\bauthor{\bsnm{{Mantelet, G.}}},
\bauthor{\bsnm{{Marchal, O.}}},
\bauthor{\bsnm{{Marchant, J. M.}}},
\bauthor{\bsnm{{Marconi, M.}}},
\bauthor{\bsnm{{Marie, J.}}},
\bauthor{\bsnm{{Marinoni, S.}}},
\bauthor{\bsnm{{Marrese, P. M.}}},
\bauthor{\bsnm{{Marschalk\'o, G.}}},
\bauthor{\bsnm{{Marshall, D. J.}}},
\bauthor{\bsnm{{Mart\'{\i}n-Fleitas, J. M.}}},
\bauthor{\bsnm{{Martino, M.}}},
\bauthor{\bsnm{{Mary, N.}}},
\bauthor{\bsnm{{Matijevic, G.}}},
\bauthor{\bsnm{{Mazeh, T.}}},
\bauthor{\bsnm{{McMillan, P. J.}}},
\bauthor{\bsnm{{Messina, S.}}},
\bauthor{\bsnm{{Mestre, A.}}},
\bauthor{\bsnm{{Michalik, D.}}},
\bauthor{\bsnm{{Millar, N. R.}}},
\bauthor{\bsnm{{Miranda, B. M. H.}}},
\bauthor{\bsnm{{Molina, D.}}},
\bauthor{\bsnm{{Molinaro, R.}}},
\bauthor{\bsnm{{Molinaro, M.}}},
\bauthor{\bsnm{{Moln\'ar, L.}}},
\bauthor{\bsnm{{Moniez, M.}}},
\bauthor{\bsnm{{Montegriffo, P.}}},
\bauthor{\bsnm{{Monteiro, D.}}},
\bauthor{\bsnm{{Mor, R.}}},
\bauthor{\bsnm{{Mora, A.}}},
\bauthor{\bsnm{{Morbidelli, R.}}},
\bauthor{\bsnm{{Morel, T.}}},
\bauthor{\bsnm{{Morgenthaler, S.}}},
\bauthor{\bsnm{{Morley, T.}}},
\bauthor{\bsnm{{Morris, D.}}},
\bauthor{\bsnm{{Mulone, A. F.}}},
\bauthor{\bsnm{{Muraveva, T.}}},
\bauthor{\bsnm{{Musella, I.}}},
\bauthor{\bsnm{{Narbonne, J.}}},
\bauthor{\bsnm{{Nelemans, G.}}},
\bauthor{\bsnm{{Nicastro, L.}}},
\bauthor{\bsnm{{Noval, L.}}},
\bauthor{\bsnm{{Ord\'enovic, C.}}},
\bauthor{\bsnm{{Ordieres-Mer\'e, J.}}},
\bauthor{\bsnm{{Osborne, P.}}},
\bauthor{\bsnm{{Pagani, C.}}},
\bauthor{\bsnm{{Pagano, I.}}},
\bauthor{\bsnm{{Pailler, F.}}},
\bauthor{\bsnm{{Palacin, H.}}},
\bauthor{\bsnm{{Palaversa, L.}}},
\bauthor{\bsnm{{Parsons, P.}}},
\bauthor{\bsnm{{Paulsen, T.}}},
\bauthor{\bsnm{{Pecoraro, M.}}},
\bauthor{\bsnm{{Pedrosa, R.}}},
\bauthor{\bsnm{{Pentik\"ainen, H.}}},
\bauthor{\bsnm{{Pereira, J.}}},
\bauthor{\bsnm{{Pichon, B.}}},
\bauthor{\bsnm{{Piersimoni, A. M.}}},
\bauthor{\bsnm{{Pineau, F.-X.}}},
\bauthor{\bsnm{{Plachy, E.}}},
\bauthor{\bsnm{{Plum, G.}}},
\bauthor{\bsnm{{Poujoulet, E.}}},
\bauthor{\bsnm{{Prsa, A.}}},
\bauthor{\bsnm{{Pulone, L.}}},
\bauthor{\bsnm{{Ragaini, S.}}},
\bauthor{\bsnm{{Rago, S.}}},
\bauthor{\bsnm{{Rambaux, N.}}},
\bauthor{\bsnm{{Ramos-Lerate, M.}}},
\bauthor{\bsnm{{Ranalli, P.}}},
\bauthor{\bsnm{{Rauw, G.}}},
\bauthor{\bsnm{{Read, A.}}},
\bauthor{\bsnm{{Regibo, S.}}},
\bauthor{\bsnm{{Renk, F.}}},
\bauthor{\bsnm{{Reyl\'e, C.}}},
\bauthor{\bsnm{{Ribeiro, R. A.}}},
\bauthor{\bsnm{{Rimoldini, L.}}},
\bauthor{\bsnm{{Ripepi, V.}}},
\bauthor{\bsnm{{Riva, A.}}},
\bauthor{\bsnm{{Rixon, G.}}},
\bauthor{\bsnm{{Roelens, M.}}},
\bauthor{\bsnm{{Romero-G\'omez, M.}}},
\bauthor{\bsnm{{Rowell, N.}}},
\bauthor{\bsnm{{Royer, F.}}},
\bauthor{\bsnm{{Rudolph, A.}}},
\bauthor{\bsnm{{Ruiz-Dern, L.}}},
\bauthor{\bsnm{{Sadowski, G.}}},
\bauthor{\bsnm{{Sagrist\`a Sell\'es, T.}}},
\bauthor{\bsnm{{Sahlmann, J.}}},
\bauthor{\bsnm{{Salgado, J.}}},
\bauthor{\bsnm{{Salguero, E.}}},
\bauthor{\bsnm{{Sarasso, M.}}},
\bauthor{\bsnm{{Savietto, H.}}},
\bauthor{\bsnm{{Schnorhk, A.}}},
\bauthor{\bsnm{{Schultheis, M.}}},
\bauthor{\bsnm{{Sciacca, E.}}},
\bauthor{\bsnm{{Segol, M.}}},
\bauthor{\bsnm{{Segovia, J. C.}}},
\bauthor{\bsnm{{Segransan, D.}}},
\bauthor{\bsnm{{Serpell, E.}}},
\bauthor{\bsnm{{Shih, I-C.}}},
\bauthor{\bsnm{{Smareglia, R.}}},
\bauthor{\bsnm{{Smart, R. L.}}},
\bauthor{\bsnm{{Smith, C.}}},
\bauthor{\bsnm{{Solano, E.}}},
\bauthor{\bsnm{{Solitro, F.}}},
\bauthor{\bsnm{{Sordo, R.}}},
\bauthor{\bsnm{{Soria Nieto, S.}}},
\bauthor{\bsnm{{Souchay, J.}}},
\bauthor{\bsnm{{Spagna, A.}}},
\bauthor{\bsnm{{Spoto, F.}}},
\bauthor{\bsnm{{Stampa, U.}}},
\bauthor{\bsnm{{Steele, I. A.}}},
\bauthor{\bsnm{{Steidelm\"uller, H.}}},
\bauthor{\bsnm{{Stephenson, C. A.}}},
\bauthor{\bsnm{{Stoev, H.}}},
\bauthor{\bsnm{{Suess, F. F.}}},
\bauthor{\bsnm{{S\"uveges, M.}}},
\bauthor{\bsnm{{Surdej, J.}}},
\bauthor{\bsnm{{Szabados, L.}}},
\bauthor{\bsnm{{Szegedi-Elek, E.}}},
\bauthor{\bsnm{{Tapiador, D.}}},
\bauthor{\bsnm{{Taris, F.}}},
\bauthor{\bsnm{{Tauran, G.}}},
\bauthor{\bsnm{{Taylor, M. B.}}},
\bauthor{\bsnm{{Teixeira, R.}}},
\bauthor{\bsnm{{Terrett, D.}}},
\bauthor{\bsnm{{Tingley, B.}}},
\bauthor{\bsnm{{Trager, S. C.}}},
\bauthor{\bsnm{{Turon, C.}}},
\bauthor{\bsnm{{Ulla, A.}}},
\bauthor{\bsnm{{Utrilla, E.}}},
\bauthor{\bsnm{{Valentini, G.}}},
\bauthor{\bsnm{{van Elteren, A.}}},
\bauthor{\bsnm{{Van Hemelryck, E.}}},
\bauthor{\bsnm{{van Leeuwen, M.}}},
\bauthor{\bsnm{{Varadi, M.}}},
\bauthor{\bsnm{{Vecchiato, A.}}},
\bauthor{\bsnm{{Veljanoski, J.}}},
\bauthor{\bsnm{{Via, T.}}},
\bauthor{\bsnm{{Vicente, D.}}},
\bauthor{\bsnm{{Vogt, S.}}},
\bauthor{\bsnm{{Voss, H.}}},
\bauthor{\bsnm{{Votruba, V.}}},
\bauthor{\bsnm{{Voutsinas, S.}}},
\bauthor{\bsnm{{Walmsley, G.}}},
\bauthor{\bsnm{{Weiler, M.}}},
\bauthor{\bsnm{{Weingrill, K.}}},
\bauthor{\bsnm{{Werner, D.}}},
\bauthor{\bsnm{{Wevers, T.}}},
\bauthor{\bsnm{{Whitehead, G.}}},
\bauthor{\bsnm{{Wyrzykowski, L.}}},
\bauthor{\bsnm{{Yoldas, A.}}},
\bauthor{\bsnm{{Zerjal, M.}}},
\bauthor{\bsnm{{Zucker, S.}}},
\bauthor{\bsnm{{Zurbach, C.}}},
\bauthor{\bsnm{{Zwitter, T.}}},
\bauthor{\bsnm{{Alecu, A.}}},
\bauthor{\bsnm{{Allen, M.}}},
\bauthor{\bsnm{{Allende Prieto, C.}}},
\bauthor{\bsnm{{Amorim, A.}}},
\bauthor{\bsnm{{Anglada-Escud\'e, G.}}},
\bauthor{\bsnm{{Arsenijevic, V.}}},
\bauthor{\bsnm{{Azaz, S.}}},
\bauthor{\bsnm{{Balm, P.}}},
\bauthor{\bsnm{{Beck, M.}}},
\bauthor{\bsnm{{Bernstein, H.-H.}}},
\bauthor{\bsnm{{Bigot, L.}}},
\bauthor{\bsnm{{Bijaoui, A.}}},
\bauthor{\bsnm{{Blasco, C.}}},
\bauthor{\bsnm{{Bonfigli, M.}}},
\bauthor{\bsnm{{Bono, G.}}},
\bauthor{\bsnm{{Boudreault, S.}}},
\bauthor{\bsnm{{Bressan, A.}}},
\bauthor{\bsnm{{Brown, S.}}},
\bauthor{\bsnm{{Brunet, P.-M.}}},
\bauthor{\bsnm{{Bunclark, P.}}},
\bauthor{\bsnm{{Buonanno, R.}}},
\bauthor{\bsnm{{Butkevich, A. G.}}},
\bauthor{\bsnm{{Carret, C.}}},
\bauthor{\bsnm{{Carrion, C.}}},
\bauthor{\bsnm{{Chemin, L.}}},
\bauthor{\bsnm{{Ch\'ereau, F.}}},
\bauthor{\bsnm{{Corcione, L.}}},
\bauthor{\bsnm{{Darmigny, E.}}},
\bauthor{\bsnm{{de Boer, K. S.}}},
\bauthor{\bsnm{{de Teodoro, P.}}},
\bauthor{\bsnm{{de Zeeuw, P. T.}}},
\bauthor{\bsnm{{Delle Luche, C.}}},
\bauthor{\bsnm{{Domingues, C. D.}}},
\bauthor{\bsnm{{Dubath, P.}}},
\bauthor{\bsnm{{Fodor, F.}}},
\bauthor{\bsnm{{Fr\'ezouls, B.}}},
\bauthor{\bsnm{{Fries, A.}}},
\bauthor{\bsnm{{Fustes, D.}}},
\bauthor{\bsnm{{Fyfe, D.}}},
\bauthor{\bsnm{{Gallardo, E.}}},
\bauthor{\bsnm{{Gallegos, J.}}},
\bauthor{\bsnm{{Gardiol, D.}}},
\bauthor{\bsnm{{Gebran, M.}}},
\bauthor{\bsnm{{Gomboc, A.}}},
\bauthor{\bsnm{{G\'omez, A.}}},
\bauthor{\bsnm{{Grux, E.}}},
\bauthor{\bsnm{{Gueguen, A.}}},
\bauthor{\bsnm{{Heyrovsky, A.}}},
\bauthor{\bsnm{{Hoar, J.}}},
\bauthor{\bsnm{{Iannicola, G.}}},
\bauthor{\bsnm{{Isasi Parache, Y.}}},
\bauthor{\bsnm{{Janotto, A.-M.}}},
\bauthor{\bsnm{{Joliet, E.}}},
\bauthor{\bsnm{{Jonckheere, A.}}},
\bauthor{\bsnm{{Keil, R.}}},
\bauthor{\bsnm{{Kim, D.-W.}}},
\bauthor{\bsnm{{Klagyivik, P.}}},
\bauthor{\bsnm{{Klar, J.}}},
\bauthor{\bsnm{{Knude, J.}}},
\bauthor{\bsnm{{Kochukhov, O.}}},
\bauthor{\bsnm{{Kolka, I.}}},
\bauthor{\bsnm{{Kos, J.}}},
\bauthor{\bsnm{{Kutka, A.}}},
\bauthor{\bsnm{{Lainey, V.}}},
\bauthor{\bsnm{{LeBouquin, D.}}},
\bauthor{\bsnm{{Liu, C.}}},
\bauthor{\bsnm{{Loreggia, D.}}},
\bauthor{\bsnm{{Makarov, V. V.}}},
\bauthor{\bsnm{{Marseille, M. G.}}},
\bauthor{\bsnm{{Martayan, C.}}},
\bauthor{\bsnm{{Martinez-Rubi, O.}}},
\bauthor{\bsnm{{Massart, B.}}},
\bauthor{\bsnm{{Meynadier, F.}}},
\bauthor{\bsnm{{Mignot, S.}}},
\bauthor{\bsnm{{Munari, U.}}},
\bauthor{\bsnm{{Nguyen, A.-T.}}},
\bauthor{\bsnm{{Nordlander, T.}}},
\bauthor{\bsnm{{Ocvirk, P.}}},
\bauthor{\bsnm{{O\'{}Flaherty, K. S.}}},
\bauthor{\bsnm{{Olias Sanz, A.}}},
\bauthor{\bsnm{{Ortiz, P.}}},
\bauthor{\bsnm{{Osorio, J.}}},
\bauthor{\bsnm{{Oszkiewicz, D.}}},
\bauthor{\bsnm{{Ouzounis, A.}}},
\bauthor{\bsnm{{Palmer, M.}}},
\bauthor{\bsnm{{Park, P.}}},
\bauthor{\bsnm{{Pasquato, E.}}},
\bauthor{\bsnm{{Peltzer, C.}}},
\bauthor{\bsnm{{Peralta, J.}}},
\bauthor{\bsnm{{P\'eturaud, F.}}},
\bauthor{\bsnm{{Pieniluoma, T.}}},
\bauthor{\bsnm{{Pigozzi, E.}}},
\bauthor{\bsnm{{Poels, J.}}},
\bauthor{\bsnm{{Prat, G.}}},
\bauthor{\bsnm{{Prod\'{}homme, T.}}},
\bauthor{\bsnm{{Raison, F.}}},
\bauthor{\bsnm{{Rebordao, J. M.}}},
\bauthor{\bsnm{{Risquez, D.}}},
\bauthor{\bsnm{{Rocca-Volmerange, B.}}},
\bauthor{\bsnm{{Rosen, S.}}},
\bauthor{\bsnm{{Ruiz-Fuertes, M. I.}}},
\bauthor{\bsnm{{Russo, F.}}},
\bauthor{\bsnm{{Sembay, S.}}},
\bauthor{\bsnm{{Serraller Vizcaino, I.}}},
\bauthor{\bsnm{{Short, A.}}},
\bauthor{\bsnm{{Siebert, A.}}},
\bauthor{\bsnm{{Silva, H.}}},
\bauthor{\bsnm{{Sinachopoulos, D.}}},
\bauthor{\bsnm{{Slezak, E.}}},
\bauthor{\bsnm{{Soffel, M.}}},
\bauthor{\bsnm{{Sosnowska, D.}}},
\bauthor{\bsnm{{Straizys, V.}}},
\bauthor{\bsnm{{ter Linden, M.}}},
\bauthor{\bsnm{{Terrell, D.}}},
\bauthor{\bsnm{{Theil, S.}}},
\bauthor{\bsnm{{Tiede, C.}}},
\bauthor{\bsnm{{Troisi, L.}}},
\bauthor{\bsnm{{Tsalmantza, P.}}},
\bauthor{\bsnm{{Tur, D.}}},
\bauthor{\bsnm{{Vaccari, M.}}},
\bauthor{\bsnm{{Vachier, F.}}},
\bauthor{\bsnm{{Valles, P.}}},
\bauthor{\bsnm{{Van Hamme, W.}}},
\bauthor{\bsnm{{Veltz, L.}}},
\bauthor{\bsnm{{Virtanen, J.}}},
\bauthor{\bsnm{{Wallut, J.-M.}}},
\bauthor{\bsnm{{Wichmann, R.}}},
\bauthor{\bsnm{{Wilkinson, M. I.}}},
\bauthor{\bsnm{{Ziaeepour, H.}}},
\bauthor{\bsnm{{Zschocke, S.}}}:
\batitle{The gaia mission}.
\bjtitle{\aap}
\bvolume{595},
\bfpage{1}
(\byear{2016}).
\doiurl{10.1051/0004-6361/201629272}
\end{barticle}
\endbibitem

\bibitem{Mayor2003}
\begin{barticle}
\bauthor{\bsnm{{Mayor}}, \binits{M.}},
\bauthor{\bsnm{{Pepe}}, \binits{F.}},
\bauthor{\bsnm{{Queloz}}, \binits{D.}},
\bauthor{\bsnm{{Bouchy}}, \binits{F.}},
\bauthor{\bsnm{{Rupprecht}}, \binits{G.}},
\bauthor{\bsnm{{Lo Curto}}, \binits{G.}},
\bauthor{\bsnm{{Avila}}, \binits{G.}},
\bauthor{\bsnm{{Benz}}, \binits{W.}},
\bauthor{\bsnm{{Bertaux}}, \binits{J.-L.}},
\bauthor{\bsnm{{Bonfils}}, \binits{X.}},
\bauthor{\bsnm{{Dall}}, \binits{T.}},
\bauthor{\bsnm{{Dekker}}, \binits{H.}},
\bauthor{\bsnm{{Delabre}}, \binits{B.}},
\bauthor{\bsnm{{Eckert}}, \binits{W.}},
\bauthor{\bsnm{{Fleury}}, \binits{M.}},
\bauthor{\bsnm{{Gilliotte}}, \binits{A.}},
\bauthor{\bsnm{{Gojak}}, \binits{D.}},
\bauthor{\bsnm{{Guzman}}, \binits{J.C.}},
\bauthor{\bsnm{{Kohler}}, \binits{D.}},
\bauthor{\bsnm{{Lizon}}, \binits{J.-L.}},
\bauthor{\bsnm{{Longinotti}}, \binits{A.}},
\bauthor{\bsnm{{Lovis}}, \binits{C.}},
\bauthor{\bsnm{{Megevand}}, \binits{D.}},
\bauthor{\bsnm{{Pasquini}}, \binits{L.}},
\bauthor{\bsnm{{Reyes}}, \binits{J.}},
\bauthor{\bsnm{{Sivan}}, \binits{J.-P.}},
\bauthor{\bsnm{{Sosnowska}}, \binits{D.}},
\bauthor{\bsnm{{Soto}}, \binits{R.}},
\bauthor{\bsnm{{Udry}}, \binits{S.}},
\bauthor{\bsnm{{van Kesteren}}, \binits{A.}},
\bauthor{\bsnm{{Weber}}, \binits{L.}},
\bauthor{\bsnm{{Weilenmann}}, \binits{U.}}:
\batitle{{Setting New Standards with HARPS}}.
\bjtitle{The Messenger}
\bvolume{114},
\bfpage{20}--\blpage{24}
(\byear{2003})
\end{barticle}
\endbibitem

\bibitem{Pollacco2006}
\begin{barticle}
\bauthor{\bsnm{{Pollacco}}, \binits{D.L.}},
\bauthor{\bsnm{{Skillen}}, \binits{I.}},
\bauthor{\bsnm{{Collier Cameron}}, \binits{A.}},
\bauthor{\bsnm{{Christian}}, \binits{D.J.}},
\bauthor{\bsnm{{Hellier}}, \binits{C.}},
\bauthor{\bsnm{{Irwin}}, \binits{J.}},
\bauthor{\bsnm{{Lister}}, \binits{T.A.}},
\bauthor{\bsnm{{Street}}, \binits{R.A.}},
\bauthor{\bsnm{{West}}, \binits{R.G.}},
\bauthor{\bsnm{{Anderson}}, \binits{D.R.}},
\bauthor{\bsnm{{Clarkson}}, \binits{W.I.}},
\bauthor{\bsnm{{Deeg}}, \binits{H.}},
\bauthor{\bsnm{{Enoch}}, \binits{B.}},
\bauthor{\bsnm{{Evans}}, \binits{A.}},
\bauthor{\bsnm{{Fitzsimmons}}, \binits{A.}},
\bauthor{\bsnm{{Haswell}}, \binits{C.A.}},
\bauthor{\bsnm{{Hodgkin}}, \binits{S.}},
\bauthor{\bsnm{{Horne}}, \binits{K.}},
\bauthor{\bsnm{{Kane}}, \binits{S.R.}},
\bauthor{\bsnm{{Keenan}}, \binits{F.P.}},
\bauthor{\bsnm{{Maxted}}, \binits{P.F.L.}},
\bauthor{\bsnm{{Norton}}, \binits{A.J.}},
\bauthor{\bsnm{{Osborne}}, \binits{J.}},
\bauthor{\bsnm{{Parley}}, \binits{N.R.}},
\bauthor{\bsnm{{Ryans}}, \binits{R.S.I.}},
\bauthor{\bsnm{{Smalley}}, \binits{B.}},
\bauthor{\bsnm{{Wheatley}}, \binits{P.J.}},
\bauthor{\bsnm{{Wilson}}, \binits{D.M.}}:
\batitle{{The WASP Project and the SuperWASP Cameras}}.
\bjtitle{\pasp}
\bvolume{118}(\bissue{848}),
\bfpage{1407}--\blpage{1418}
(\byear{2006})
{\href{https://arxiv.org/abs/astro-ph/0608454}{{arXiv:astro-ph/0608454}}}
{[astro-ph]}.
\doiurl{10.1086/508556}
\end{barticle}
\endbibitem

\bibitem{pepper2007}
\begin{barticle}
\bauthor{\bsnm{{Pepper}}, \binits{J.}},
\bauthor{\bsnm{{Pogge}}, \binits{R.W.}},
\bauthor{\bsnm{{DePoy}}, \binits{D.L.}},
\bauthor{\bsnm{{Marshall}}, \binits{J.L.}},
\bauthor{\bsnm{{Stanek}}, \binits{K.Z.}},
\bauthor{\bsnm{{Stutz}}, \binits{A.M.}},
\bauthor{\bsnm{{Poindexter}}, \binits{S.}},
\bauthor{\bsnm{{Siverd}}, \binits{R.}},
\bauthor{\bsnm{{O'Brien}}, \binits{T.P.}},
\bauthor{\bsnm{{Trueblood}}, \binits{M.}},
\bauthor{\bsnm{{Trueblood}}, \binits{P.}}:
\batitle{{The Kilodegree Extremely Little Telescope (KELT): A Small Robotic
  Telescope for Large-Area Synoptic Surveys}}.
\bjtitle{\pasp}
\bvolume{119}(\bissue{858}),
\bfpage{923}--\blpage{935}
(\byear{2007})
{\href{https://arxiv.org/abs/0704.0460}{{arXiv:0704.0460}}}
{[astro-ph]}.
\doiurl{10.1086/521836}
\end{barticle}
\endbibitem

\bibitem{Udalski2015}
\begin{barticle}
\bauthor{\bsnm{{Udalski}}, \binits{A.}},
\bauthor{\bsnm{{Szyma{\'n}ski}}, \binits{M.K.}},
\bauthor{\bsnm{{Szyma{\'n}ski}}, \binits{G.}}:
\batitle{{OGLE-IV: Fourth Phase of the Optical Gravitational Lensing
  Experiment}}.
\bjtitle{\actaa}
\bvolume{65}(\bissue{1}),
\bfpage{1}--\blpage{38}
(\byear{2015})
{\href{https://arxiv.org/abs/1504.05966}{{arXiv:1504.05966}}}
{[astro-ph.SR]}
\end{barticle}
\endbibitem

\bibitem{Seager2000}
\begin{barticle}
\bauthor{\bsnm{{Seager}}, \binits{S.}},
\bauthor{\bsnm{{Sasselov}}, \binits{D.D.}}:
\batitle{{Theoretical Transmission Spectra during Extrasolar Giant Planet
  Transits}}.
\bjtitle{\apj}
\bvolume{537}(\bissue{2}),
\bfpage{916}--\blpage{921}
(\byear{2000})
{\href{https://arxiv.org/abs/astro-ph/9912241}{{arXiv:astro-ph/9912241}}}
{[astro-ph]}.
\doiurl{10.1086/309088}
\end{barticle}
\endbibitem

\bibitem{Charbonneau2005}
\begin{barticle}
\bauthor{\bsnm{Charbonneau}, \binits{D.}},
\bauthor{\bsnm{Allen}, \binits{L.E.}},
\bauthor{\bsnm{Megeath}, \binits{S.T.}},
\bauthor{\bsnm{Torres}, \binits{G.}},
\bauthor{\bsnm{Alonso}, \binits{R.}},
\bauthor{\bsnm{Brown}, \binits{T.M.}},
\bauthor{\bsnm{Gilliland}, \binits{R.L.}},
\bauthor{\bsnm{Latham}, \binits{D.W.}},
\bauthor{\bsnm{Mandushev}, \binits{G.}},
\bauthor{\bsnm{O'Donovan}, \binits{F.T.}},
\bauthor{\bsnm{Sozzetti}, \binits{A.}}:
\batitle{{Detection of Thermal Emission from an Extrasolar Planet}}.
\bjtitle{\apj}
\bvolume{626}(\bissue{1}),
\bfpage{523}--\blpage{529}
(\byear{2005})
{\href{https://arxiv.org/abs/astro-ph/0503457}{{arXiv:astro-ph/0503457}}}
{[astro-ph]}.
\doiurl{10.1086/429991}
\end{barticle}
\endbibitem

\bibitem{Tinetti2007}
\begin{barticle}
\bauthor{\bsnm{Tinetti}, \binits{G.}},
\bauthor{\bsnm{Vidal-Madjar}, \binits{A.}},
\bauthor{\bsnm{Liang}, \binits{M.}},
\bauthor{\bsnm{Beaulieu}, \binits{J.-P.}},
\bauthor{\bsnm{Yung}, \binits{Y.}},
\bauthor{\bsnm{Carey}, \binits{S.}},
\bauthor{\bsnm{Barber}, \binits{R.}},
\bauthor{\bsnm{Tennyson}, \binits{J.}},
\bauthor{\bsnm{Ribas}, \binits{I.}},
\bauthor{\bsnm{Allard}, \binits{N.}},
\bauthor{\bsnm{Ballester}, \binits{G.}},
\bauthor{\bsnm{Sing}, \binits{D.}},
\bauthor{\bsnm{Selsis}, \binits{F.}}:
\batitle{Water vapour in the atmosphere of a transiting extrasolar planet}.
\bjtitle{Nature}
\bvolume{448},
\bfpage{169}--\blpage{71}
(\byear{2007}).
\doiurl{10.1038/nature06002}
\end{barticle}
\endbibitem

\bibitem{Madhusudhan2012}
\begin{barticle}
\bauthor{\bsnm{{Madhusudhan}}, \binits{N.}},
\bauthor{\bsnm{{Lee}}, \binits{K.K.M.}},
\bauthor{\bsnm{{Mousis}}, \binits{O.}}:
\batitle{{A Possible Carbon-rich Interior in Super-Earth 55 Cancri e}}.
\bjtitle{\apjl}
\bvolume{759}(\bissue{2}),
\bfpage{40}
(\byear{2012})
{\href{https://arxiv.org/abs/1210.2720}{{arXiv:1210.2720}}}
{[astro-ph.EP]}.
\doiurl{10.1088/2041-8205/759/2/L40}
\end{barticle}
\endbibitem

\bibitem{Tinetti2013}
\begin{barticle}
\bauthor{\bsnm{{Tinetti}}, \binits{G.}},
\bauthor{\bsnm{{Encrenaz}}, \binits{T.}},
\bauthor{\bsnm{{Coustenis}}, \binits{A.}}:
\batitle{{Spectroscopy of planetary atmospheres in our Galaxy}}.
\bjtitle{\aapr}
\bvolume{21},
\bfpage{63}
(\byear{2013}).
\doiurl{10.1007/s00159-013-0063-6}
\end{barticle}
\endbibitem

\bibitem{Kreidberg201}
\begin{barticle}
\bauthor{\bsnm{Kreidberg}, \binits{L.}},
\bauthor{\bsnm{Bean}, \binits{J.L.}},
\bauthor{\bsnm{D{\'{e}}sert}, \binits{J.-M.}},
\bauthor{\bsnm{Benneke}, \binits{B.}},
\bauthor{\bsnm{Deming}, \binits{D.}},
\bauthor{\bsnm{Stevenson}, \binits{K.B.}},
\bauthor{\bsnm{Seager}, \binits{S.}},
\bauthor{\bsnm{Berta-Thompson}, \binits{Z.}},
\bauthor{\bsnm{Seifahrt}, \binits{A.}},
\bauthor{\bsnm{Homeier}, \binits{D.}}:
\batitle{{Clouds in the atmosphere of the super-Earth exoplanet GJ1214b}}.
\bjtitle{\nat}
\bvolume{505}(\bissue{7481}),
\bfpage{69}--\blpage{72}
(\byear{2014})
{\href{https://arxiv.org/abs/1401.0022}{{arXiv:1401.0022}}}
{[astro-ph.EP]}.
\doiurl{10.1038/nature12888}
\end{barticle}
\endbibitem

\bibitem{Sing2016}
\begin{barticle}
\bauthor{\bsnm{Sing}, \binits{D.K.}},
\bauthor{\bsnm{Fortney}, \binits{J.J.}},
\bauthor{\bsnm{Nikolov}, \binits{N.}},
\bauthor{\bsnm{Wakeford}, \binits{H.R.}},
\bauthor{\bsnm{Kataria}, \binits{T.}},
\bauthor{\bsnm{Evans}, \binits{T.M.}},
\bauthor{\bsnm{Aigrain}, \binits{S.}},
\bauthor{\bsnm{Ballester}, \binits{G.E.}},
\bauthor{\bsnm{Burrows}, \binits{A.S.}},
\bauthor{\bsnm{Deming}, \binits{D.}},
\bauthor{\bsnm{D{\'{e}}sert}, \binits{J.-M.}},
\bauthor{\bsnm{Gibson}, \binits{N.P.}},
\bauthor{\bsnm{Henry}, \binits{G.W.}},
\bauthor{\bsnm{Huitson}, \binits{C.M.}},
\bauthor{\bsnm{Knutson}, \binits{H.A.}},
\bauthor{\bsnm{{Lecavelier Des Etangs}}, \binits{A.}},
\bauthor{\bsnm{Pont}, \binits{F.}},
\bauthor{\bsnm{Showman}, \binits{A.P.}},
\bauthor{\bsnm{Vidal-Madjar}, \binits{A.}},
\bauthor{\bsnm{Williamson}, \binits{M.H.}},
\bauthor{\bsnm{Wilson}, \binits{P.A.}}:
\batitle{{A continuum from clear to cloudy hot-Jupiter exoplanets without
  primordial water depletion}}.
\bjtitle{\nat}
\bvolume{529}(\bissue{7584}),
\bfpage{59}--\blpage{62}
(\byear{2016})
{\href{https://arxiv.org/abs/1512.04341}{{arXiv:1512.04341}}}
{[astro-ph.EP]}.
\doiurl{10.1038/nature16068}
\end{barticle}
\endbibitem

\bibitem{Line2016}
\begin{barticle}
\bauthor{\bsnm{{Line}}, \binits{M.R.}},
\bauthor{\bsnm{{Stevenson}}, \binits{K.B.}},
\bauthor{\bsnm{{Bean}}, \binits{J.}},
\bauthor{\bsnm{{Desert}}, \binits{J.-M.}},
\bauthor{\bsnm{{Fortney}}, \binits{J.J.}},
\bauthor{\bsnm{{Kreidberg}}, \binits{L.}},
\bauthor{\bsnm{{Madhusudhan}}, \binits{N.}},
\bauthor{\bsnm{{Showman}}, \binits{A.P.}},
\bauthor{\bsnm{{Diamond-Lowe}}, \binits{H.}}:
\batitle{{No Thermal Inversion and a Solar Water Abundance for the Hot Jupiter
  HD 209458b from HST/WFC3 Spectroscopy}}.
\bjtitle{\aj}
\bvolume{152}(\bissue{6}),
\bfpage{203}
(\byear{2016})
{\href{https://arxiv.org/abs/1605.08810}{{arXiv:1605.08810}}}
{[astro-ph.EP]}.
\doiurl{10.3847/0004-6256/152/6/203}
\end{barticle}
\endbibitem

\bibitem{Tsiaras2018}
\begin{barticle}
\bauthor{\bsnm{{Tsiaras}}, \binits{A.}},
\bauthor{\bsnm{{Waldmann}}, \binits{I.P.}},
\bauthor{\bsnm{{Zingales}}, \binits{T.}},
\bauthor{\bsnm{{Rocchetto}}, \binits{M.}},
\bauthor{\bsnm{{Morello}}, \binits{G.}},
\bauthor{\bsnm{{Damiano}}, \binits{M.}},
\bauthor{\bsnm{{Karpouzas}}, \binits{K.}},
\bauthor{\bsnm{{Tinetti}}, \binits{G.}},
\bauthor{\bsnm{{McKemmish}}, \binits{L.K.}},
\bauthor{\bsnm{{Tennyson}}, \binits{J.}},
\bauthor{\bsnm{{Yurchenko}}, \binits{S.N.}}:
\batitle{{A Population Study of Gaseous Exoplanets}}.
\bjtitle{\aj}
\bvolume{155}(\bissue{4}),
\bfpage{156}
(\byear{2018})
{\href{https://arxiv.org/abs/1704.05413}{{arXiv:1704.05413}}}
{[astro-ph.EP]}.
\doiurl{10.3847/1538-3881/aaaf75}
\end{barticle}
\endbibitem

\bibitem{Evans_2018}
\begin{barticle}
\bauthor{\bsnm{{Evans}}, \binits{T.M.}},
\bauthor{\bsnm{{Sing}}, \binits{D.K.}},
\bauthor{\bsnm{{Goyal}}, \binits{J.M.}},
\bauthor{\bsnm{{Nikolov}}, \binits{N.}},
\bauthor{\bsnm{{Marley}}, \binits{M.S.}},
\bauthor{\bsnm{{Zahnle}}, \binits{K.}},
\bauthor{\bsnm{{Henry}}, \binits{G.W.}},
\bauthor{\bsnm{{Barstow}}, \binits{J.K.}},
\bauthor{\bsnm{{Alam}}, \binits{M.K.}},
\bauthor{\bsnm{{Sanz-Forcada}}, \binits{J.}},
\bauthor{\bsnm{{Kataria}}, \binits{T.}},
\bauthor{\bsnm{{Lewis}}, \binits{N.K.}},
\bauthor{\bsnm{{Lavvas}}, \binits{P.}},
\bauthor{\bsnm{{Ballester}}, \binits{G.E.}},
\bauthor{\bsnm{{Ben-Jaffel}}, \binits{L.}},
\bauthor{\bsnm{{Blumenthal}}, \binits{S.D.}},
\bauthor{\bsnm{{Bourrier}}, \binits{V.}},
\bauthor{\bsnm{{Drummond}}, \binits{B.}},
\bauthor{\bsnm{{Garc{\'\i}a Mu{\~n}oz}}, \binits{A.}},
\bauthor{\bsnm{{L{\'o}pez-Morales}}, \binits{M.}},
\bauthor{\bsnm{{Tremblin}}, \binits{P.}},
\bauthor{\bsnm{{Ehrenreich}}, \binits{D.}},
\bauthor{\bsnm{{Wakeford}}, \binits{H.R.}},
\bauthor{\bsnm{{Buchhave}}, \binits{L.A.}},
\bauthor{\bsnm{{Lecavelier des Etangs}}, \binits{A.}},
\bauthor{\bsnm{{H{\'e}brard}}, \binits{{\'E}.}},
\bauthor{\bsnm{{Williamson}}, \binits{M.H.}}:
\batitle{{An Optical Transmission Spectrum for the Ultra-hot Jupiter WASP-121b
  Measured with the Hubble Space Telescope}}.
\bjtitle{\aj}
\bvolume{156}(\bissue{6}),
\bfpage{283}
(\byear{2018})
{\href{https://arxiv.org/abs/1810.10969}{{arXiv:1810.10969}}}
{[astro-ph.EP]}.
\doiurl{10.3847/1538-3881/aaebff}
\end{barticle}
\endbibitem

\bibitem{pinhas2019}
\begin{barticle}
\bauthor{\bsnm{Pinhas}, \binits{A.}},
\bauthor{\bsnm{Madhusudhan}, \binits{N.}},
\bauthor{\bsnm{Gandhi}, \binits{S.}},
\bauthor{\bsnm{MacDonald}, \binits{R.}}:
\batitle{{H2O abundances and cloud properties in ten hot giant exoplanets}}.
\bjtitle{Monthly Notices of the Royal Astronomical Society}
\bvolume{482}(\bissue{2}),
\bfpage{1485}--\blpage{1498}
(\byear{2018})
{\href{https://arxiv.org/abs/https://academic.oup.com/mnras/article-pdf/482/2/1485/26288856/sty2544.pdf}{{https://academic.oup.com/mnras/article-pdf/482/2/1485/26288856/sty2544.pdf}}}.
\doiurl{10.1093/mnras/sty2544}
\end{barticle}
\endbibitem

\bibitem{Welbanks2019}
\begin{barticle}
\bauthor{\bsnm{Welbanks}, \binits{L.}},
\bauthor{\bsnm{Madhusudhan}, \binits{N.}},
\bauthor{\bsnm{Allard}, \binits{N.F.}},
\bauthor{\bsnm{Hubeny}, \binits{I.}},
\bauthor{\bsnm{Spiegelman}, \binits{F.}},
\bauthor{\bsnm{Leininger}, \binits{T.}}:
\batitle{Mass–metallicity trends in transiting exoplanets from atmospheric
  abundances of h2o, na, and k}.
\bjtitle{\apjl}
\bvolume{887}(\bissue{1}),
\bfpage{20}
(\byear{2019}).
\doiurl{10.3847/2041-8213/ab5a89}
\end{barticle}
\endbibitem

\bibitem{Mikal_Evans2020}
\begin{barticle}
\bauthor{\bsnm{{Mikal-Evans}}, \binits{T.}},
\bauthor{\bsnm{{Sing}}, \binits{D.K.}},
\bauthor{\bsnm{{Kataria}}, \binits{T.}},
\bauthor{\bsnm{{Wakeford}}, \binits{H.R.}},
\bauthor{\bsnm{{Mayne}}, \binits{N.J.}},
\bauthor{\bsnm{{Lewis}}, \binits{N.K.}},
\bauthor{\bsnm{{Barstow}}, \binits{J.K.}},
\bauthor{\bsnm{{Spake}}, \binits{J.J.}}:
\batitle{{Confirmation of water emission in the dayside spectrum of the
  ultrahot Jupiter WASP-121b}}.
\bjtitle{\mnras}
\bvolume{496}(\bissue{2}),
\bfpage{1638}--\blpage{1644}
(\byear{2020})
{\href{https://arxiv.org/abs/2005.09631}{{arXiv:2005.09631}}}
{[astro-ph.EP]}.
\doiurl{10.1093/mnras/staa1628}
\end{barticle}
\endbibitem

\bibitem{Pluriel2020A}
\begin{barticle}
\bauthor{\bsnm{{Pluriel}}, \binits{W.}},
\bauthor{\bsnm{{Zingales}}, \binits{T.}},
\bauthor{\bsnm{{Leconte}}, \binits{J.}},
\bauthor{\bsnm{{Parmentier}}, \binits{V.}}:
\batitle{{Strong biases in retrieved atmospheric composition caused by
  day-night chemical heterogeneities}}.
\bjtitle{\aap}
\bvolume{636},
\bfpage{66}
(\byear{2020})
{\href{https://arxiv.org/abs/2003.05943}{{arXiv:2003.05943}}}
{[astro-ph.EP]}.
\doiurl{10.1051/0004-6361/202037678}
\end{barticle}
\endbibitem

\bibitem{ares1}
\begin{barticle}
\bauthor{\bsnm{Edwards}, \binits{B.}},
\bauthor{\bsnm{Changeat}, \binits{Q.}},
\bauthor{\bsnm{Baeyens}, \binits{R.}},
\bauthor{\bsnm{Tsiaras}, \binits{A.}},
\bauthor{\bsnm{Al-Refaie}, \binits{A.}},
\bauthor{\bsnm{Taylor}, \binits{J.}},
\bauthor{\bsnm{Yip}, \binits{K.H.}},
\bauthor{\bsnm{Bieger}, \binits{M.F.}},
\bauthor{\bsnm{Blain}, \binits{D.}},
\bauthor{\bsnm{Gressier}, \binits{A.}},
\bauthor{\bsnm{Guilluy}, \binits{G.}},
\bauthor{\bsnm{Jaziri}, \binits{A.Y.}},
\bauthor{\bsnm{Kiefer}, \binits{F.}},
\bauthor{\bsnm{Modirrousta-Galian}, \binits{D.}},
\bauthor{\bsnm{Morvan}, \binits{M.}},
\bauthor{\bsnm{Mugnai}, \binits{L.V.}},
\bauthor{\bsnm{Pluriel}, \binits{W.}},
\bauthor{\bsnm{Poveda}, \binits{M.}},
\bauthor{\bsnm{Skaf}, \binits{N.}},
\bauthor{\bsnm{Whiteford}, \binits{N.}},
\bauthor{\bsnm{Wright}, \binits{S.}},
\bauthor{\bsnm{Zingales}, \binits{T.}},
\bauthor{\bsnm{Charnay}, \binits{B.}},
\bauthor{\bsnm{Drossart}, \binits{P.}},
\bauthor{\bsnm{Leconte}, \binits{J.}},
\bauthor{\bsnm{Venot}, \binits{O.}},
\bauthor{\bsnm{Waldmann}, \binits{I.}},
\bauthor{\bsnm{Beaulieu}, \binits{J.-P.}}:
\batitle{{ARES I: WASP-76 b, A Tale of Two HST Spectra}}.
\bjtitle{\aj}
\bvolume{160}(\bissue{1}),
\bfpage{8}
(\byear{2020}).
\doiurl{10.3847/1538-3881/AB9225}
\end{barticle}
\endbibitem

\bibitem{ares2}
\begin{barticle}
\bauthor{\bsnm{Skaf}, \binits{N.}},
\bauthor{\bsnm{Bieger}, \binits{M.F.}},
\bauthor{\bsnm{Edwards}, \binits{B.}},
\bauthor{\bsnm{Changeat}, \binits{Q.}},
\bauthor{\bsnm{Morvan}, \binits{M.}},
\bauthor{\bsnm{Kiefer}, \binits{F.}},
\bauthor{\bsnm{Blain}, \binits{D.}},
\bauthor{\bsnm{Zingales}, \binits{T.}},
\bauthor{\bsnm{Poveda}, \binits{M.}},
\bauthor{\bsnm{Al-Refaie}, \binits{A.}},
\bauthor{\bsnm{Baeyens}, \binits{R.}},
\bauthor{\bsnm{Gressier}, \binits{A.}},
\bauthor{\bsnm{Guilluy}, \binits{G.}},
\bauthor{\bsnm{Jaziri}, \binits{A.Y.}},
\bauthor{\bsnm{Modirrousta-Galian}, \binits{D.}},
\bauthor{\bsnm{Mugnai}, \binits{L.V.}},
\bauthor{\bsnm{Pluriel}, \binits{W.}},
\bauthor{\bsnm{Whiteford}, \binits{N.}},
\bauthor{\bsnm{Wright}, \binits{S.}},
\bauthor{\bsnm{Yip}, \binits{K.H.}},
\bauthor{\bsnm{Charnay}, \binits{B.}},
\bauthor{\bsnm{Leconte}, \binits{J.}},
\bauthor{\bsnm{Drossart}, \binits{P.}},
\bauthor{\bsnm{Tsiaras}, \binits{A.}},
\bauthor{\bsnm{Venot}, \binits{O.}},
\bauthor{\bsnm{Waldmann}, \binits{I.}},
\bauthor{\bsnm{Beaulieu}, \binits{J.-P.}}:
\batitle{{ARES. II. Characterizing the Hot Jupiters WASP-127 b, WASP-79 b, and
  WASP-62b with the Hubble Space Telescope}}.
\bjtitle{\aj}
\bvolume{160}(\bissue{3}),
\bfpage{109}
(\byear{2020})
{\href{https://arxiv.org/abs/2005.09615}{{arXiv:2005.09615}}}
{[astro-ph.EP]}.
\doiurl{10.3847/1538-3881/ab94a3}
\end{barticle}
\endbibitem

\bibitem{ares3}
\begin{barticle}
\bauthor{\bsnm{Pluriel}, \binits{W.}},
\bauthor{\bsnm{Whiteford}, \binits{N.}},
\bauthor{\bsnm{Edwards}, \binits{B.}},
\bauthor{\bsnm{Changeat}, \binits{Q.}},
\bauthor{\bsnm{Yip}, \binits{K.H.}},
\bauthor{\bsnm{Baeyens}, \binits{R.}},
\bauthor{\bsnm{Al-Refaie}, \binits{A.}},
\bauthor{\bsnm{{Fabienne Bieger}}, \binits{M.}},
\bauthor{\bsnm{Blain}, \binits{D.}},
\bauthor{\bsnm{Gressier}, \binits{A.}},
\bauthor{\bsnm{Guilluy}, \binits{G.}},
\bauthor{\bsnm{{Yassin Jaziri}}, \binits{A.}},
\bauthor{\bsnm{Kiefer}, \binits{F.}},
\bauthor{\bsnm{Modirrousta-Galian}, \binits{D.}},
\bauthor{\bsnm{Morvan}, \binits{M.}},
\bauthor{\bsnm{Mugnai}, \binits{L.V.}},
\bauthor{\bsnm{Poveda}, \binits{M.}},
\bauthor{\bsnm{Skaf}, \binits{N.}},
\bauthor{\bsnm{Zingales}, \binits{T.}},
\bauthor{\bsnm{Wright}, \binits{S.}},
\bauthor{\bsnm{Charnay}, \binits{B.}},
\bauthor{\bsnm{Drossart}, \binits{P.}},
\bauthor{\bsnm{Leconte}, \binits{J.}},
\bauthor{\bsnm{Tsiaras}, \binits{A.}},
\bauthor{\bsnm{Venot}, \binits{O.}},
\bauthor{\bsnm{Waldmann}, \binits{I.}},
\bauthor{\bsnm{Beaulieu}, \binits{J.-P.}},
\bauthor{\bsnm{Bieger}, \binits{M.F.}},
\bauthor{\bsnm{Blain}, \binits{D.}},
\bauthor{\bsnm{Gressier}, \binits{A.}},
\bauthor{\bsnm{Guilluy}, \binits{G.}},
\bauthor{\bsnm{Jaziri}, \binits{A.Y.}},
\bauthor{\bsnm{Kiefer}, \binits{F.}},
\bauthor{\bsnm{Modirrousta-Galian}, \binits{D.}},
\bauthor{\bsnm{Morvan}, \binits{M.}},
\bauthor{\bsnm{Mugnai}, \binits{L.V.}},
\bauthor{\bsnm{Poveda}, \binits{M.}},
\bauthor{\bsnm{Skaf}, \binits{N.}},
\bauthor{\bsnm{Zingales}, \binits{T.}},
\bauthor{\bsnm{Wright}, \binits{S.}},
\bauthor{\bsnm{Charnay}, \binits{B.}},
\bauthor{\bsnm{Drossart}, \binits{P.}},
\bauthor{\bsnm{Leconte}, \binits{J.}},
\bauthor{\bsnm{Tsiaras}, \binits{A.}},
\bauthor{\bsnm{Venot}, \binits{O.}},
\bauthor{\bsnm{Waldmann}, \binits{I.}},
\bauthor{\bsnm{Beaulieu}, \binits{J.-P.}}:
\batitle{{ARES. III. Unveiling the Two Faces of KELT-7 b with HST WFC3}}.
\bjtitle{\aj}
\bvolume{160}(\bissue{3}),
\bfpage{112}
(\byear{2020})
{\href{https://arxiv.org/abs/2006.14199}{{arXiv:2006.14199}}}
{[astro-ph.EP]}.
\doiurl{10.3847/1538-3881/aba000}
\end{barticle}
\endbibitem

\bibitem{ares4}
\begin{barticle}
\bauthor{\bsnm{Guilluy}, \binits{G.}},
\bauthor{\bsnm{Gressier}, \binits{A.}},
\bauthor{\bsnm{Wright}, \binits{S.}},
\bauthor{\bsnm{Santerne}, \binits{A.}},
\bauthor{\bsnm{Jaziri}, \binits{A.Y.}},
\bauthor{\bsnm{Edwards}, \binits{B.}},
\bauthor{\bsnm{Changeat}, \binits{Q.}},
\bauthor{\bsnm{Modirrousta-Galian}, \binits{D.}},
\bauthor{\bsnm{Skaf}, \binits{N.}},
\bauthor{\bsnm{Al-Refaie}, \binits{A.}},
\bauthor{\bsnm{Baeyens}, \binits{R.}},
\bauthor{\bsnm{Bieger}, \binits{M.F.}},
\bauthor{\bsnm{Blain}, \binits{D.}},
\bauthor{\bsnm{Kiefer}, \binits{F.}},
\bauthor{\bsnm{Morvan}, \binits{M.}},
\bauthor{\bsnm{Mugnai}, \binits{L.V.}},
\bauthor{\bsnm{Pluriel}, \binits{W.}},
\bauthor{\bsnm{Poveda}, \binits{M.}},
\bauthor{\bsnm{Zingales}, \binits{T.}},
\bauthor{\bsnm{Whiteford}, \binits{N.}},
\bauthor{\bsnm{Yip}, \binits{K.H.}},
\bauthor{\bsnm{Charnay}, \binits{B.}},
\bauthor{\bsnm{Leconte}, \binits{J.}},
\bauthor{\bsnm{Drossart}, \binits{P.}},
\bauthor{\bsnm{Sozzetti}, \binits{A.}},
\bauthor{\bsnm{Marcq}, \binits{E.}},
\bauthor{\bsnm{Tsiaras}, \binits{A.}},
\bauthor{\bsnm{Venot}, \binits{O.}},
\bauthor{\bsnm{Waldmann}, \binits{I.}},
\bauthor{\bsnm{Beaulieu}, \binits{J.-P.}}:
\batitle{{ARES IV: Probing the Atmospheres of the Two Warm Small Planets HD
  106315c and HD 3167c with the HST/WFC3 Camera}}.
\bjtitle{\aj}
\bvolume{161}(\bissue{1}),
\bfpage{19}
(\byear{2021})
{\href{https://arxiv.org/abs/2011.03221}{{arXiv:2011.03221}}}
{[astro-ph.EP]}.
\doiurl{10.3847/1538-3881/abc3c8}
\end{barticle}
\endbibitem

\bibitem{ares5}
\begin{barticle}
\bauthor{\bsnm{Mugnai}, \binits{L.V.}},
\bauthor{\bsnm{Modirrousta-Galian}, \binits{D.}},
\bauthor{\bsnm{Edwards}, \binits{B.}},
\bauthor{\bsnm{Changeat}, \binits{Q.}},
\bauthor{\bsnm{Bouwman}, \binits{J.}},
\bauthor{\bsnm{Morello}, \binits{G.}},
\bauthor{\bsnm{Al-Refaie}, \binits{A.}},
\bauthor{\bsnm{Baeyens}, \binits{R.}},
\bauthor{\bsnm{Bieger}, \binits{M.F.}},
\bauthor{\bsnm{Blain}, \binits{D.}},
\bauthor{\bsnm{Gressier}, \binits{A.}},
\bauthor{\bsnm{Guilluy}, \binits{G.}},
\bauthor{\bsnm{Jaziri}, \binits{Y.}},
\bauthor{\bsnm{Kiefer}, \binits{F.}},
\bauthor{\bsnm{Morvan}, \binits{M.}},
\bauthor{\bsnm{Pluriel}, \binits{W.}},
\bauthor{\bsnm{Poveda}, \binits{M.}},
\bauthor{\bsnm{Skaf}, \binits{N.}},
\bauthor{\bsnm{Whiteford}, \binits{N.}},
\bauthor{\bsnm{Wright}, \binits{S.}},
\bauthor{\bsnm{Yip}, \binits{K.H.}},
\bauthor{\bsnm{Zingales}, \binits{T.}},
\bauthor{\bsnm{Charnay}, \binits{B.}},
\bauthor{\bsnm{Drossart}, \binits{P.}},
\bauthor{\bsnm{Leconte}, \binits{J.}},
\bauthor{\bsnm{Venot}, \binits{O.}},
\bauthor{\bsnm{Waldmann}, \binits{I.}},
\bauthor{\bsnm{Beaulieu}, \binits{J.-P.}}:
\batitle{{ARES. V. No Evidence For Molecular Absorption in the HST WFC3
  Spectrum of GJ 1132 b}}.
\bjtitle{\aj}
\bvolume{161}(\bissue{6}),
\bfpage{284}
(\byear{2021})
{\href{https://arxiv.org/abs/2104.01873}{{arXiv:2104.01873}}}
{[astro-ph.EP]}.
\doiurl{10.3847/1538-3881/abf3c3}
\end{barticle}
\endbibitem

\bibitem{PC_Changeat2022}
\begin{barticle}
\bauthor{\bsnm{Changeat}, \binits{Q.}}:
\batitle{On spectroscopic phase-curve retrievals: H2 dissociation and thermal
  inversion in the atmosphere of the ultrahot jupiter wasp-103 b}.
\bjtitle{\aj}
\bvolume{163}(\bissue{3}),
\bfpage{106}
(\byear{2022}).
\doiurl{10.3847/1538-3881/ac4475}
\end{barticle}
\endbibitem

\bibitem{Encrenaz2015}
\begin{barticle}
\bauthor{\bsnm{Encrenaz}, \binits{T.}},
\bauthor{\bsnm{Tinetti}, \binits{G.}},
\bauthor{\bsnm{Tessenyi}, \binits{M.}},
\bauthor{\bsnm{Drossart}, \binits{P.}},
\bauthor{\bsnm{Hartogh}, \binits{P.}},
\bauthor{\bsnm{Coustenis}, \binits{A.}}:
\batitle{{Transit spectroscopy of exoplanets from space: how to optimize the
  wavelength coverage and spectral resolving power}}.
\bjtitle{\expa}
\bvolume{40}(\bissue{2-3}),
\bfpage{523}--\blpage{543}
(\byear{2015}).
\doiurl{10.1007/s10686-014-9415-0}
\end{barticle}
\endbibitem

\bibitem{Tsiaras2019}
\begin{botherref}
\oauthor{\bsnm{Tsiaras}, \binits{A.}},
\oauthor{\bsnm{Waldmann}, \binits{I.P.}},
\oauthor{\bsnm{Tinetti}, \binits{G.}},
\oauthor{\bsnm{Tennyson}, \binits{J.}},
\oauthor{\bsnm{Yurchenko}, \binits{S.N.}}:
{Water vapour in the atmosphere of the habitable-zone eight-Earth-mass planet
  K2-18 b}.
\nastro
\textbf{3}
(2019).
\doiurl{10.1038/s41550-019-0878-9}
\end{botherref}
\endbibitem

\bibitem{5keyChangeat2022}
\begin{barticle}
\bauthor{\bsnm{Changeat}, \binits{Q.}},
\bauthor{\bsnm{Edwards}, \binits{B.}},
\bauthor{\bsnm{Al-Refaie}, \binits{A.F.}},
\bauthor{\bsnm{Tsiaras}, \binits{A.}},
\bauthor{\bsnm{Skinner}, \binits{J.W.}},
\bauthor{\bsnm{Cho}, \binits{J.Y.K.}},
\bauthor{\bsnm{Yip}, \binits{K.H.}},
\bauthor{\bsnm{Anisman}, \binits{L.}},
\bauthor{\bsnm{Ikoma}, \binits{M.}},
\bauthor{\bsnm{Bieger}, \binits{M.F.}},
\bauthor{\bsnm{Venot}, \binits{O.}},
\bauthor{\bsnm{Shibata}, \binits{S.}},
\bauthor{\bsnm{Waldmann}, \binits{I.P.}},
\bauthor{\bsnm{Tinetti}, \binits{G.}}:
\batitle{Five key exoplanet questions answered via the analysis of 25
  hot-jupiter atmospheres in eclipse}.
\bjtitle{\apjs}
\bvolume{260}(\bissue{1}),
\bfpage{3}
(\byear{2022}).
\doiurl{10.3847/1538-4365/ac5cc2}
\end{barticle}
\endbibitem

\bibitem{Greene2016}
\begin{barticle}
\bauthor{\bsnm{{Greene}}, \binits{T.P.}},
\bauthor{\bsnm{{Line}}, \binits{M.R.}},
\bauthor{\bsnm{{Montero}}, \binits{C.}},
\bauthor{\bsnm{{Fortney}}, \binits{J.J.}},
\bauthor{\bsnm{{Lustig-Yaeger}}, \binits{J.}},
\bauthor{\bsnm{{Luther}}, \binits{K.}}:
\batitle{{Characterizing Transiting Exoplanet Atmospheres with JWST}}.
\bjtitle{\apj}
\bvolume{817}(\bissue{1}),
\bfpage{17}
(\byear{2016})
{\href{https://arxiv.org/abs/1511.05528}{{arXiv:1511.05528}}}
{[astro-ph.EP]}.
\doiurl{10.3847/0004-637X/817/1/17}
\end{barticle}
\endbibitem

\bibitem{ERS_WASP39b_JWST_NIRISS}
\begin{botherref}
\oauthor{\bsnm{Feinstein}, \binits{A.D.}},
\oauthor{\bsnm{Radica}, \binits{M.}},
\oauthor{\bsnm{Welbanks}, \binits{L.}},
\oauthor{\bsnm{Murray}, \binits{C.A.}},
\oauthor{\bsnm{Ohno}, \binits{K.}},
\oauthor{\bsnm{Coulombe}, \binits{L.-P.}},
\oauthor{\bsnm{Espinoza}, \binits{N.}},
\oauthor{\bsnm{Bean}, \binits{J.L.}},
\oauthor{\bsnm{Teske}, \binits{J.K.}},
\oauthor{\bsnm{Benneke}, \binits{B.}},
\oauthor{\bsnm{Line}, \binits{M.R.}},
\oauthor{\bsnm{Rustamkulov}, \binits{Z.}},
\oauthor{\bsnm{Saba}, \binits{A.}},
\oauthor{\bsnm{Tsiaras}, \binits{A.}},
\oauthor{\bsnm{Barstow}, \binits{J.K.}},
\oauthor{\bsnm{Fortney}, \binits{J.J.}},
\oauthor{\bsnm{Gao}, \binits{P.}},
\oauthor{\bsnm{Knutson}, \binits{H.A.}},
\oauthor{\bsnm{MacDonald}, \binits{R.J.}},
\oauthor{\bsnm{Mikal-Evans}, \binits{T.}},
\oauthor{\bsnm{Rackham}, \binits{B.V.}},
\oauthor{\bsnm{Taylor}, \binits{J.}},
\oauthor{\bsnm{Parmentier}, \binits{V.}},
\oauthor{\bsnm{Batalha}, \binits{N.M.}},
\oauthor{\bsnm{Berta-Thompson}, \binits{Z.K.}},
\oauthor{\bsnm{Carter}, \binits{A.L.}},
\oauthor{\bsnm{Changeat}, \binits{Q.}},
\oauthor{\bsnm{Santos}, \binits{L.A.D.}},
\oauthor{\bsnm{Gibson}, \binits{N.P.}},
\oauthor{\bsnm{Goyal}, \binits{J.M.}},
\oauthor{\bsnm{Kreidberg}, \binits{L.}},
\oauthor{\bsnm{López-Morales}, \binits{M.}},
\oauthor{\bsnm{Lothringer}, \binits{J.D.}},
\oauthor{\bsnm{Miguel}, \binits{Y.}},
\oauthor{\bsnm{Molaverdikhani}, \binits{K.}},
\oauthor{\bsnm{Moran}, \binits{S.E.}},
\oauthor{\bsnm{Morello}, \binits{G.}},
\oauthor{\bsnm{Mukherjee}, \binits{S.}},
\oauthor{\bsnm{Sing}, \binits{D.K.}},
\oauthor{\bsnm{Stevenson}, \binits{K.B.}},
\oauthor{\bsnm{Wakeford}, \binits{H.R.}},
\oauthor{\bsnm{Ahrer}, \binits{E.-M.}},
\oauthor{\bsnm{Alam}, \binits{M.K.}},
\oauthor{\bsnm{Alderson}, \binits{L.}},
\oauthor{\bsnm{Allen}, \binits{N.H.}},
\oauthor{\bsnm{Batalha}, \binits{N.E.}},
\oauthor{\bsnm{Bell}, \binits{T.J.}},
\oauthor{\bsnm{Blecic}, \binits{J.}},
\oauthor{\bsnm{Brande}, \binits{J.}},
\oauthor{\bsnm{Caceres}, \binits{C.}},
\oauthor{\bsnm{Casewell}, \binits{S.L.}},
\oauthor{\bsnm{Chubb}, \binits{K.L.}},
\oauthor{\bsnm{Crossfield}, \binits{I.J.M.}},
\oauthor{\bsnm{Crouzet}, \binits{N.}},
\oauthor{\bsnm{Cubillos}, \binits{P.E.}},
\oauthor{\bsnm{Decin}, \binits{L.}},
\oauthor{\bsnm{Désert}, \binits{J.-M.}},
\oauthor{\bsnm{Harrington}, \binits{J.}},
\oauthor{\bsnm{Heng}, \binits{K.}},
\oauthor{\bsnm{Henning}, \binits{T.}},
\oauthor{\bsnm{Iro}, \binits{N.}},
\oauthor{\bsnm{Kempton}, \binits{E.M.-R.}},
\oauthor{\bsnm{Kendrew}, \binits{S.}},
\oauthor{\bsnm{Kirk}, \binits{J.}},
\oauthor{\bsnm{Krick}, \binits{J.}},
\oauthor{\bsnm{Lagage}, \binits{P.-O.}},
\oauthor{\bsnm{Lendl}, \binits{M.}},
\oauthor{\bsnm{Mancini}, \binits{L.}},
\oauthor{\bsnm{Mansfield}, \binits{M.}},
\oauthor{\bsnm{May}, \binits{E.M.}},
\oauthor{\bsnm{Mayne}, \binits{N.J.}},
\oauthor{\bsnm{Nikolov}, \binits{N.K.}},
\oauthor{\bsnm{Palle}, \binits{E.}},
\oauthor{\bparticle{de~la} \bsnm{Roche}, \binits{D.J.M.P.d.}},
\oauthor{\bsnm{Piaulet}, \binits{C.}},
\oauthor{\bsnm{Powell}, \binits{D.}},
\oauthor{\bsnm{Redfield}, \binits{S.}},
\oauthor{\bsnm{Rogers}, \binits{L.K.}},
\oauthor{\bsnm{Roman}, \binits{M.T.}},
\oauthor{\bsnm{Roy}, \binits{P.-A.}},
\oauthor{\bsnm{Nixon}, \binits{M.C.}},
\oauthor{\bsnm{Schlawin}, \binits{E.}},
\oauthor{\bsnm{Tan}, \binits{X.}},
\oauthor{\bsnm{Tremblin}, \binits{P.}},
\oauthor{\bsnm{Turner}, \binits{J.D.}},
\oauthor{\bsnm{Venot}, \binits{O.}},
\oauthor{\bsnm{Waalkes}, \binits{W.C.}},
\oauthor{\bsnm{Wheatley}, \binits{P.J.}},
\oauthor{\bsnm{Zhang}, \binits{X.}}:
Early Release Science of the exoplanet WASP-39b with JWST NIRISS.
arXiv
(2022).
\doiurl{10.48550/ARXIV.2211.10493}.
\url{https://arxiv.org/abs/2211.10493}
\end{botherref}
\endbibitem

\bibitem{ERS_WASP39b_JWST_NIRCam}
\begin{botherref}
\oauthor{\bsnm{Ahrer}, \binits{E.-M.}},
\oauthor{\bsnm{Stevenson}, \binits{K.B.}},
\oauthor{\bsnm{Mansfield}, \binits{M.}},
\oauthor{\bsnm{Moran}, \binits{S.E.}},
\oauthor{\bsnm{Brande}, \binits{J.}},
\oauthor{\bsnm{Morello}, \binits{G.}},
\oauthor{\bsnm{Murray}, \binits{C.A.}},
\oauthor{\bsnm{Nikolov}, \binits{N.K.}},
\oauthor{\bparticle{de~la} \bsnm{Roche}, \binits{D.J.M.P.d.}},
\oauthor{\bsnm{Schlawin}, \binits{E.}},
\oauthor{\bsnm{Wheatley}, \binits{P.J.}},
\oauthor{\bsnm{Zieba}, \binits{S.}},
\oauthor{\bsnm{Batalha}, \binits{N.E.}},
\oauthor{\bsnm{Damiano}, \binits{M.}},
\oauthor{\bsnm{Goyal}, \binits{J.M.}},
\oauthor{\bsnm{Lendl}, \binits{M.}},
\oauthor{\bsnm{Lothringer}, \binits{J.D.}},
\oauthor{\bsnm{Mukherjee}, \binits{S.}},
\oauthor{\bsnm{Ohno}, \binits{K.}},
\oauthor{\bsnm{Batalha}, \binits{N.M.}},
\oauthor{\bsnm{Battley}, \binits{M.P.}},
\oauthor{\bsnm{Bean}, \binits{J.L.}},
\oauthor{\bsnm{Beatty}, \binits{T.G.}},
\oauthor{\bsnm{Benneke}, \binits{B.}},
\oauthor{\bsnm{Berta-Thompson}, \binits{Z.K.}},
\oauthor{\bsnm{Carter}, \binits{A.L.}},
\oauthor{\bsnm{Cubillos}, \binits{P.E.}},
\oauthor{\bsnm{Daylan}, \binits{T.}},
\oauthor{\bsnm{Espinoza}, \binits{N.}},
\oauthor{\bsnm{Gao}, \binits{P.}},
\oauthor{\bsnm{Gibson}, \binits{N.P.}},
\oauthor{\bsnm{Gill}, \binits{S.}},
\oauthor{\bsnm{Harrington}, \binits{J.}},
\oauthor{\bsnm{Hu}, \binits{R.}},
\oauthor{\bsnm{Kreidberg}, \binits{L.}},
\oauthor{\bsnm{Lewis}, \binits{N.K.}},
\oauthor{\bsnm{Line}, \binits{M.R.}},
\oauthor{\bsnm{López-Morales}, \binits{M.}},
\oauthor{\bsnm{Parmentier}, \binits{V.}},
\oauthor{\bsnm{Powell}, \binits{D.K.}},
\oauthor{\bsnm{Sing}, \binits{D.K.}},
\oauthor{\bsnm{Tsai}, \binits{S.-M.}},
\oauthor{\bsnm{Wakeford}, \binits{H.R.}},
\oauthor{\bsnm{Welbanks}, \binits{L.}},
\oauthor{\bsnm{Alam}, \binits{M.K.}},
\oauthor{\bsnm{Alderson}, \binits{L.}},
\oauthor{\bsnm{Allen}, \binits{N.H.}},
\oauthor{\bsnm{Anderson}, \binits{D.R.}},
\oauthor{\bsnm{Barstow}, \binits{J.K.}},
\oauthor{\bsnm{Bayliss}, \binits{D.}},
\oauthor{\bsnm{Bell}, \binits{T.J.}},
\oauthor{\bsnm{Blecic}, \binits{J.}},
\oauthor{\bsnm{Bryant}, \binits{E.M.}},
\oauthor{\bsnm{Burleigh}, \binits{M.R.}},
\oauthor{\bsnm{Carone}, \binits{L.}},
\oauthor{\bsnm{Casewell}, \binits{S.L.}},
\oauthor{\bsnm{Changeat}, \binits{Q.}},
\oauthor{\bsnm{Chubb}, \binits{K.L.}},
\oauthor{\bsnm{Crossfield}, \binits{I.J.M.}},
\oauthor{\bsnm{Crouzet}, \binits{N.}},
\oauthor{\bsnm{Decin}, \binits{L.}},
\oauthor{\bsnm{Désert}, \binits{J.-M.}},
\oauthor{\bsnm{Feinstein}, \binits{A.D.}},
\oauthor{\bsnm{Flagg}, \binits{L.}},
\oauthor{\bsnm{Fortney}, \binits{J.J.}},
\oauthor{\bsnm{Gizis}, \binits{J.E.}},
\oauthor{\bsnm{Heng}, \binits{K.}},
\oauthor{\bsnm{Iro}, \binits{N.}},
\oauthor{\bsnm{Kempton}, \binits{E.M.-R.}},
\oauthor{\bsnm{Kendrew}, \binits{S.}},
\oauthor{\bsnm{Kirk}, \binits{J.}},
\oauthor{\bsnm{Knutson}, \binits{H.A.}},
\oauthor{\bsnm{Komacek}, \binits{T.D.}},
\oauthor{\bsnm{Lagage}, \binits{P.-O.}},
\oauthor{\bsnm{Leconte}, \binits{J.}},
\oauthor{\bsnm{Lustig-Yaeger}, \binits{J.}},
\oauthor{\bsnm{MacDonald}, \binits{R.J.}},
\oauthor{\bsnm{Mancini}, \binits{L.}},
\oauthor{\bsnm{May}, \binits{E.M.}},
\oauthor{\bsnm{Mayne}, \binits{N.J.}},
\oauthor{\bsnm{Miguel}, \binits{Y.}},
\oauthor{\bsnm{Mikal-Evans}, \binits{T.}},
\oauthor{\bsnm{Molaverdikhani}, \binits{K.}},
\oauthor{\bsnm{Palle}, \binits{E.}},
\oauthor{\bsnm{Piaulet}, \binits{C.}},
\oauthor{\bsnm{Rackham}, \binits{B.V.}},
\oauthor{\bsnm{Redfield}, \binits{S.}},
\oauthor{\bsnm{Rogers}, \binits{L.K.}},
\oauthor{\bsnm{Roy}, \binits{P.-A.}},
\oauthor{\bsnm{Rustamkulov}, \binits{Z.}},
\oauthor{\bsnm{Shkolnik}, \binits{E.L.}},
\oauthor{\bsnm{Sotzen}, \binits{K.S.}},
\oauthor{\bsnm{Taylor}, \binits{J.}},
\oauthor{\bsnm{Tremblin}, \binits{P.}},
\oauthor{\bsnm{Tucker}, \binits{G.S.}},
\oauthor{\bsnm{Turner}, \binits{J.D.}},
\oauthor{\bparticle{de} \bsnm{Val-Borro}, \binits{M.}},
\oauthor{\bsnm{Venot}, \binits{O.}},
\oauthor{\bsnm{Zhang}, \binits{X.}}:
Early Release Science of the exoplanet WASP-39b with JWST NIRCam.
arXiv
(2022).
\doiurl{10.48550/ARXIV.2211.10489}.
\url{https://arxiv.org/abs/2211.10489}
\end{botherref}
\endbibitem

\bibitem{ERS_WASP39b_JWST_NIRSpec_PRISM}
\begin{botherref}
\oauthor{\bsnm{Rustamkulov}, \binits{Z.}},
\oauthor{\bsnm{Sing}, \binits{D.K.}},
\oauthor{\bsnm{Mukherjee}, \binits{S.}},
\oauthor{\bsnm{May}, \binits{E.M.}},
\oauthor{\bsnm{Kirk}, \binits{J.}},
\oauthor{\bsnm{Schlawin}, \binits{E.}},
\oauthor{\bsnm{Line}, \binits{M.R.}},
\oauthor{\bsnm{Piaulet}, \binits{C.}},
\oauthor{\bsnm{Carter}, \binits{A.L.}},
\oauthor{\bsnm{Batalha}, \binits{N.E.}},
\oauthor{\bsnm{Goyal}, \binits{J.M.}},
\oauthor{\bsnm{López-Morales}, \binits{M.}},
\oauthor{\bsnm{Lothringer}, \binits{J.D.}},
\oauthor{\bsnm{MacDonald}, \binits{R.J.}},
\oauthor{\bsnm{Moran}, \binits{S.E.}},
\oauthor{\bsnm{Stevenson}, \binits{K.B.}},
\oauthor{\bsnm{Wakeford}, \binits{H.R.}},
\oauthor{\bsnm{Espinoza}, \binits{N.}},
\oauthor{\bsnm{Bean}, \binits{J.L.}},
\oauthor{\bsnm{Batalha}, \binits{N.M.}},
\oauthor{\bsnm{Benneke}, \binits{B.}},
\oauthor{\bsnm{Berta-Thompson}, \binits{Z.K.}},
\oauthor{\bsnm{Crossfield}, \binits{I.J.M.}},
\oauthor{\bsnm{Gao}, \binits{P.}},
\oauthor{\bsnm{Kreidberg}, \binits{L.}},
\oauthor{\bsnm{Powell}, \binits{D.K.}},
\oauthor{\bsnm{Cubillos}, \binits{P.E.}},
\oauthor{\bsnm{Gibson}, \binits{N.P.}},
\oauthor{\bsnm{Leconte}, \binits{J.}},
\oauthor{\bsnm{Molaverdikhani}, \binits{K.}},
\oauthor{\bsnm{Nikolov}, \binits{N.K.}},
\oauthor{\bsnm{Parmentier}, \binits{V.}},
\oauthor{\bsnm{Roy}, \binits{P.}},
\oauthor{\bsnm{Taylor}, \binits{J.}},
\oauthor{\bsnm{Turner}, \binits{J.D.}},
\oauthor{\bsnm{Wheatley}, \binits{P.J.}},
\oauthor{\bsnm{Aggarwal}, \binits{K.}},
\oauthor{\bsnm{Ahrer}, \binits{E.}},
\oauthor{\bsnm{Alam}, \binits{M.K.}},
\oauthor{\bsnm{Alderson}, \binits{L.}},
\oauthor{\bsnm{Allen}, \binits{N.H.}},
\oauthor{\bsnm{Banerjee}, \binits{A.}},
\oauthor{\bsnm{Barat}, \binits{S.}},
\oauthor{\bsnm{Barrado}, \binits{D.}},
\oauthor{\bsnm{Barstow}, \binits{J.K.}},
\oauthor{\bsnm{Bell}, \binits{T.J.}},
\oauthor{\bsnm{Blecic}, \binits{J.}},
\oauthor{\bsnm{Brande}, \binits{J.}},
\oauthor{\bsnm{Casewell}, \binits{S.}},
\oauthor{\bsnm{Changeat}, \binits{Q.}},
\oauthor{\bsnm{Chubb}, \binits{K.L.}},
\oauthor{\bsnm{Crouzet}, \binits{N.}},
\oauthor{\bsnm{Daylan}, \binits{T.}},
\oauthor{\bsnm{Decin}, \binits{L.}},
\oauthor{\bsnm{Désert}, \binits{J.}},
\oauthor{\bsnm{Mikal-Evans}, \binits{T.}},
\oauthor{\bsnm{Feinstein}, \binits{A.D.}},
\oauthor{\bsnm{Flagg}, \binits{L.}},
\oauthor{\bsnm{Fortney}, \binits{J.J.}},
\oauthor{\bsnm{Harrington}, \binits{J.}},
\oauthor{\bsnm{Heng}, \binits{K.}},
\oauthor{\bsnm{Hong}, \binits{Y.}},
\oauthor{\bsnm{Hu}, \binits{R.}},
\oauthor{\bsnm{Iro}, \binits{N.}},
\oauthor{\bsnm{Kataria}, \binits{T.}},
\oauthor{\bsnm{Kempton}, \binits{E.M.-R.}},
\oauthor{\bsnm{Krick}, \binits{J.}},
\oauthor{\bsnm{Lendl}, \binits{M.}},
\oauthor{\bsnm{Lillo-Box}, \binits{J.}},
\oauthor{\bsnm{Louca}, \binits{A.}},
\oauthor{\bsnm{Lustig-Yaeger}, \binits{J.}},
\oauthor{\bsnm{Mancini}, \binits{L.}},
\oauthor{\bsnm{Mansfield}, \binits{M.}},
\oauthor{\bsnm{Mayne}, \binits{N.J.}},
\oauthor{\bsnm{Miguel}, \binits{Y.}},
\oauthor{\bsnm{Morello}, \binits{G.}},
\oauthor{\bsnm{Ohno}, \binits{K.}},
\oauthor{\bsnm{Palle}, \binits{E.}},
\oauthor{\bparticle{de~la} \bsnm{Roche}, \binits{D.J.M.P.d.}},
\oauthor{\bsnm{Rackham}, \binits{B.V.}},
\oauthor{\bsnm{Radica}, \binits{M.}},
\oauthor{\bsnm{Ramos-Rosado}, \binits{L.}},
\oauthor{\bsnm{Redfield}, \binits{S.}},
\oauthor{\bsnm{Rogers}, \binits{L.K.}},
\oauthor{\bsnm{Shkolnik}, \binits{E.L.}},
\oauthor{\bsnm{Southworth}, \binits{J.}},
\oauthor{\bsnm{Teske}, \binits{J.}},
\oauthor{\bsnm{Tremblin}, \binits{P.}},
\oauthor{\bsnm{Tucker}, \binits{G.S.}},
\oauthor{\bsnm{Venot}, \binits{O.}},
\oauthor{\bsnm{Waalkes}, \binits{W.C.}},
\oauthor{\bsnm{Welbanks}, \binits{L.}},
\oauthor{\bsnm{Zhang}, \binits{X.}},
\oauthor{\bsnm{Zieba}, \binits{S.}}:
Early Release Science of the exoplanet WASP-39b with JWST NIRSpec PRISM.
arXiv
(2022).
\doiurl{10.48550/ARXIV.2211.10487}.
\url{https://arxiv.org/abs/2211.10487}
\end{botherref}
\endbibitem

\bibitem{ERS_WASP39b_JWST_NIRSpec_G395H}
\begin{botherref}
\oauthor{\bsnm{Alderson}, \binits{L.}},
\oauthor{\bsnm{Wakeford}, \binits{H.R.}},
\oauthor{\bsnm{Alam}, \binits{M.K.}},
\oauthor{\bsnm{Batalha}, \binits{N.E.}},
\oauthor{\bsnm{Lothringer}, \binits{J.D.}},
\oauthor{\bsnm{Redai}, \binits{J.A.}},
\oauthor{\bsnm{Barat}, \binits{S.}},
\oauthor{\bsnm{Brande}, \binits{J.}},
\oauthor{\bsnm{Damiano}, \binits{M.}},
\oauthor{\bsnm{Daylan}, \binits{T.}},
\oauthor{\bsnm{Espinoza}, \binits{N.}},
\oauthor{\bsnm{Flagg}, \binits{L.}},
\oauthor{\bsnm{Goyal}, \binits{J.M.}},
\oauthor{\bsnm{Grant}, \binits{D.}},
\oauthor{\bsnm{Hu}, \binits{R.}},
\oauthor{\bsnm{Inglis}, \binits{J.}},
\oauthor{\bsnm{Lee}, \binits{E.K.H.}},
\oauthor{\bsnm{Mikal-Evans}, \binits{T.}},
\oauthor{\bsnm{Ramos-Rosado}, \binits{L.}},
\oauthor{\bsnm{Roy}, \binits{P.-A.}},
\oauthor{\bsnm{Wallack}, \binits{N.L.}},
\oauthor{\bsnm{Batalha}, \binits{N.M.}},
\oauthor{\bsnm{Bean}, \binits{J.L.}},
\oauthor{\bsnm{Benneke}, \binits{B.}},
\oauthor{\bsnm{Berta-Thompson}, \binits{Z.K.}},
\oauthor{\bsnm{Carter}, \binits{A.L.}},
\oauthor{\bsnm{Changeat}, \binits{Q.}},
\oauthor{\bsnm{Colón}, \binits{K.D.}},
\oauthor{\bsnm{Crossfield}, \binits{I.J.M.}},
\oauthor{\bsnm{Désert}, \binits{J.-M.}},
\oauthor{\bsnm{Foreman-Mackey}, \binits{D.}},
\oauthor{\bsnm{Gibson}, \binits{N.P.}},
\oauthor{\bsnm{Kreidberg}, \binits{L.}},
\oauthor{\bsnm{Line}, \binits{M.R.}},
\oauthor{\bsnm{López-Morales}, \binits{M.}},
\oauthor{\bsnm{Molaverdikhani}, \binits{K.}},
\oauthor{\bsnm{Moran}, \binits{S.E.}},
\oauthor{\bsnm{Morello}, \binits{G.}},
\oauthor{\bsnm{Moses}, \binits{J.I.}},
\oauthor{\bsnm{Mukherjee}, \binits{S.}},
\oauthor{\bsnm{Schlawin}, \binits{E.}},
\oauthor{\bsnm{Sing}, \binits{D.K.}},
\oauthor{\bsnm{Stevenson}, \binits{K.B.}},
\oauthor{\bsnm{Taylor}, \binits{J.}},
\oauthor{\bsnm{Aggarwal}, \binits{K.}},
\oauthor{\bsnm{Ahrer}, \binits{E.-M.}},
\oauthor{\bsnm{Allen}, \binits{N.H.}},
\oauthor{\bsnm{Barstow}, \binits{J.K.}},
\oauthor{\bsnm{Bell}, \binits{T.J.}},
\oauthor{\bsnm{Blecic}, \binits{J.}},
\oauthor{\bsnm{Casewell}, \binits{S.L.}},
\oauthor{\bsnm{Chubb}, \binits{K.L.}},
\oauthor{\bsnm{Crouzet}, \binits{N.}},
\oauthor{\bsnm{Cubillos}, \binits{P.E.}},
\oauthor{\bsnm{Decin}, \binits{L.}},
\oauthor{\bsnm{Feinstein}, \binits{A.D.}},
\oauthor{\bsnm{Fortney}, \binits{J.J.}},
\oauthor{\bsnm{Harrington}, \binits{J.}},
\oauthor{\bsnm{Heng}, \binits{K.}},
\oauthor{\bsnm{Iro}, \binits{N.}},
\oauthor{\bsnm{Kempton}, \binits{E.M.-R.}},
\oauthor{\bsnm{Kirk}, \binits{J.}},
\oauthor{\bsnm{Knutson}, \binits{H.A.}},
\oauthor{\bsnm{Krick}, \binits{J.}},
\oauthor{\bsnm{Leconte}, \binits{J.}},
\oauthor{\bsnm{Lendl}, \binits{M.}},
\oauthor{\bsnm{MacDonald}, \binits{R.J.}},
\oauthor{\bsnm{Mancini}, \binits{L.}},
\oauthor{\bsnm{Mansfield}, \binits{M.}},
\oauthor{\bsnm{May}, \binits{E.M.}},
\oauthor{\bsnm{Mayne}, \binits{N.J.}},
\oauthor{\bsnm{Miguel}, \binits{Y.}},
\oauthor{\bsnm{Nikolov}, \binits{N.K.}},
\oauthor{\bsnm{Ohno}, \binits{K.}},
\oauthor{\bsnm{Palle}, \binits{E.}},
\oauthor{\bsnm{Parmentier}, \binits{V.}},
\oauthor{\bparticle{de~la} \bsnm{Roche}, \binits{D.J.M.P.d.}},
\oauthor{\bsnm{Piaulet}, \binits{C.}},
\oauthor{\bsnm{Powell}, \binits{D.}},
\oauthor{\bsnm{Rackham}, \binits{B.V.}},
\oauthor{\bsnm{Redfield}, \binits{S.}},
\oauthor{\bsnm{Rogers}, \binits{L.K.}},
\oauthor{\bsnm{Rustamkulov}, \binits{Z.}},
\oauthor{\bsnm{Tan}, \binits{X.}},
\oauthor{\bsnm{Tremblin}, \binits{P.}},
\oauthor{\bsnm{Tsai}, \binits{S.-M.}},
\oauthor{\bsnm{Turner}, \binits{J.D.}},
\oauthor{\bparticle{de} \bsnm{Val-Borro}, \binits{M.}},
\oauthor{\bsnm{Venot}, \binits{O.}},
\oauthor{\bsnm{Welbanks}, \binits{L.}},
\oauthor{\bsnm{Wheatley}, \binits{P.J.}},
\oauthor{\bsnm{Zhang}, \binits{X.}}:
Early Release Science of the Exoplanet WASP-39b with JWST NIRSpec G395H.
arXiv
(2022).
\doiurl{10.48550/ARXIV.2211.10488}.
\url{https://arxiv.org/abs/2211.10488}
\end{botherref}
\endbibitem

\bibitem{SO2_JWST_WASP39b}
\begin{botherref}
\oauthor{\bsnm{Tsai}, \binits{S.-M.}},
\oauthor{\bsnm{Lee}, \binits{E.K.H.}},
\oauthor{\bsnm{Powell}, \binits{D.}},
\oauthor{\bsnm{Gao}, \binits{P.}},
\oauthor{\bsnm{Zhang}, \binits{X.}},
\oauthor{\bsnm{Moses}, \binits{J.}},
\oauthor{\bsnm{Hébrard}, \binits{E.}},
\oauthor{\bsnm{Venot}, \binits{O.}},
\oauthor{\bsnm{Parmentier}, \binits{V.}},
\oauthor{\bsnm{Jordan}, \binits{S.}},
\oauthor{\bsnm{Hu}, \binits{R.}},
\oauthor{\bsnm{Alam}, \binits{M.K.}},
\oauthor{\bsnm{Alderson}, \binits{L.}},
\oauthor{\bsnm{Batalha}, \binits{N.M.}},
\oauthor{\bsnm{Bean}, \binits{J.L.}},
\oauthor{\bsnm{Benneke}, \binits{B.}},
\oauthor{\bsnm{Bierson}, \binits{C.J.}},
\oauthor{\bsnm{Brady}, \binits{R.P.}},
\oauthor{\bsnm{Carone}, \binits{L.}},
\oauthor{\bsnm{Carter}, \binits{A.L.}},
\oauthor{\bsnm{Chubb}, \binits{K.L.}},
\oauthor{\bsnm{Inglis}, \binits{J.}},
\oauthor{\bsnm{Leconte}, \binits{J.}},
\oauthor{\bsnm{Lopez-Morales}, \binits{M.}},
\oauthor{\bsnm{Miguel}, \binits{Y.}},
\oauthor{\bsnm{Molaverdikhani}, \binits{K.}},
\oauthor{\bsnm{Rustamkulov}, \binits{Z.}},
\oauthor{\bsnm{Sing}, \binits{D.K.}},
\oauthor{\bsnm{Stevenson}, \binits{K.B.}},
\oauthor{\bsnm{Wakeford}, \binits{H.R.}},
\oauthor{\bsnm{Yang}, \binits{J.}},
\oauthor{\bsnm{Aggarwal}, \binits{K.}},
\oauthor{\bsnm{Baeyens}, \binits{R.}},
\oauthor{\bsnm{Barat}, \binits{S.}},
\oauthor{\bsnm{Borro}, \binits{M.d.V.}},
\oauthor{\bsnm{Daylan}, \binits{T.}},
\oauthor{\bsnm{Fortney}, \binits{J.J.}},
\oauthor{\bsnm{France}, \binits{K.}},
\oauthor{\bsnm{Goyal}, \binits{J.M.}},
\oauthor{\bsnm{Grant}, \binits{D.}},
\oauthor{\bsnm{Kirk}, \binits{J.}},
\oauthor{\bsnm{Kreidberg}, \binits{L.}},
\oauthor{\bsnm{Louca}, \binits{A.}},
\oauthor{\bsnm{Moran}, \binits{S.E.}},
\oauthor{\bsnm{Mukherjee}, \binits{S.}},
\oauthor{\bsnm{Nasedkin}, \binits{E.}},
\oauthor{\bsnm{Ohno}, \binits{K.}},
\oauthor{\bsnm{Rackham}, \binits{B.V.}},
\oauthor{\bsnm{Redfield}, \binits{S.}},
\oauthor{\bsnm{Taylor}, \binits{J.}},
\oauthor{\bsnm{Tremblin}, \binits{P.}},
\oauthor{\bsnm{Visscher}, \binits{C.}},
\oauthor{\bsnm{Wallack}, \binits{N.L.}},
\oauthor{\bsnm{Welbanks}, \binits{L.}},
\oauthor{\bsnm{Youngblood}, \binits{A.}},
\oauthor{\bsnm{Ahrer}, \binits{E.-M.}},
\oauthor{\bsnm{Batalha}, \binits{N.E.}},
\oauthor{\bsnm{Behr}, \binits{P.}},
\oauthor{\bsnm{Berta-Thompson}, \binits{Z.K.}},
\oauthor{\bsnm{Blecic}, \binits{J.}},
\oauthor{\bsnm{Casewell}, \binits{S.L.}},
\oauthor{\bsnm{Crossfield}, \binits{I.J.M.}},
\oauthor{\bsnm{Crouzet}, \binits{N.}},
\oauthor{\bsnm{Cubillos}, \binits{P.E.}},
\oauthor{\bsnm{Decin}, \binits{L.}},
\oauthor{\bsnm{Désert}, \binits{J.-M.}},
\oauthor{\bsnm{Feinstein}, \binits{A.D.}},
\oauthor{\bsnm{Gibson}, \binits{N.P.}},
\oauthor{\bsnm{Harrington}, \binits{J.}},
\oauthor{\bsnm{Heng}, \binits{K.}},
\oauthor{\bsnm{Henning}, \binits{T.}},
\oauthor{\bsnm{Kempton}, \binits{E.M.-R.}},
\oauthor{\bsnm{Krick}, \binits{J.}},
\oauthor{\bsnm{Lagage}, \binits{P.-O.}},
\oauthor{\bsnm{Lendl}, \binits{M.}},
\oauthor{\bsnm{Line}, \binits{M.}},
\oauthor{\bsnm{Lothringer}, \binits{J.D.}},
\oauthor{\bsnm{Mansfield}, \binits{M.}},
\oauthor{\bsnm{Mayne}, \binits{N.J.}},
\oauthor{\bsnm{Mikal-Evans}, \binits{T.}},
\oauthor{\bsnm{Palle}, \binits{E.}},
\oauthor{\bsnm{Schlawin}, \binits{E.}},
\oauthor{\bsnm{Shorttle}, \binits{O.}},
\oauthor{\bsnm{Wheatley}, \binits{P.J.}},
\oauthor{\bsnm{Yurchenko}, \binits{S.N.}}:
Direct Evidence of Photochemistry in an Exoplanet Atmosphere.
arXiv
(2022).
\doiurl{10.48550/ARXIV.2211.10490}.
\url{https://arxiv.org/abs/2211.10490}
\end{botherref}
\endbibitem

\bibitem{Tinetti2018}
\begin{barticle}
\bauthor{\bsnm{Tinetti}, \binits{G.}},
\bauthor{\bsnm{Drossart}, \binits{P.}},
\bauthor{\bsnm{Eccleston}, \binits{P.}},
\bauthor{\bsnm{Hartogh}, \binits{P.}},
\bauthor{\bsnm{Heske}, \binits{A.}},
\bauthor{\bsnm{Leconte}, \binits{J.}},
\bauthor{\bsnm{Micela}, \binits{G.}},
\bauthor{\bsnm{Ollivier}, \binits{M.}},
\bauthor{\bsnm{Pilbratt}, \binits{G.}},
\bauthor{\bsnm{Puig}, \binits{L.}},
\bauthor{\bsnm{Turrini}, \binits{D.}},
\bauthor{\bsnm{Vandenbussche}, \binits{B.}},
\bauthor{\bsnm{Wolkenberg}, \binits{P.}},
\bauthor{\bsnm{Beaulieu}, \binits{J.-P.}},
\bauthor{\bsnm{Buchave}, \binits{L.A.}},
\bauthor{\bsnm{Ferus}, \binits{M.}},
\bauthor{\bsnm{Griffin}, \binits{M.}},
\bauthor{\bsnm{Guedel}, \binits{M.}},
\bauthor{\bsnm{Justtanont}, \binits{K.}},
\bauthor{\bsnm{Lagage}, \binits{P.-O.}},
\bauthor{\bsnm{Machado}, \binits{P.}},
\bauthor{\bsnm{Malaguti}, \binits{G.}},
\bauthor{\bsnm{Min}, \binits{M.}},
\bauthor{\bsnm{N{\o}rgaard-Nielsen}, \binits{H.U.}},
\bauthor{\bsnm{Rataj}, \binits{M.}},
\bauthor{\bsnm{Ray}, \binits{T.}},
\bauthor{\bsnm{Ribas}, \binits{I.}},
\bauthor{\bsnm{Swain}, \binits{M.}},
\bauthor{\bsnm{Szabo}, \binits{R.}},
\bauthor{\bsnm{Werner}, \binits{S.}},
\bauthor{\bsnm{Barstow}, \binits{J.}},
\bauthor{\bsnm{Burleigh}, \binits{M.}},
\bauthor{\bsnm{Cho}, \binits{J.}},
\bauthor{\bparticle{du} \bsnm{Foresto}, \binits{V.C.}},
\bauthor{\bsnm{Coustenis}, \binits{A.}},
\bauthor{\bsnm{Decin}, \binits{L.}},
\bauthor{\bsnm{Encrenaz}, \binits{T.}},
\bauthor{\bsnm{Galand}, \binits{M.}},
\bauthor{\bsnm{Gillon}, \binits{M.}},
\bauthor{\bsnm{Helled}, \binits{R.}},
\bauthor{\bsnm{Morales}, \binits{J.C.}},
\bauthor{\bsnm{Mu{\~{n}}oz}, \binits{A.G.}},
\bauthor{\bsnm{Moneti}, \binits{A.}},
\bauthor{\bsnm{Pagano}, \binits{I.}},
\bauthor{\bsnm{Pascale}, \binits{E.}},
\bauthor{\bsnm{Piccioni}, \binits{G.}},
\bauthor{\bsnm{Pinfield}, \binits{D.}},
\bauthor{\bsnm{Sarkar}, \binits{S.}},
\bauthor{\bsnm{Selsis}, \binits{F.}},
\bauthor{\bsnm{Tennyson}, \binits{J.}},
\bauthor{\bsnm{Triaud}, \binits{A.}},
\bauthor{\bsnm{Venot}, \binits{O.}},
\bauthor{\bsnm{Waldmann}, \binits{I.}},
\bauthor{\bsnm{Waltham}, \binits{D.}},
\bauthor{\bsnm{Wright}, \binits{G.}},
\bauthor{\bsnm{Amiaux}, \binits{J.}},
\bauthor{\bsnm{Augu{\`{e}}res}, \binits{J.-L.}},
\bauthor{\bsnm{Berth{\'{e}}}, \binits{M.}},
\bauthor{\bsnm{Bezawada}, \binits{N.}},
\bauthor{\bsnm{Bishop}, \binits{G.}},
\bauthor{\bsnm{Bowles}, \binits{N.}},
\bauthor{\bsnm{Coffey}, \binits{D.}},
\bauthor{\bsnm{Colom{\'{e}}}, \binits{J.}},
\bauthor{\bsnm{Crook}, \binits{M.}},
\bauthor{\bsnm{Crouzet}, \binits{P.-E.}},
\bauthor{\bsnm{{Da Peppo}}, \binits{V.}},
\bauthor{\bsnm{Sanz}, \binits{I.E.}},
\bauthor{\bsnm{Focardi}, \binits{M.}},
\bauthor{\bsnm{Frericks}, \binits{M.}},
\bauthor{\bsnm{Hunt}, \binits{T.}},
\bauthor{\bsnm{Kohley}, \binits{R.}},
\bauthor{\bsnm{Middleton}, \binits{K.}},
\bauthor{\bsnm{Morgante}, \binits{G.}},
\bauthor{\bsnm{Ottensamer}, \binits{R.}},
\bauthor{\bsnm{Pace}, \binits{E.}},
\bauthor{\bsnm{Pearson}, \binits{C.}},
\bauthor{\bsnm{Stamper}, \binits{R.}},
\bauthor{\bsnm{Symonds}, \binits{K.}},
\bauthor{\bsnm{Rengel}, \binits{M.}},
\bauthor{\bsnm{Renotte}, \binits{E.}},
\bauthor{\bsnm{Ade}, \binits{P.}},
\bauthor{\bsnm{Affer}, \binits{L.}},
\bauthor{\bsnm{Alard}, \binits{C.}},
\bauthor{\bsnm{Allard}, \binits{N.}},
\bauthor{\bsnm{Altieri}, \binits{F.}},
\bauthor{\bsnm{Andr{\'{e}}}, \binits{Y.}},
\bauthor{\bsnm{Arena}, \binits{C.}},
\bauthor{\bsnm{Argyriou}, \binits{I.}},
\bauthor{\bsnm{Aylward}, \binits{A.}},
\bauthor{\bsnm{Baccani}, \binits{C.}},
\bauthor{\bsnm{Bakos}, \binits{G.}},
\bauthor{\bsnm{Banaszkiewicz}, \binits{M.}},
\bauthor{\bsnm{Barlow}, \binits{M.}},
\bauthor{\bsnm{Batista}, \binits{V.}},
\bauthor{\bsnm{Bellucci}, \binits{G.}},
\bauthor{\bsnm{Benatti}, \binits{S.}},
\bauthor{\bsnm{Bernardi}, \binits{P.}},
\bauthor{\bsnm{B{\'{e}}zard}, \binits{B.}},
\bauthor{\bsnm{Blecka}, \binits{M.}},
\bauthor{\bsnm{Bolmont}, \binits{E.}},
\bauthor{\bsnm{Bonfond}, \binits{B.}},
\bauthor{\bsnm{Bonito}, \binits{R.}},
\bauthor{\bsnm{Bonomo}, \binits{A.S.}},
\bauthor{\bsnm{Brucato}, \binits{J.R.}},
\bauthor{\bsnm{Brun}, \binits{A.S.}},
\bauthor{\bsnm{Bryson}, \binits{I.}},
\bauthor{\bsnm{Bujwan}, \binits{W.}},
\bauthor{\bsnm{Casewell}, \binits{S.}},
\bauthor{\bsnm{Charnay}, \binits{B.}},
\bauthor{\bsnm{Pestellini}, \binits{C.C.}},
\bauthor{\bsnm{Chen}, \binits{G.}},
\bauthor{\bsnm{Ciaravella}, \binits{A.}},
\bauthor{\bsnm{Claudi}, \binits{R.}},
\bauthor{\bsnm{Cl{\'{e}}dassou}, \binits{R.}},
\bauthor{\bsnm{Damasso}, \binits{M.}},
\bauthor{\bsnm{Damiano}, \binits{M.}},
\bauthor{\bsnm{Danielski}, \binits{C.}},
\bauthor{\bsnm{Deroo}, \binits{P.}},
\bauthor{\bsnm{{Di Giorgio}}, \binits{A.M.}},
\bauthor{\bsnm{Dominik}, \binits{C.}},
\bauthor{\bsnm{Doublier}, \binits{V.}},
\bauthor{\bsnm{Doyle}, \binits{S.}},
\bauthor{\bsnm{Doyon}, \binits{R.}},
\bauthor{\bsnm{Drummond}, \binits{B.}},
\bauthor{\bsnm{Duong}, \binits{B.}},
\bauthor{\bsnm{Eales}, \binits{S.}},
\bauthor{\bsnm{Edwards}, \binits{B.}},
\bauthor{\bsnm{Farina}, \binits{M.}},
\bauthor{\bsnm{Flaccomio}, \binits{E.}},
\bauthor{\bsnm{Fletcher}, \binits{L.}},
\bauthor{\bsnm{Forget}, \binits{F.}},
\bauthor{\bsnm{Fossey}, \binits{S.}},
\bauthor{\bsnm{Fr{\"{a}}nz}, \binits{M.}},
\bauthor{\bsnm{Fujii}, \binits{Y.}},
\bauthor{\bsnm{Garc{\'{i}}a-Piquer}, \binits{{\'{A}}.}},
\bauthor{\bsnm{Gear}, \binits{W.}},
\bauthor{\bsnm{Geoffray}, \binits{H.}},
\bauthor{\bsnm{G{\'{e}}rard}, \binits{J.C.}},
\bauthor{\bsnm{Gesa}, \binits{L.}},
\bauthor{\bsnm{Gomez}, \binits{H.}},
\bauthor{\bsnm{{Graczyk Rafa{\l}and Griffith}}, \binits{C.}},
\bauthor{\bsnm{Grodent}, \binits{D.}},
\bauthor{\bsnm{Guarcello}, \binits{M.G.}},
\bauthor{\bsnm{Gustin}, \binits{J.}},
\bauthor{\bsnm{Hamano}, \binits{K.}},
\bauthor{\bsnm{Hargrave}, \binits{P.}},
\bauthor{\bsnm{Hello}, \binits{Y.}},
\bauthor{\bsnm{Heng}, \binits{K.}},
\bauthor{\bsnm{Herrero}, \binits{E.}},
\bauthor{\bsnm{Hornstrup}, \binits{A.}},
\bauthor{\bsnm{Hubert}, \binits{B.}},
\bauthor{\bsnm{Ida}, \binits{S.}},
\bauthor{\bsnm{Ikoma}, \binits{M.}},
\bauthor{\bsnm{Iro}, \binits{N.}},
\bauthor{\bsnm{Irwin}, \binits{P.}},
\bauthor{\bsnm{Jarchow}, \binits{C.}},
\bauthor{\bsnm{Jaubert}, \binits{J.}},
\bauthor{\bsnm{Jones}, \binits{H.}},
\bauthor{\bsnm{Julien}, \binits{Q.}},
\bauthor{\bsnm{Kameda}, \binits{S.}},
\bauthor{\bsnm{Kerschbaum}, \binits{F.}},
\bauthor{\bsnm{Kervella}, \binits{P.}},
\bauthor{\bsnm{Koskinen}, \binits{T.}},
\bauthor{\bsnm{Krijger}, \binits{M.}},
\bauthor{\bsnm{Krupp}, \binits{N.}},
\bauthor{\bsnm{Lafarga}, \binits{M.}},
\bauthor{\bsnm{Landini}, \binits{F.}},
\bauthor{\bsnm{Lellouch}, \binits{E.}},
\bauthor{\bsnm{Leto}, \binits{G.}},
\bauthor{\bsnm{Luntzer}, \binits{A.}},
\bauthor{\bsnm{Rank-L{\"{u}}ftinger}, \binits{T.}},
\bauthor{\bsnm{Maggio}, \binits{A.}},
\bauthor{\bsnm{Maldonado}, \binits{J.}},
\bauthor{\bsnm{Maillard}, \binits{J.-P.}},
\bauthor{\bsnm{Mall}, \binits{U.}},
\bauthor{\bsnm{Marquette}, \binits{J.-B.}},
\bauthor{\bsnm{Mathis}, \binits{S.}},
\bauthor{\bsnm{Maxted}, \binits{P.}},
\bauthor{\bsnm{Matsuo}, \binits{T.}},
\bauthor{\bsnm{Medvedev}, \binits{A.}},
\bauthor{\bsnm{Miguel}, \binits{Y.}},
\bauthor{\bsnm{Minier}, \binits{V.}},
\bauthor{\bsnm{Morello}, \binits{G.}},
\bauthor{\bsnm{Mura}, \binits{A.}},
\bauthor{\bsnm{Narita}, \binits{N.}},
\bauthor{\bsnm{Nascimbeni}, \binits{V.}},
\bauthor{\bsnm{{Nguyen Tong}}, \binits{N.}},
\bauthor{\bsnm{Noce}, \binits{V.}},
\bauthor{\bsnm{Oliva}, \binits{F.}},
\bauthor{\bsnm{Palle}, \binits{E.}},
\bauthor{\bsnm{Palmer}, \binits{P.}},
\bauthor{\bsnm{Pancrazzi}, \binits{M.}},
\bauthor{\bsnm{Papageorgiou}, \binits{A.}},
\bauthor{\bsnm{Parmentier}, \binits{V.}},
\bauthor{\bsnm{Perger}, \binits{M.}},
\bauthor{\bsnm{Petralia}, \binits{A.}},
\bauthor{\bsnm{Pezzuto}, \binits{S.}},
\bauthor{\bsnm{Pierrehumbert}, \binits{R.}},
\bauthor{\bsnm{Pillitteri}, \binits{I.}},
\bauthor{\bsnm{Piotto}, \binits{G.}},
\bauthor{\bsnm{Pisano}, \binits{G.}},
\bauthor{\bsnm{Prisinzano}, \binits{L.}},
\bauthor{\bsnm{Radioti}, \binits{A.}},
\bauthor{\bsnm{R{\'{e}}ess}, \binits{J.-M.}},
\bauthor{\bsnm{Rezac}, \binits{L.}},
\bauthor{\bsnm{Rocchetto}, \binits{M.}},
\bauthor{\bsnm{Rosich}, \binits{A.}},
\bauthor{\bsnm{Sanna}, \binits{N.}},
\bauthor{\bsnm{Santerne}, \binits{A.}},
\bauthor{\bsnm{Savini}, \binits{G.}},
\bauthor{\bsnm{Scandariato}, \binits{G.}},
\bauthor{\bsnm{Sicardy}, \binits{B.}},
\bauthor{\bsnm{Sierra}, \binits{C.}},
\bauthor{\bsnm{Sindoni}, \binits{G.}},
\bauthor{\bsnm{Skup}, \binits{K.}},
\bauthor{\bsnm{Snellen}, \binits{I.}},
\bauthor{\bsnm{Sobiecki}, \binits{M.}},
\bauthor{\bsnm{Soret}, \binits{L.}},
\bauthor{\bsnm{Sozzetti}, \binits{A.}},
\bauthor{\bsnm{Stiepen}, \binits{A.}},
\bauthor{\bsnm{Strugarek}, \binits{A.}},
\bauthor{\bsnm{Taylor}, \binits{J.}},
\bauthor{\bsnm{Taylor}, \binits{W.}},
\bauthor{\bsnm{Terenzi}, \binits{L.}},
\bauthor{\bsnm{Tessenyi}, \binits{M.}},
\bauthor{\bsnm{Tsiaras}, \binits{A.}},
\bauthor{\bsnm{Tucker}, \binits{C.}},
\bauthor{\bsnm{Valencia}, \binits{D.}},
\bauthor{\bsnm{Vasisht}, \binits{G.}},
\bauthor{\bsnm{Vazan}, \binits{A.}},
\bauthor{\bsnm{Vilardell}, \binits{F.}},
\bauthor{\bsnm{Vinatier}, \binits{S.}},
\bauthor{\bsnm{Viti}, \binits{S.}},
\bauthor{\bsnm{Waters}, \binits{R.}},
\bauthor{\bsnm{Wawer}, \binits{P.}},
\bauthor{\bsnm{Wawrzaszek}, \binits{A.}},
\bauthor{\bsnm{Whitworth}, \binits{A.}},
\bauthor{\bsnm{Yung}, \binits{Y.L.}},
\bauthor{\bsnm{Yurchenko}, \binits{S.N.}},
\bauthor{\bsnm{Osorio}, \binits{M.R.Z.}},
\bauthor{\bsnm{Zellem}, \binits{R.}},
\bauthor{\bsnm{Zingales}, \binits{T.}},
\bauthor{\bsnm{Zwart}, \binits{F.}}:
\batitle{{A chemical survey of exoplanets with ARIEL}}.
\bjtitle{\expa}
\bvolume{46}(\bissue{1}),
\bfpage{135}--\blpage{209}
(\byear{2018}).
\doiurl{10.1007/s10686-018-9598-x}
\end{barticle}
\endbibitem

\bibitem{Edwards2019}
\begin{barticle}
\bauthor{\bsnm{Edwards}, \binits{B.}},
\bauthor{\bsnm{Mugnai}, \binits{L.}},
\bauthor{\bsnm{Tinetti}, \binits{G.}},
\bauthor{\bsnm{Pascale}, \binits{E.}},
\bauthor{\bsnm{Sarkar}, \binits{S.}}:
\batitle{{An Updated Study of Potential Targets for Ariel}}.
\bjtitle{\aj}
\bvolume{157}(\bissue{6}),
\bfpage{242}
(\byear{2019})
{\href{https://arxiv.org/abs/1905.04959}{{arXiv:1905.04959}}}
{[astro-ph.EP]}.
\doiurl{10.3847/1538-3881/ab1cb9}
\end{barticle}
\endbibitem

\bibitem{Mugnai2021a}
\begin{barticle}
\bauthor{\bsnm{Mugnai}, \binits{L.V.}},
\bauthor{\bsnm{Al-Refaie}, \binits{A.}},
\bauthor{\bsnm{Bocchieri}, \binits{A.}},
\bauthor{\bsnm{Changeat}, \binits{Q.}},
\bauthor{\bsnm{Pascale}, \binits{E.}},
\bauthor{\bsnm{Tinetti}, \binits{G.}}:
\batitle{{Alfnoor: Assessing the Information Content of Ariel's Low-resolution
  Spectra with Planetary Population Studies}}.
\bjtitle{\aj}
\bvolume{162},
\bfpage{288}
(\byear{2021}).
\doiurl{10.3847/1538-3881/ac2e92}
\end{barticle}
\endbibitem

\bibitem{Wall2012}
\begin{bbook}
\bauthor{\bsnm{{Wall}}, \binits{J.V.}},
\bauthor{\bsnm{{Jenkins}}, \binits{C.R.}}:
\bbtitle{Practical Statistics for Astronomers},
(\byear{2012})
\end{bbook}
\endbibitem

\bibitem{Changeat2020b}
\begin{barticle}
\bauthor{\bsnm{Changeat}, \binits{Q.}},
\bauthor{\bsnm{Al-Refaie}, \binits{A.}},
\bauthor{\bsnm{Mugnai}, \binits{L.V.}},
\bauthor{\bsnm{Edwards}, \binits{B.}},
\bauthor{\bsnm{Waldmann}, \binits{I.P.}},
\bauthor{\bsnm{Pascale}, \binits{E.}},
\bauthor{\bsnm{Tinetti}, \binits{G.}}:
\batitle{{Alfnoor: A Retrieval Simulation of the Ariel Target List}}.
\bjtitle{\aj}
\bvolume{160}(\bissue{2}),
\bfpage{80}
(\byear{2020}).
\doiurl{10.3847/1538-3881/ab9a53}
\end{barticle}
\endbibitem

\bibitem{refaie2021}
\begin{barticle}
\bauthor{\bsnm{Al-Refaie}, \binits{A.F.}},
\bauthor{\bsnm{Changeat}, \binits{Q.}},
\bauthor{\bsnm{Waldmann}, \binits{I.P.}},
\bauthor{\bsnm{Tinetti}, \binits{G.}}:
\batitle{{TauREx 3: A Fast, Dynamic, and Extendable Framework for Retrievals}}.
\bjtitle{\apj}
\bvolume{917}(\bissue{1}),
\bfpage{37}
(\byear{2021}).
\doiurl{10.3847/1538-4357/ac0252}
\end{barticle}
\endbibitem

\bibitem{Mugnai2020a}
\begin{barticle}
\bauthor{\bsnm{Mugnai}, \binits{L.V.}},
\bauthor{\bsnm{Pascale}, \binits{E.}},
\bauthor{\bsnm{Edwards}, \binits{B.}},
\bauthor{\bsnm{Papageorgiou}, \binits{A.}},
\bauthor{\bsnm{Sarkar}, \binits{S.}}:
\batitle{{ArielRad: the Ariel radiometric model}}.
\bjtitle{\expa}
\bvolume{50}(\bissue{2-3}),
\bfpage{303}--\blpage{328}
(\byear{2020}).
\doiurl{10.1007/s10686-020-09676-7}
\end{barticle}
\endbibitem

\bibitem{Abel2011}
\begin{barticle}
\bauthor{\bsnm{Abel}, \binits{M.}},
\bauthor{\bsnm{Frommhold}, \binits{L.}},
\bauthor{\bsnm{Li}, \binits{X.}},
\bauthor{\bsnm{Hunt}, \binits{K.L.C.}}:
\batitle{Collision-induced absorption by h2 pairs: From hundreds to thousands
  of kelvin}.
\bjtitle{The Journal of Physical Chemistry A}
\bvolume{115}(\bissue{25}),
\bfpage{6805}--\blpage{6812}
(\byear{2011})
{\href{https://arxiv.org/abs/https://doi.org/10.1021/jp109441f}{{https://doi.org/10.1021/jp109441f}}}.
\doiurl{10.1021/jp109441f}.
\bcomment{PMID: 21207941}
\end{barticle}
\endbibitem

\bibitem{Fletcher2018}
\begin{barticle}
\bauthor{\bsnm{{Fletcher}}, \binits{L.N.}},
\bauthor{\bsnm{{Gustafsson}}, \binits{M.}},
\bauthor{\bsnm{{Orton}}, \binits{G.S.}}:
\batitle{{Hydrogen Dimers in Giant-planet Infrared Spectra}}.
\bjtitle{\apjs}
\bvolume{235}(\bissue{1}),
\bfpage{24}
(\byear{2018})
{\href{https://arxiv.org/abs/1712.02813}{{arXiv:1712.02813}}}
{[astro-ph.EP]}.
\doiurl{10.3847/1538-4365/aaa07a}
\end{barticle}
\endbibitem

\bibitem{Abel2012}
\begin{barticle}
\bauthor{\bsnm{Abel}, \binits{M.}},
\bauthor{\bsnm{Frommhold}, \binits{L.}},
\bauthor{\bsnm{Li}, \binits{X.}},
\bauthor{\bsnm{Hunt}, \binits{K.L.C.}}:
\batitle{Infrared absorption by collisional h2-he complexes at temperatures up
  to 9000 k and frequencies from 0 to 20,000 cm(-1)}.
\bjtitle{The Journal of Chemical Physics}
\bvolume{136}(\bissue{4}),
\bfpage{044319}
(\byear{2012})
{\href{https://arxiv.org/abs/https://doi.org/10.1063/1.3676405}{{https://doi.org/10.1063/1.3676405}}}.
\doiurl{10.1063/1.3676405}
\end{barticle}
\endbibitem

\bibitem{Barton2017}
\begin{barticle}
\bauthor{\bsnm{{Barton}}, \binits{E.J.}},
\bauthor{\bsnm{{Hill}}, \binits{C.}},
\bauthor{\bsnm{{Yurchenko}}, \binits{S.N.}},
\bauthor{\bsnm{{Tennyson}}, \binits{J.}},
\bauthor{\bsnm{{Dudaryonok}}, \binits{A.S.}},
\bauthor{\bsnm{{Lavrentieva}}, \binits{N.N.}}:
\batitle{{Pressure-dependent water absorption cross sections for exoplanets and
  other atmospheres}}.
\bjtitle{\jqsrt}
\bvolume{187},
\bfpage{453}--\blpage{460}
(\byear{2017})
{\href{https://arxiv.org/abs/1610.09008}{{arXiv:1610.09008}}}
{[astro-ph.EP]}.
\doiurl{10.1016/j.jqsrt.2016.10.024}
\end{barticle}
\endbibitem

\bibitem{Polyansky2018}
\begin{barticle}
\bauthor{\bsnm{{Polyansky}}, \binits{O.L.}},
\bauthor{\bsnm{{Kyuberis}}, \binits{A.A.}},
\bauthor{\bsnm{{Zobov}}, \binits{N.F.}},
\bauthor{\bsnm{{Tennyson}}, \binits{J.}},
\bauthor{\bsnm{{Yurchenko}}, \binits{S.N.}},
\bauthor{\bsnm{{Lodi}}, \binits{L.}}:
\batitle{{ExoMol molecular line lists XXX: a complete high-accuracy line list
  for water}}.
\bjtitle{\mnras}
\bvolume{480}(\bissue{2}),
\bfpage{2597}--\blpage{2608}
(\byear{2018})
{\href{https://arxiv.org/abs/1807.04529}{{arXiv:1807.04529}}}
{[astro-ph.EP]}.
\doiurl{10.1093/mnras/sty1877}
\end{barticle}
\endbibitem

\bibitem{Hill2013}
\begin{barticle}
\bauthor{\bsnm{Hill}, \binits{C.}},
\bauthor{\bsnm{Yurchenko}, \binits{S.N.}},
\bauthor{\bsnm{Tennyson}, \binits{J.}}:
\batitle{Temperature-dependent molecular absorption cross sections for
  exoplanets and other atmospheres}.
\bjtitle{\icarus}
\bvolume{226}(\bissue{2}),
\bfpage{1673}--\blpage{1677}
(\byear{2013}).
\doiurl{10.1016/j.icarus.2012.07.028}
\end{barticle}
\endbibitem

\bibitem{Yurchenko2014}
\begin{barticle}
\bauthor{\bsnm{{Yurchenko}}, \binits{S.N.}},
\bauthor{\bsnm{{Tennyson}}, \binits{J.}}:
\batitle{{ExoMol line lists - IV. The rotation-vibration spectrum of methane up
  to 1500 K}}.
\bjtitle{\mnras}
\bvolume{440}(\bissue{2}),
\bfpage{1649}--\blpage{1661}
(\byear{2014})
{\href{https://arxiv.org/abs/1401.4852}{{arXiv:1401.4852}}}
{[astro-ph.EP]}.
\doiurl{10.1093/mnras/stu326}
\end{barticle}
\endbibitem

\bibitem{Rothman2010}
\begin{barticle}
\bauthor{\bsnm{{Rothman}}, \binits{L.S.}},
\bauthor{\bsnm{{Gordon}}, \binits{I.E.}},
\bauthor{\bsnm{{Barber}}, \binits{R.J.}},
\bauthor{\bsnm{{Dothe}}, \binits{H.}},
\bauthor{\bsnm{{Gamache}}, \binits{R.R.}},
\bauthor{\bsnm{{Goldman}}, \binits{A.}},
\bauthor{\bsnm{{Perevalov}}, \binits{V.I.}},
\bauthor{\bsnm{{Tashkun}}, \binits{S.A.}},
\bauthor{\bsnm{{Tennyson}}, \binits{J.}}:
\batitle{{HITEMP, the high-temperature molecular spectroscopic database}}.
\bjtitle{\jqsrt}
\bvolume{111},
\bfpage{2139}--\blpage{2150}
(\byear{2010}).
\doiurl{10.1016/j.jqsrt.2010.05.001}
\end{barticle}
\endbibitem

\bibitem{Yurchenko2011}
\begin{barticle}
\bauthor{\bsnm{Yurchenko}, \binits{S.N.}},
\bauthor{\bsnm{Barber}, \binits{R.J.}},
\bauthor{\bsnm{Tennyson}, \binits{J.}}:
\batitle{{A variationally computed line list for hot NH3}}.
\bjtitle{Monthly Notices of the Royal Astronomical Society}
\bvolume{413}(\bissue{3}),
\bfpage{1828}--\blpage{1834}
(\byear{2011})
{\href{https://arxiv.org/abs/https://academic.oup.com/mnras/article-pdf/413/3/1828/2883744/mnras0413-1828.pdf}{{https://academic.oup.com/mnras/article-pdf/413/3/1828/2883744/mnras0413-1828.pdf}}}.
\doiurl{10.1111/j.1365-2966.2011.18261.x}
\end{barticle}
\endbibitem

\bibitem{Tennyson2012}
\begin{barticle}
\bauthor{\bsnm{Tennyson}, \binits{J.}},
\bauthor{\bsnm{Yurchenko}, \binits{S.N.}}:
\batitle{{ExoMol: molecular line lists for exoplanet and other atmospheres}}.
\bjtitle{Monthly Notices of the Royal Astronomical Society}
\bvolume{425}(\bissue{1}),
\bfpage{21}--\blpage{33}
(\byear{2012})
{\href{https://arxiv.org/abs/https://academic.oup.com/mnras/article-pdf/425/1/21/3182616/425-1-21.pdf}{{https://academic.oup.com/mnras/article-pdf/425/1/21/3182616/425-1-21.pdf}}}.
\doiurl{10.1111/j.1365-2966.2012.21440.x}
\end{barticle}
\endbibitem

\bibitem{Li_2015}
\begin{barticle}
\bauthor{\bsnm{{Li}}, \binits{G.}},
\bauthor{\bsnm{{Gordon}}, \binits{I.E.}},
\bauthor{\bsnm{{Rothman}}, \binits{L.S.}},
\bauthor{\bsnm{{Tan}}, \binits{Y.}},
\bauthor{\bsnm{{Hu}}, \binits{S.-M.}},
\bauthor{\bsnm{{Kassi}}, \binits{S.}},
\bauthor{\bsnm{{Campargue}}, \binits{A.}},
\bauthor{\bsnm{{Medvedev}}, \binits{E.S.}}:
\batitle{{Rovibrational Line Lists for Nine Isotopologues of the CO Molecule in
  the X $^{1}${\ensuremath{\Sigma}}$^{+}$ Ground Electronic State}}.
\bjtitle{\apjs}
\bvolume{216}(\bissue{1}),
\bfpage{15}
(\byear{2015}).
\doiurl{10.1088/0067-0049/216/1/15}
\end{barticle}
\endbibitem

\bibitem{10.1093/mnras/stw1133}
\begin{barticle}
\bauthor{\bsnm{Azzam}, \binits{A.A.A.}},
\bauthor{\bsnm{Tennyson}, \binits{J.}},
\bauthor{\bsnm{Yurchenko}, \binits{S.N.}},
\bauthor{\bsnm{Naumenko}, \binits{O.V.}}:
\batitle{{ExoMol molecular line lists – XVI. The rotation–vibration
  spectrum of hot H2S}}.
\bjtitle{Monthly Notices of the Royal Astronomical Society}
\bvolume{460}(\bissue{4}),
\bfpage{4063}--\blpage{4074}
(\byear{2016})
{\href{https://arxiv.org/abs/https://academic.oup.com/mnras/article-pdf/460/4/4063/13773124/stw1133.pdf}{{https://academic.oup.com/mnras/article-pdf/460/4/4063/13773124/stw1133.pdf}}}.
\doiurl{10.1093/mnras/stw1133}
\end{barticle}
\endbibitem

\bibitem{10.1093/mnras/stt2011}
\begin{barticle}
\bauthor{\bsnm{Barber}, \binits{R.J.}},
\bauthor{\bsnm{Strange}, \binits{J.K.}},
\bauthor{\bsnm{Hill}, \binits{C.}},
\bauthor{\bsnm{Polyansky}, \binits{O.L.}},
\bauthor{\bsnm{Mellau}, \binits{G.C.}},
\bauthor{\bsnm{Yurchenko}, \binits{S.N.}},
\bauthor{\bsnm{Tennyson}, \binits{J.}}:
\batitle{{ExoMol line lists – III. An improved hot rotation-vibration line
  list for HCN and HNC}}.
\bjtitle{Monthly Notices of the Royal Astronomical Society}
\bvolume{437}(\bissue{2}),
\bfpage{1828}--\blpage{1835}
(\byear{2013})
{\href{https://arxiv.org/abs/https://academic.oup.com/mnras/article-pdf/437/2/1828/3885942/stt2011.pdf}{{https://academic.oup.com/mnras/article-pdf/437/2/1828/3885942/stt2011.pdf}}}.
\doiurl{10.1093/mnras/stt2011}
\end{barticle}
\endbibitem

\bibitem{changeat2020d}
\begin{barticle}
\bauthor{\bsnm{{Changeat}}, \binits{Q.}},
\bauthor{\bsnm{{Keyte}}, \binits{L.}},
\bauthor{\bsnm{{Waldmann}}, \binits{I.P.}},
\bauthor{\bsnm{{Tinetti}}, \binits{G.}}:
\batitle{{Impact of Planetary Mass Uncertainties on Exoplanet Atmospheric
  Retrievals}}.
\bjtitle{\apj}
\bvolume{896}(\bissue{2}),
\bfpage{107}
(\byear{2020})
{\href{https://arxiv.org/abs/1908.06305}{{arXiv:1908.06305}}}
{[astro-ph.EP]}.
\doiurl{10.3847/1538-4357/ab8f8b}
\end{barticle}
\endbibitem

\bibitem{feroz2009}
\begin{barticle}
\bauthor{\bsnm{Feroz}, \binits{F.}},
\bauthor{\bsnm{Hobson}, \binits{M.P.}},
\bauthor{\bsnm{Bridges}, \binits{M.}}:
\batitle{{MultiNest: an efficient and robust Bayesian inference tool for
  cosmology and particle physics}}.
\bjtitle{Monthly Notices of the Royal Astronomical Society}
\bvolume{398}(\bissue{4}),
\bfpage{1601}--\blpage{1614}
(\byear{2009})
{\href{https://arxiv.org/abs/https://academic.oup.com/mnras/article-pdf/398/4/1601/3039078/mnras0398-1601.pdf}{{https://academic.oup.com/mnras/article-pdf/398/4/1601/3039078/mnras0398-1601.pdf}}}.
\doiurl{10.1111/j.1365-2966.2009.14548.x}
\end{barticle}
\endbibitem

\bibitem{Buchner2021}
\begin{botherref}
\oauthor{\bsnm{{Buchner}}, \binits{J.}}:
{Nested Sampling Methods}.
arXiv e-prints,
2101--09675
(2021)
{\href{https://arxiv.org/abs/2101.09675}{{arXiv:2101.09675}}}
{[stat.CO]}
\end{botherref}
\endbibitem

\bibitem{kass1995}
\begin{barticle}
\bauthor{\bsnm{Kass}, \binits{R.E.}},
\bauthor{\bsnm{Raftery}, \binits{A.E.}}:
\batitle{Bayes factors}.
\bjtitle{Journal of the American Statistical Association}
\bvolume{90}(\bissue{430}),
\bfpage{773}--\blpage{795}
(\byear{1995})
{\href{https://arxiv.org/abs/https://www.tandfonline.com/doi/pdf/10.1080/01621459.1995.10476572}{{https://www.tandfonline.com/doi/pdf/10.1080/01621459.1995.10476572}}}.
\doiurl{10.1080/01621459.1995.10476572}
\end{barticle}
\endbibitem

\bibitem{Jenkins2011}
\begin{barticle}
\bauthor{\bsnm{Jenkins}, \binits{C.R.}},
\bauthor{\bsnm{Peacock}, \binits{J.A.}}:
\batitle{{The power of Bayesian evidence in astronomy}}.
\bjtitle{Monthly Notices of the Royal Astronomical Society}
\bvolume{413}(\bissue{4}),
\bfpage{2895}--\blpage{2905}
(\byear{2011})
{\href{https://arxiv.org/abs/https://academic.oup.com/mnras/article-pdf/413/4/2895/2882506/mnras0413-2895.pdf}{{https://academic.oup.com/mnras/article-pdf/413/4/2895/2882506/mnras0413-2895.pdf}}}.
\doiurl{10.1111/j.1365-2966.2011.18361.x}
\end{barticle}
\endbibitem

\bibitem{Sanders1963}
\begin{barticle}
\bauthor{\bsnm{Sanders}, \binits{F.}}:
\batitle{{On Subjective Probability Forecasting}}.
\bjtitle{Journal of Applied Meteorology and Climatology}
\bvolume{2}(\bissue{2}),
\bfpage{191}--\blpage{201}
(\byear{1963}).
\doiurl{10.1175/1520-0450(1963)002<0191:OSPF>2.0.CO;2}
\end{barticle}
\endbibitem

\bibitem{WILKS2019369}
\begin{bchapter}
\bauthor{\bsnm{Wilks}, \binits{D.S.}}:
\bctitle{Chapter 9 - forecast verification}.
In: \beditor{\bsnm{Wilks}, \binits{D.S.}} (ed.)
\bbtitle{Statistical Methods in the Atmospheric Sciences (Fourth Edition)},
\bedition{Fourth edition} edn.,
pp. \bfpage{369}--\blpage{483}.
\bpublisher{Elsevier}, \blocation{???}
(\byear{2019}).
\doiurl{10.1016/B978-0-12-815823-4.00009-2}.
\burl{https://www.sciencedirect.com/science/article/pii/B9780128158234000092}
\end{bchapter}
\endbibitem

\bibitem{scikit-learn}
\begin{barticle}
\bauthor{\bsnm{Pedregosa}, \binits{F.}},
\bauthor{\bsnm{Varoquaux}, \binits{G.}},
\bauthor{\bsnm{Gramfort}, \binits{A.}},
\bauthor{\bsnm{Michel}, \binits{V.}},
\bauthor{\bsnm{Thirion}, \binits{B.}},
\bauthor{\bsnm{Grisel}, \binits{O.}},
\bauthor{\bsnm{Blondel}, \binits{M.}},
\bauthor{\bsnm{Prettenhofer}, \binits{P.}},
\bauthor{\bsnm{Weiss}, \binits{R.}},
\bauthor{\bsnm{Dubourg}, \binits{V.}},
\bauthor{\bsnm{Vanderplas}, \binits{J.}},
\bauthor{\bsnm{Passos}, \binits{A.}},
\bauthor{\bsnm{Cournapeau}, \binits{D.}},
\bauthor{\bsnm{Brucher}, \binits{M.}},
\bauthor{\bsnm{Perrot}, \binits{M.}},
\bauthor{\bsnm{Duchesnay}, \binits{E.}}:
\batitle{{Scikit-learn: Machine Learning in Python}}.
\bjtitle{Journal of Machine Learning Research}
\bvolume{12},
\bfpage{2825}--\blpage{2830}
(\byear{2011})
\end{barticle}
\endbibitem

\bibitem{BrierScore1950}
\begin{barticle}
\bauthor{\bsnm{Brier}, \binits{G.W.}}:
\batitle{{Verification of forecasts expressed in terms of probability}}.
\bjtitle{Monthly Weather Review}
\bvolume{78}(\bissue{1}),
\bfpage{1}--\blpage{3}
(\byear{1950}).
\doiurl{10.1175/1520-0493(1950)078<0001:VOFEIT>2.0.CO;2}
\end{barticle}
\endbibitem

\bibitem{Platt1999}
\begin{barticle}
\bauthor{\bsnm{Platt}, \binits{J.}}, \betal:
\batitle{Probabilistic outputs for support vector machines and comparisons to
  regularized likelihood methods}.
\bjtitle{Advances in large margin classifiers}
\bvolume{10}(\bissue{3}),
\bfpage{61}--\blpage{74}
(\byear{1999})
\end{barticle}
\endbibitem

\bibitem{isotonic2002}
\begin{barticle}
\bauthor{\bsnm{Zadrozny}, \binits{B.}},
\bauthor{\bsnm{Elkan}, \binits{C.}}:
\batitle{Transforming classifier scores into accurate multiclass probability
  estimates}.
\bjtitle{Proceedings of the ACM SIGKDD International Conference on Knowledge
  Discovery and Data Mining}
(\byear{2002}).
\doiurl{10.1145/775047.775151}
\end{barticle}
\endbibitem

\bibitem{Niculescu-Mizil2005}
\begin{bchapter}
\bauthor{\bsnm{Niculescu-Mizil}, \binits{A.}},
\bauthor{\bsnm{Caruana}, \binits{R.}}:
\bctitle{{Predicting good probabilities with supervised learning}}.
In: \bbtitle{ICML 2005 - Proceedings of the 22nd International Conference on
  Machine Learning},
pp. \bfpage{625}--\blpage{632}
(\byear{2005}).
\doiurl{10.1145/1102351.1102430}
\end{bchapter}
\endbibitem

\bibitem{Press2007}
\begin{bbook}
\bauthor{\bsnm{{Press}}, \binits{W.H.}},
\bauthor{\bsnm{{Teukolsky}}, \binits{S.A.}},
\bauthor{\bsnm{{Vetterling}}, \binits{W.T.}},
\bauthor{\bsnm{{Flannery}}, \binits{B.P.}}:
\bbtitle{Numerical Recipes in C. The Art of Scientific Computing},
(\byear{1992})
\end{bbook}
\endbibitem

\bibitem{Trotta2008}
\begin{barticle}
\bauthor{\bsnm{{Trotta}}, \binits{R.}}:
\batitle{{Bayes in the sky: Bayesian inference and model selection in
  cosmology}}.
\bjtitle{Contemporary Physics}
\bvolume{49}(\bissue{2}),
\bfpage{71}--\blpage{104}
(\byear{2008})
{\href{https://arxiv.org/abs/0803.4089}{{arXiv:0803.4089}}}
{[astro-ph]}.
\doiurl{10.1080/00107510802066753}
\end{barticle}
\endbibitem

\bibitem{Oreshenko_2017}
\begin{barticle}
\bauthor{\bsnm{Oreshenko}, \binits{M.}},
\bauthor{\bsnm{Lavie}, \binits{B.}},
\bauthor{\bsnm{Grimm}, \binits{S.L.}},
\bauthor{\bsnm{Tsai}, \binits{S.-M.}},
\bauthor{\bsnm{Malik}, \binits{M.}},
\bauthor{\bsnm{Demory}, \binits{B.-O.}},
\bauthor{\bsnm{Mordasini}, \binits{C.}},
\bauthor{\bsnm{Alibert}, \binits{Y.}},
\bauthor{\bsnm{Benz}, \binits{W.}},
\bauthor{\bsnm{Quanz}, \binits{S.P.}},
\bauthor{\bsnm{Trotta}, \binits{R.}},
\bauthor{\bsnm{Heng}, \binits{K.}}:
\batitle{Retrieval analysis of the emission spectrum of {WASP}-12b: Sensitivity
  of outcomes to prior assumptions and implications for formation history}.
\bjtitle{\apj}
\bvolume{847}(\bissue{1}),
\bfpage{3}
(\byear{2017}).
\doiurl{10.3847/2041-8213/aa8acf}
\end{barticle}
\endbibitem

\bibitem{astropy}
\begin{barticle}
\bauthor{\bsnm{{Astropy Collaboration}}},
\bauthor{\bsnm{{Price-Whelan}}, \binits{A.M.}},
\bauthor{\bsnm{{Sip{\H{o}}cz}}, \binits{B.M.}},
\bauthor{\bsnm{{G{\"u}nther}}, \binits{H.M.}},
\bauthor{\bsnm{{Lim}}, \binits{P.L.}},
\bauthor{\bsnm{{Crawford}}, \binits{S.M.}},
\bauthor{\bsnm{{Conseil}}, \binits{S.}},
\bauthor{\bsnm{{Shupe}}, \binits{D.L.}},
\bauthor{\bsnm{{Craig}}, \binits{M.W.}},
\bauthor{\bsnm{{Dencheva}}, \binits{N.}},
\bauthor{\bsnm{{Ginsburg}}, \binits{A.}},
\bauthor{\bsnm{{VanderPlas}}, \binits{J.T.}},
\bauthor{\bsnm{{Bradley}}, \binits{L.D.}},
\bauthor{\bsnm{{P{\'e}rez-Su{\'a}rez}}, \binits{D.}},
\bauthor{\bsnm{{de Val-Borro}}, \binits{M.}},
\bauthor{\bsnm{{Aldcroft}}, \binits{T.L.}},
\bauthor{\bsnm{{Cruz}}, \binits{K.L.}},
\bauthor{\bsnm{{Robitaille}}, \binits{T.P.}},
\bauthor{\bsnm{{Tollerud}}, \binits{E.J.}},
\bauthor{\bsnm{{Ardelean}}, \binits{C.}},
\bauthor{\bsnm{{Babej}}, \binits{T.}},
\bauthor{\bsnm{{Bach}}, \binits{Y.P.}},
\bauthor{\bsnm{{Bachetti}}, \binits{M.}},
\bauthor{\bsnm{{Bakanov}}, \binits{A.V.}},
\bauthor{\bsnm{{Bamford}}, \binits{S.P.}},
\bauthor{\bsnm{{Barentsen}}, \binits{G.}},
\bauthor{\bsnm{{Barmby}}, \binits{P.}},
\bauthor{\bsnm{{Baumbach}}, \binits{A.}},
\bauthor{\bsnm{{Berry}}, \binits{K.L.}},
\bauthor{\bsnm{{Biscani}}, \binits{F.}},
\bauthor{\bsnm{{Boquien}}, \binits{M.}},
\bauthor{\bsnm{{Bostroem}}, \binits{K.A.}},
\bauthor{\bsnm{{Bouma}}, \binits{L.G.}},
\bauthor{\bsnm{{Brammer}}, \binits{G.B.}},
\bauthor{\bsnm{{Bray}}, \binits{E.M.}},
\bauthor{\bsnm{{Breytenbach}}, \binits{H.}},
\bauthor{\bsnm{{Buddelmeijer}}, \binits{H.}},
\bauthor{\bsnm{{Burke}}, \binits{D.J.}},
\bauthor{\bsnm{{Calderone}}, \binits{G.}},
\bauthor{\bsnm{{Cano Rodr{\'\i}guez}}, \binits{J.L.}},
\bauthor{\bsnm{{Cara}}, \binits{M.}},
\bauthor{\bsnm{{Cardoso}}, \binits{J.V.M.}},
\bauthor{\bsnm{{Cheedella}}, \binits{S.}},
\bauthor{\bsnm{{Copin}}, \binits{Y.}},
\bauthor{\bsnm{{Corrales}}, \binits{L.}},
\bauthor{\bsnm{{Crichton}}, \binits{D.}},
\bauthor{\bsnm{{D'Avella}}, \binits{D.}},
\bauthor{\bsnm{{Deil}}, \binits{C.}},
\bauthor{\bsnm{{Depagne}}, \binits{{\'E}.}},
\bauthor{\bsnm{{Dietrich}}, \binits{J.P.}},
\bauthor{\bsnm{{Donath}}, \binits{A.}},
\bauthor{\bsnm{{Droettboom}}, \binits{M.}},
\bauthor{\bsnm{{Earl}}, \binits{N.}},
\bauthor{\bsnm{{Erben}}, \binits{T.}},
\bauthor{\bsnm{{Fabbro}}, \binits{S.}},
\bauthor{\bsnm{{Ferreira}}, \binits{L.A.}},
\bauthor{\bsnm{{Finethy}}, \binits{T.}},
\bauthor{\bsnm{{Fox}}, \binits{R.T.}},
\bauthor{\bsnm{{Garrison}}, \binits{L.H.}},
\bauthor{\bsnm{{Gibbons}}, \binits{S.L.J.}},
\bauthor{\bsnm{{Goldstein}}, \binits{D.A.}},
\bauthor{\bsnm{{Gommers}}, \binits{R.}},
\bauthor{\bsnm{{Greco}}, \binits{J.P.}},
\bauthor{\bsnm{{Greenfield}}, \binits{P.}},
\bauthor{\bsnm{{Groener}}, \binits{A.M.}},
\bauthor{\bsnm{{Grollier}}, \binits{F.}},
\bauthor{\bsnm{{Hagen}}, \binits{A.}},
\bauthor{\bsnm{{Hirst}}, \binits{P.}},
\bauthor{\bsnm{{Homeier}}, \binits{D.}},
\bauthor{\bsnm{{Horton}}, \binits{A.J.}},
\bauthor{\bsnm{{Hosseinzadeh}}, \binits{G.}},
\bauthor{\bsnm{{Hu}}, \binits{L.}},
\bauthor{\bsnm{{Hunkeler}}, \binits{J.S.}},
\bauthor{\bsnm{{Ivezi{\'c}}}, \binits{{\v{Z}}.}},
\bauthor{\bsnm{{Jain}}, \binits{A.}},
\bauthor{\bsnm{{Jenness}}, \binits{T.}},
\bauthor{\bsnm{{Kanarek}}, \binits{G.}},
\bauthor{\bsnm{{Kendrew}}, \binits{S.}},
\bauthor{\bsnm{{Kern}}, \binits{N.S.}},
\bauthor{\bsnm{{Kerzendorf}}, \binits{W.E.}},
\bauthor{\bsnm{{Khvalko}}, \binits{A.}},
\bauthor{\bsnm{{King}}, \binits{J.}},
\bauthor{\bsnm{{Kirkby}}, \binits{D.}},
\bauthor{\bsnm{{Kulkarni}}, \binits{A.M.}},
\bauthor{\bsnm{{Kumar}}, \binits{A.}},
\bauthor{\bsnm{{Lee}}, \binits{A.}},
\bauthor{\bsnm{{Lenz}}, \binits{D.}},
\bauthor{\bsnm{{Littlefair}}, \binits{S.P.}},
\bauthor{\bsnm{{Ma}}, \binits{Z.}},
\bauthor{\bsnm{{Macleod}}, \binits{D.M.}},
\bauthor{\bsnm{{Mastropietro}}, \binits{M.}},
\bauthor{\bsnm{{McCully}}, \binits{C.}},
\bauthor{\bsnm{{Montagnac}}, \binits{S.}},
\bauthor{\bsnm{{Morris}}, \binits{B.M.}},
\bauthor{\bsnm{{Mueller}}, \binits{M.}},
\bauthor{\bsnm{{Mumford}}, \binits{S.J.}},
\bauthor{\bsnm{{Muna}}, \binits{D.}},
\bauthor{\bsnm{{Murphy}}, \binits{N.A.}},
\bauthor{\bsnm{{Nelson}}, \binits{S.}},
\bauthor{\bsnm{{Nguyen}}, \binits{G.H.}},
\bauthor{\bsnm{{Ninan}}, \binits{J.P.}},
\bauthor{\bsnm{{N{\"o}the}}, \binits{M.}},
\bauthor{\bsnm{{Ogaz}}, \binits{S.}},
\bauthor{\bsnm{{Oh}}, \binits{S.}},
\bauthor{\bsnm{{Parejko}}, \binits{J.K.}},
\bauthor{\bsnm{{Parley}}, \binits{N.}},
\bauthor{\bsnm{{Pascual}}, \binits{S.}},
\bauthor{\bsnm{{Patil}}, \binits{R.}},
\bauthor{\bsnm{{Patil}}, \binits{A.A.}},
\bauthor{\bsnm{{Plunkett}}, \binits{A.L.}},
\bauthor{\bsnm{{Prochaska}}, \binits{J.X.}},
\bauthor{\bsnm{{Rastogi}}, \binits{T.}},
\bauthor{\bsnm{{Reddy Janga}}, \binits{V.}},
\bauthor{\bsnm{{Sabater}}, \binits{J.}},
\bauthor{\bsnm{{Sakurikar}}, \binits{P.}},
\bauthor{\bsnm{{Seifert}}, \binits{M.}},
\bauthor{\bsnm{{Sherbert}}, \binits{L.E.}},
\bauthor{\bsnm{{Sherwood-Taylor}}, \binits{H.}},
\bauthor{\bsnm{{Shih}}, \binits{A.Y.}},
\bauthor{\bsnm{{Sick}}, \binits{J.}},
\bauthor{\bsnm{{Silbiger}}, \binits{M.T.}},
\bauthor{\bsnm{{Singanamalla}}, \binits{S.}},
\bauthor{\bsnm{{Singer}}, \binits{L.P.}},
\bauthor{\bsnm{{Sladen}}, \binits{P.H.}},
\bauthor{\bsnm{{Sooley}}, \binits{K.A.}},
\bauthor{\bsnm{{Sornarajah}}, \binits{S.}},
\bauthor{\bsnm{{Streicher}}, \binits{O.}},
\bauthor{\bsnm{{Teuben}}, \binits{P.}},
\bauthor{\bsnm{{Thomas}}, \binits{S.W.}},
\bauthor{\bsnm{{Tremblay}}, \binits{G.R.}},
\bauthor{\bsnm{{Turner}}, \binits{J.E.H.}},
\bauthor{\bsnm{{Terr{\'o}n}}, \binits{V.}},
\bauthor{\bsnm{{van Kerkwijk}}, \binits{M.H.}},
\bauthor{\bsnm{{de la Vega}}, \binits{A.}},
\bauthor{\bsnm{{Watkins}}, \binits{L.L.}},
\bauthor{\bsnm{{Weaver}}, \binits{B.A.}},
\bauthor{\bsnm{{Whitmore}}, \binits{J.B.}},
\bauthor{\bsnm{{Woillez}}, \binits{J.}},
\bauthor{\bsnm{{Zabalza}}, \binits{V.}},
\bauthor{\bsnm{{Astropy Contributors}}}:
\batitle{{The Astropy Project: Building an Open-science Project and Status of
  the v2.0 Core Package}}.
\bjtitle{\aj}
\bvolume{156}(\bissue{3}),
\bfpage{123}
(\byear{2018})
{\href{https://arxiv.org/abs/1801.02634}{{arXiv:1801.02634}}}
{[astro-ph.IM]}.
\doiurl{10.3847/1538-3881/aabc4f}
\end{barticle}
\endbibitem

\bibitem{hdf5_collette}
\begin{botherref}
\oauthor{\bsnm{{Collette}}, \binits{A.}},
\oauthor{\bsnm{{Caswell}}, \binits{T.A.}},
\oauthor{\bsnm{{Tocknell}}, \binits{J.}},
\oauthor{\bsnm{{Kluyver}}, \binits{T.}},
\oauthor{\bsnm{{Dale}}, \binits{D.}},
\oauthor{\bsnm{{Scopatz}}, \binits{A.}},
\oauthor{\bsnm{{Jelenak}}, \binits{A.}},
\oauthor{\bsnm{{Valls}}, \binits{V.}},
\oauthor{\bsnm{{Kofoed Pedersen}}, \binits{U.}},
\oauthor{\bsnm{{Raspaud}}, \binits{M.}},
\oauthor{\bsnm{{jakirkham}}},
\oauthor{\bsnm{{Parsons}}, \binits{A.}},
\oauthor{\bsnm{{jialin}}},
\oauthor{\bsnm{{Chan}}, \binits{L.}},
\oauthor{\bsnm{{Paramonov}}, \binits{A.}},
\oauthor{\bsnm{{Hole}}, \binits{L.}},
\oauthor{\bsnm{{Feng}}, \binits{Y.}},
\oauthor{\bsnm{{Johnson}}, \binits{S.R.}},
\oauthor{\bsnm{{Brucher}}, \binits{M.}},
\oauthor{\bsnm{{Teichmann}}, \binits{M.}},
\oauthor{\bsnm{{Vaillant}}, \binits{G.A.}},
\oauthor{\bsnm{{de Buyl}}, \binits{P.}},
\oauthor{\bsnm{{Hinsen}}, \binits{K.}},
\oauthor{\bsnm{{Huebl}}, \binits{A.}},
\oauthor{\bsnm{{VINCENT}}, \binits{T.}},
\oauthor{\bsnm{{Dietz}}, \binits{M.}},
\oauthor{\bsnm{{Rathgeber}}, \binits{F.}},
\oauthor{\bsnm{{Billington}}, \binits{C.}},
\oauthor{\bsnm{{Kieffer}}, \binits{J.}},
\oauthor{\bsnm{{Wright}}, \binits{G.}}:
{h5py/h5py: 2.10.0}.
Zenodo
(2019).
\doiurl{10.5281/zenodo.3401726}
\end{botherref}
\endbibitem

\bibitem{Hunter_matplotlib}
\begin{barticle}
\bauthor{\bsnm{{Hunter}}, \binits{J.D.}}:
\batitle{{Matplotlib: A 2D Graphics Environment}}.
\bjtitle{Computing in Science and Engineering}
\bvolume{9}(\bissue{3}),
\bfpage{90}--\blpage{95}
(\byear{2007}).
\doiurl{10.1109/MCSE.2007.55}
\end{barticle}
\endbibitem

\bibitem{oliphant_numpy}
\begin{barticle}
\bauthor{\bsnm{Harris}, \binits{C.R.}},
\bauthor{\bsnm{Millman}, \binits{K.J.}},
\bauthor{\bparticle{van~der} \bsnm{Walt}, \binits{S.J.}},
\bauthor{\bsnm{Gommers}, \binits{R.}},
\bauthor{\bsnm{Virtanen}, \binits{P.}},
\bauthor{\bsnm{Cournapeau}, \binits{D.}},
\bauthor{\bsnm{Wieser}, \binits{E.}},
\bauthor{\bsnm{Taylor}, \binits{J.}},
\bauthor{\bsnm{Berg}, \binits{S.}},
\bauthor{\bsnm{Smith}, \binits{N.J.}},
\bauthor{\bsnm{Kern}, \binits{R.}},
\bauthor{\bsnm{Picus}, \binits{M.}},
\bauthor{\bsnm{Hoyer}, \binits{S.}},
\bauthor{\bparticle{van} \bsnm{Kerkwijk}, \binits{M.H.}},
\bauthor{\bsnm{Brett}, \binits{M.}},
\bauthor{\bsnm{Haldane}, \binits{A.}},
\bauthor{\bparticle{del} \bsnm{R{\'{i}}o}, \binits{J.F.}},
\bauthor{\bsnm{Wiebe}, \binits{M.}},
\bauthor{\bsnm{Peterson}, \binits{P.}},
\bauthor{\bsnm{G{\'{e}}rard-Marchant}, \binits{P.}},
\bauthor{\bsnm{Sheppard}, \binits{K.}},
\bauthor{\bsnm{Reddy}, \binits{T.}},
\bauthor{\bsnm{Weckesser}, \binits{W.}},
\bauthor{\bsnm{Abbasi}, \binits{H.}},
\bauthor{\bsnm{Gohlke}, \binits{C.}},
\bauthor{\bsnm{Oliphant}, \binits{T.E.}}:
\batitle{Array programming with {NumPy}}.
\bjtitle{Nature}
\bvolume{585}(\bissue{7825}),
\bfpage{357}--\blpage{362}
(\byear{2020}).
\doiurl{10.1038/s41586-020-2649-2}
\end{barticle}
\endbibitem

\end{thebibliography}

\end{document}